# Unraveling multi-state molecular dynamics in single-molecule FRET experiments

# Part I: Theory of FRET-Lines

Anders Barth[1,a,*], Oleg Opanasyuk[1,*], Thomas-Otavio Peulen[1,b,*], Suren Felekyan[1], Stanislav Kalinin[1], Hugo Sanabria[2‡], Claus A.M. Seidel[1,‡]

[1] Institut für Physikalische Chemie, Lehrstuhl für Molekulare Physikalische Chemie, Heinrich-Heine-Universität, Düsseldorf, Germany

[2] Department of Physics and Astronomy, Clemson University, Clemson, S.C., USA

[a] Present address: Department of Bionanoscience, Kavli Institute of Nanoscience, Delft University of Technology, Delft, The Netherlands

[b] Present address: Department of Bioengineering and Therapeutic Sciences, University of California, San Francisco, California, USA

*Contributed equally

[‡]Corresponding authors: cseidel@hhu.de, hsanabr@clemson.edu

**Abstract**

Conformational dynamics of biomolecules are of fundamental importance for their function. Single-molecule Förster Resonance Energy Transfer (smFRET) is a powerful approach to inform on the structure and the dynamics of labeled molecules. If the dynamics occur on the sub-millisecond timescale, capturing and quantifying conformational dynamics can be challenging by intensity-based smFRET. Multiparameter fluorescence detection (MFD) addresses this challenge by simultaneously registering intensities and fluorescence lifetimes. Together, the mean donor fluorescence lifetime and the fluorescence intensities inform on the variance, and the mean FRET efficiency tells the conformational dynamics. Here, we present a general framework that relates average fluorescence lifetimes and intensities in smFRET counting histograms. Using this framework, we show how to compute parametric relations (FRET-lines) of these observables that facilitate a graphical interpretation of experimental data, can be used to test models, identify conformational states, resolve the connectivity of states, and can be applied to unstructured systems to infer properties of polymer chains or study fast protein folding. To simplify the graphical analysis of complex kinetic networks, we derive a moment-based representation of the experimental data and show how to decouple the motion of the fluorescence labels from the conformational dynamics of the biomolecule.

**Table of Contents**

# 1 Introduction

Many experimental techniques provide information on biomolecular structural heterogeneity and can be utilized to resolve ensembles of structures through integrative modeling[1]. However, few techniques simultaneously inform on structure and dynamics from picoseconds to seconds and offer the option for live-cell and *in vivo* measurements. Current advanced fluorescence spectroscopy has a broad dynamic range and can inform on local motions (femtosecond to nanosecond timescales), chain dynamics in disordered systems (nano- to microsecond), and large-scale conformational changes (milliseconds to seconds)[2-4], and can be applied to a variety of *in vitro*, in live cells[5-8], and *in vivo* samples[9]. Thus, there is considerable interest to exploit fluorescence spectroscopic information for integrative modeling of biological processes[3, 10].

A typical fluorescence spectroscopic modality is single-molecule Förster resonance energy transfer. smFRET opened the possibility to interrogate structures directly and conformational dynamics of individual fluorescently-labeled biomolecules by the distance-dependent dipolar coupling of fluorophores[11-15], provided mechanistic insights in diverse areas of biological research and could pave the way towards dynamic structural biology[2]. Examples of biomolecular processes studied by smFRET are folding and unfolding transitions[16-19], dynamics of intrinsically disordered proteins[20-24], conformational dynamics of nucleic acids[25-28] proteins[29-33], multidomain structural rearrangements[34, 35], and membrane receptors[36, 37]. The need for accurate and precise distance information for integrative modeling motivated a previous inter-laboratory benchmark study[38] and the current effort of the smFRET community to establish standards for accurate processing of smFRET data[39]. Still, as will be exemplified in this manuscript, conformational dynamics of multi-state systems with fast exchange kinetics can be overlooked. Thus, we generalize our previous approach that jointly interprets different spectroscopic observables to detect conformational dynamics[40] to a general framework to highlight conformational dynamics and facilitate the interpretation of smFRET data of dynamic multi-state systems.

In smFRET, a broad range of fluorescence spectroscopic observables such as absorption and emission spectra[41], brightness and quantum yields[42-44], fluorescence lifetimes[45, 46] and anisotropies[47, 48] can be registered. However, the most used quantifier for FRET is the FRET efficiency, $E$, which is usually estimated by average fluorescence intensities. The FRET efficiency is the yield of the FRET process, i.e., the fraction of excited donor molecules that transfer energy to an acceptor molecule due to dipolar coupling. Besides intensities, fluorescence spectroscopy offers the anisotropy and the time-evolution information as quantifiers for FRET[11, 49-51]. Here, we provide a framework, which is simple to use that combines information from fluorescence intensities and time-resolved observables. While we focus on revealing and interpreting conformational dynamics in smFRET experiments, our framework can be applied to all FRET experiments where intensity and time-resolved information are registered simultaneously, such as fluorescence lifetime imaging (FLIM).

smFRET experiments are either performed on freely diffusing molecules or molecules tethered to surfaces. In experiments on freely diffusing molecules, the molecules are excited and detected by confocal optics with point detectors[52]. In experiments on surface-immobilized molecules, the molecules are typically excited by total internal reflection fluorescence (TIRF) and detected by cameras[53]. The readout time limits the time resolution in camera-based detection to ~1-10 ms[54]. Point detectors have a higher time-resolution paired with time-correlated single-photon counting (TCSPC) have a picosecond timing precision that enables accurate measurements of the fluorescence lifetimes. Fluorescence spectroscopy provides multidimensional observables that can be registered in parallel. A parallel spectral-, polarized-, and time-resolved registration of photons is called MFD (multiparameter fluorescence detection). Simultaneous registration of multiple fluorescence parameters by MFD has been widely applied to study the conformational dynamics of biomolecules in our and other groups[20, 52, 55-60].

Due to its time-resolution, confocal detection is particularly well-suited to study fast biomolecular dynamics. Various approaches have been developed to reveal and quantify dynamics by analyzing fluorescence intensities in confocal smFRET experiments. Different maximum likelihood approaches take advantage of the color information and the arrival time of single photons to determine kinetic rates from the unprocessed photon streams[61, 62]. An analysis of FRET efficiency histograms (FEH) of single molecules reveals and informs on single-molecule kinetics. By variation of the integration time, dynamics are identified by changes of the FEH shapes[40, 63, 64]. FEHs can be described by a combination of Gaussian distributions to reveals kinetic rate constants[65]. For more accurate analysis, the shot-noise in FEHs is explicitly accounted for in (dynamic) photon distribution analysis (PDA)[40, 66]. Alternatively, variance analysis of the FRET efficiencies of single molecules reveals heterogeneities, e.g., by comparing the average photon arrival times in the donor and FRET channels[58, 67]. In burst variance analysis (BVA), the variance of the FRET efficiency is estimated, and dynamics are detected if the variance exceeds the shot-noise limit[68]. The two-channel kernel density estimator method (FRET-2CDE filter) applies a similar approach to detect anticorrelated fluctuations of the donor and acceptor signal[69]. Finally, very fast conformational dynamics on the sub-millisecond timescale can be determined by fluorescence correlation spectroscopy[15, 70], where the donor and FRET-sensitized acceptor fluctuations signal result in a characteristic anti-correlation in the cross-correlation function[15, 33]. For robust estimation of the timescales of exchange, the contrast in FCS can be amplified in filtered-FCS by statistical filters that use spectral, lifetime, and anisotropy information registered in MFD experiments[71, 72].

Even though various analysis methods have been developed for intensity-based FRET experiments, interpreting the data of systems with fast kinetics remains challenging. Here, kinetics is considered fast if the associated exchange of states happens on a timescale comparable to or faster than the observation time (Figure 1A). In confocal experiments, the upper limit of the observation time is set by the diffusion time of a molecule in the confocal volume. The photon detection rate determines the lower limit of the observation time. In a typical confocal smFRET experiment, usually less than 500 photons are detected per single molecule in an observation time of a few milliseconds. For each molecule, the FRET efficiency, $E$, is calculated from the integrated fluorescence intensities. As most a few hundred photons are registered, only an average FRET efficiency, $E$, can be estimated reliably for each molecule, and the kinetic information is partially lost (Figure 1A).

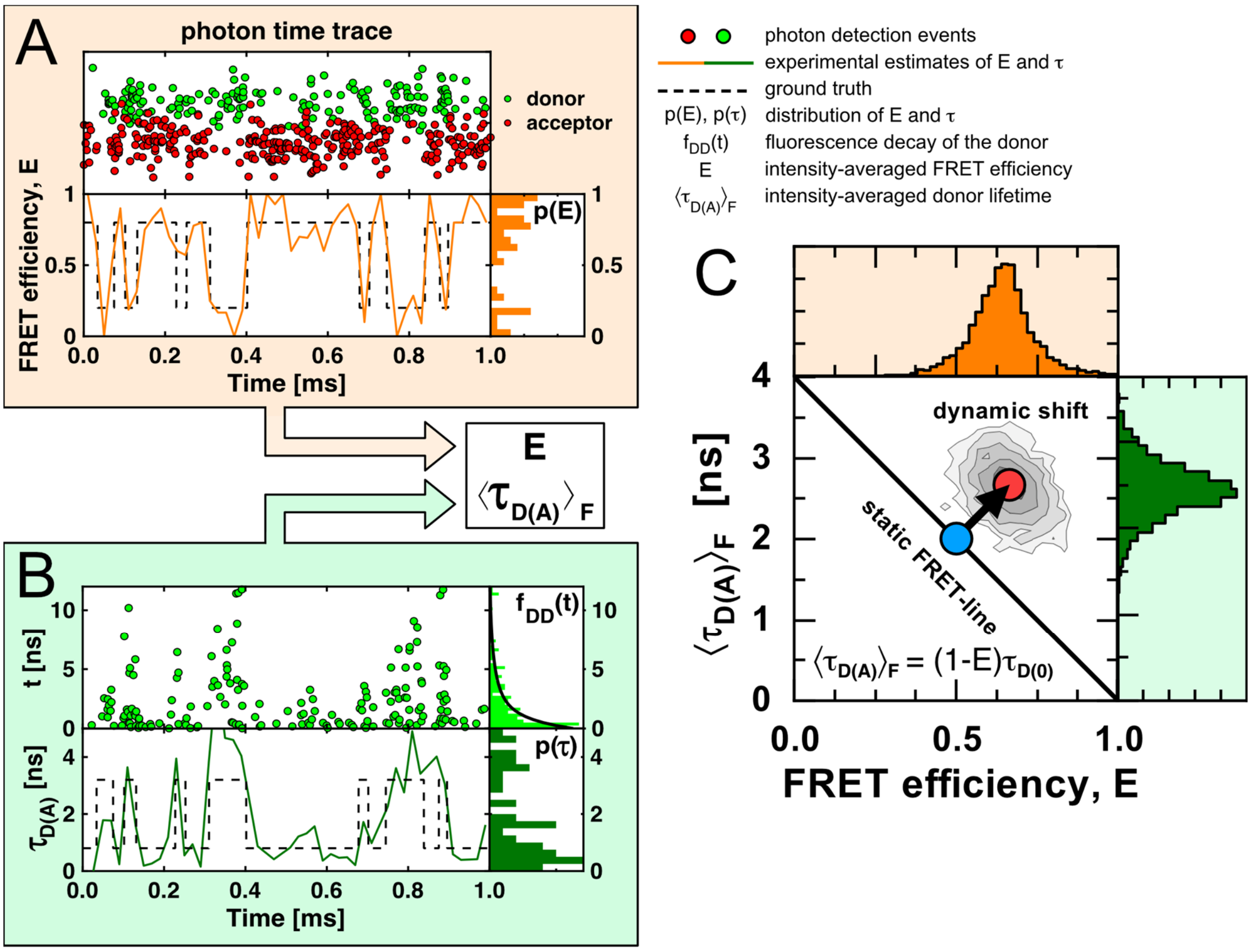

**Figure 1:** Identifying conformational dynamics and heterogeneities in single-molecule FRET. **A**) Simulated single-molecule interconverting between states with different FRET efficiencies (Dashed line). The molecule emits red and green photons (circles) registered in the donor and acceptor (FRET) channel (top). An analysis of the photons yields an estimate of the FRET efficiency (orange line) with the corresponding distribution of average FRET efficiencies visualized as a histogram to the right. **B**) Time-correlated single-photon counting additionally measures the time $\boldsymbol{t}$ since the excitation pulse for each photon (top left), from which the fluorescence lifetime of the donor, $\boldsymbol{\tau_{D(A)}}$, is estimated (bottom left). In practice, only the intensity weighted average fluorescence lifetime $\langle\boldsymbol{\tau_{D(A)}}\rangle_{\boldsymbol{F}}$ can be estimated from the histogram of delay times (top right). The time trace of the fluorescence lifetimes (green line, bottom) and the distribution of fluorescence lifetimes (bottom right) are not accessible. **C**) Single-molecule histogram of $\boldsymbol{E}$ and $\langle\boldsymbol{\tau_{D(A)}}\rangle_{\boldsymbol{F}}$. A shift from the static FRET-line defined by $\langle\boldsymbol{\tau_{D(A)}}\rangle_{\boldsymbol{F}} = (\mathbf{1}-\boldsymbol{E})\boldsymbol{\tau_{D(0)}}$ highlights heterogeneities in the FRET efficiency and indicates conformational dynamics.

In experiments with time-correlated single-photon counting in addition to the fluorescence intensity, the delay time $t$ since the excitation pulse is recorded for each registered photon. The average delay time $\langle t\rangle$ relates to the fluorescence lifetime $\tau_{D(A)}$. Significantly, $\langle t\rangle$ corresponds to the intensity weighted average fluorescence lifetime $\langle\tau_{D(A)}\rangle_F$. The fluorescence lifetime of the donor in the presence of FRET fluctuates with the FRET efficiency (Figure 1B). For a donor dye with a mono-exponential fluorescence decay and a fixed distance between the dyes, the quantities $E$, $\tau_{D(A)}$ and $\langle\tau_{D(A)}\rangle_F$ are related by:

$$E = 1 - \frac{\tau_{D(A)}}{\tau_{D(0)}}; \qquad \langle t\rangle = \langle\tau_{D(A)}\rangle_F = \tau_{D(A)}, \tag{1}$$

where $\tau_{D(0)}$ is the fluorescence lifetime of the donor in the absence of FRET. In this case, the two observables $E$ and $\langle\tau_{D(A)}\rangle_F$ follow the linear dependence: $\langle\tau_{D(A)}\rangle_F = \tau_{D(0)}(1-E)$. We call the reference line described by this relation the ideal static FRET-line, as it is valid for single molecules and ensembles with a single FRET efficiency.

When the molecule switches between different conformational states with different FRET efficiencies during the observation time, only average quantities can be estimated robustly due to the limited number of photons[73]. In this case, the FRET efficiency relates to the species average of lifetimes $\langle\tau_{D(A)}\rangle_x$. On the other hand, the intensity-weighted average fluorescence lifetime $\langle\tau_{D(A)}\rangle_F$ is determined by the donor intensity and species with a smaller FRET efficiency contribute more to the donor fluorescence decay. Therefore, the estimated average lifetime, $\langle t\rangle = \langle\tau_{D(A)}\rangle_F$, is biased towards longer fluorescence lifetimes compared to the species average $\langle\tau_{D(A)}\rangle_x$ (Figure 1B).

$$E = 1 - \frac{\langle\tau_{D(A)}\rangle_x}{\tau_{D(0)}}; \qquad \langle t\rangle = \langle\tau_{D(A)}\rangle_F > \langle\tau_{D(A)}\rangle_x. \tag{2}$$

Because $E$ and $\langle\tau_{D(A)}\rangle_F$ correspond to different averages, the pair of experimental observables $(E, \langle\tau_{D(A)}\rangle_F)$ reveals sample dynamics and heterogeneities through a deviation from the ideal behavior. In single-molecule counting histograms of $(E, \langle\tau_{D(A)}\rangle_F)$, heterogeneities are identified by a shift of populations from the reference static FRET-line (Figure 1C).

The detection and the interpretation of histograms that combine lifetime and intensity information require reference lines. There are many ways to compute other reference lines that relate a FRET efficiency, $E$, to an average fluorescence weighted lifetime, $\langle\tau_{D(A)}\rangle_F$ . We call all parametric relations between the FRET observables a "FRET-lines". FRET-lines can serve as valuable guides to interpret experimental distributions, because they relate model parameters to experimental observables. A FRET-line can help identify dynamic populations and help to understand the dynamic exchange in complex kinetic networks encoded as an experimental fluorescence fingerprint.

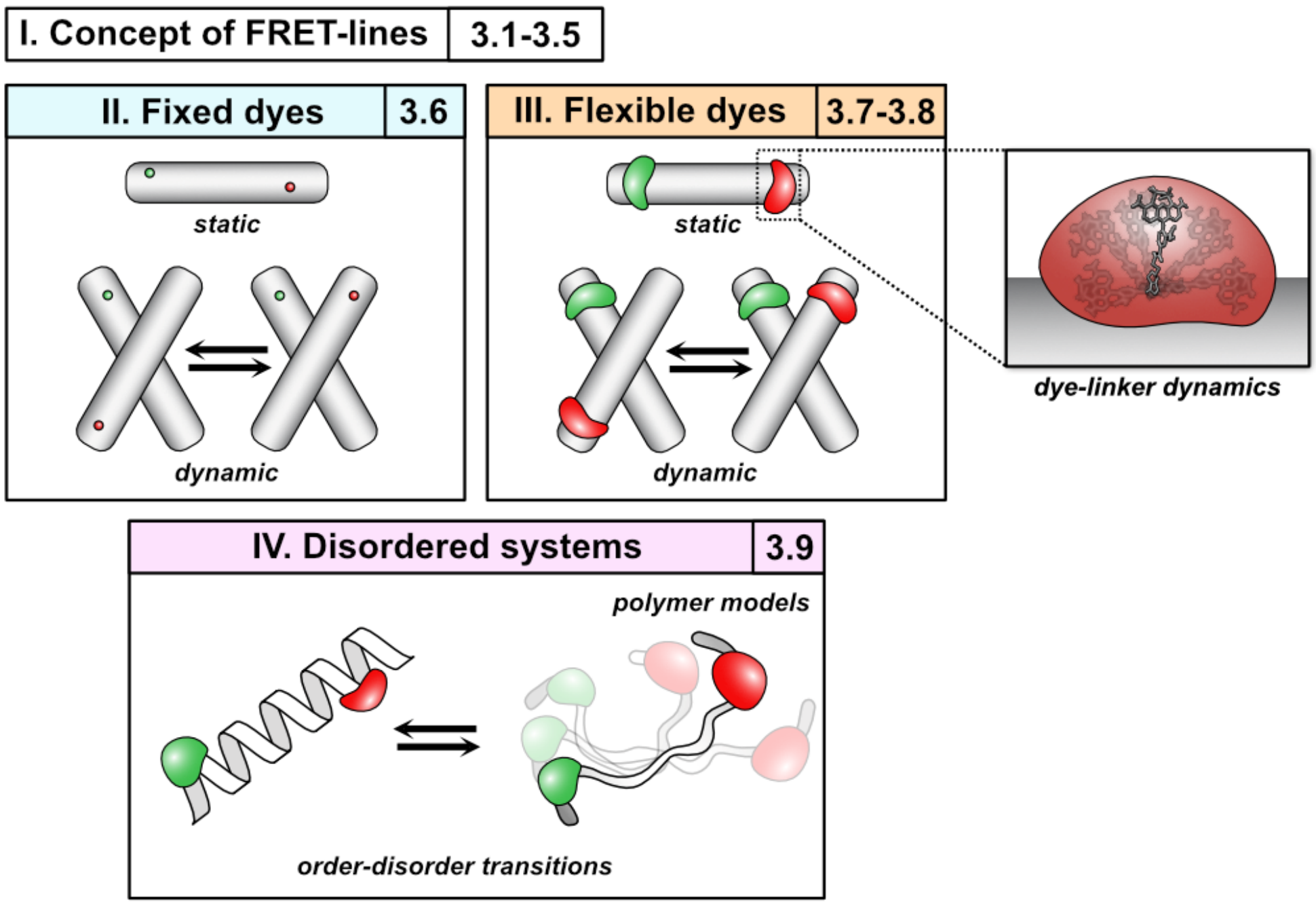

**Figure 2:** Overview of described systems. The static and dynamic FRET-lines for dyes whose position is fixed are discussed in sections 3.1-3.5. The theory is then extended to covalently coupled dyes with long (~20 Å) linkers in sections 3.6-3.7. In section 3.8, FRET-lines for disordered systems are derived using standard polymer models, and order-disorder transitions (e.g., between folded and unfolded peptide chains) are discussed.

To interpret single-molecule histograms computed using average intensities and lifetimes, we introduce the average observables and relate them to conformational heterogeneities. We give a detailed description of how to compute reference FRET-lines and give application examples (Figure 2, Concepts). Using a simple two-state system, we describe how model parameters can be recovered from the FRET-lines. Next, we present the definition of the FRET-lines and provide a rigorous framework for their calculation. We present transformations that can be applied to experimental data that directly visualize conformational heterogeneity and can be used to resolve the species population of exchanging states and generalize the concepts presented for two-state systems to multi-state systems (Figure 2, Concepts). The second part of this manuscript assembles the most relevant equations needed to interpret data of static and dynamic system for dyes that are fixed stiffly to the molecule of interest (Figure 2, Fixed dyes), and presents conformational heterogeneity caused by flexibly coupling dyes is accounted for (Figure 2, Flexible dyes). Finally, we present FRET-lines that can inform on an order-disorder transition (Figure 2, Disordered systems).

# 2 Theory

## 2.1 Förster resonance energy transfer

FRET is the non-radiative energy transfer from an excited donor (*D*) to an acceptor (*A*) fluorophore by dipolar coupling that depends strongly on the interdye distance $R_{DA}$. The rate constant of the energy transfer from $D$ to $A$, $k_{RET}$, depends on the distance between the donor and the acceptor transition dipole moments[11]:

$$k_{RET} = \frac{k_{F,D}}{\Phi_{F,D}}\left(\frac{R_0}{R_{DA}}\right)^6 = \frac{1}{\tau_{D(0)}}\left(\frac{R_0}{R_{DA}}\right)^6. \quad (3)$$

Above, $k_{F,D}$ is the radiative rate constant of the donor, $\Phi_{F,D}$ is the fluorescence quantum yield of the donor, $R_0$ is the dye-pair specific Förster radius, and $R_{DA}$ is the DA-distance. The Förster radius, $R_0$, depends on the mutual orientation of the fluorophore dipoles, captured by the orientation factor $\kappa^2$. Moreover, $R_0$ depends on the spectral overlap integral $J(\lambda)$, the refractive index of the medium, $n$, and $\Phi_{F,D}$, the quantum yield of donor fluorophore:

$$R_0^6 = \frac{9(\ln 10)}{128\pi^5 \cdot N_A} \cdot \frac{\kappa^2 \Phi_{F,D} J(\lambda)}{n^4}. \quad (4)$$

Here, $N_A$ is Avogadro's constant. The spectral overlap integral is defined by $J(\lambda) = \int f_D(\lambda)\varepsilon_A(\lambda)\lambda^4 d\lambda$, where $f_D(\lambda)$ is the normalized emission spectrum of the donor and $\varepsilon_A(\lambda)$ is the extinction of the acceptor at wavelength $\lambda$. For simplicity, we focus on dyes that reorient fast compared to the FRET rate constant. For such a case, $\kappa^2$ can be approximated by the isotropic average, $\langle\kappa^2\rangle_{\text{iso}} = 2/3$. This approximation applies to free rotating dyes that are flexibly coupled to biomolecules via long linkers[34, 38, 74, 75].

The rate constant of FRET and the resulting fluorescence lifetimes relate to its FRET efficiency, $E$, by

$$E = \frac{k_{RET}}{k_{RET} + k_{F,D} + \sum_j k_Q^{(j)}} = 1 - \frac{\tau_{D(A)}}{\tau_{D(0)}}. \quad (5)$$

Here, $\sum_j k_Q^{(j)}$ is the sum over the rate constants of all additional non-radiative pathways depopulating the excited state of the donor, and $\tau_{D(0)}$ and $\tau_{D(A)}$ are the donor fluorescence lifetimes in the absence and presence of the acceptor. The FRET efficiency is related to the interdye distance, $R_{DA}$, by[11]:

$$E = \frac{1}{1 + \left(\frac{R_{DA}}{R_0}\right)^6}. \quad (6)$$

Thus, the fluorescence lifetime of the donor in the presence of the acceptor is related to the FRET efficiency and interdye distance by:

$$\tau_{D(A)} = (1 - E)\tau_{D(0)} = \frac{\tau_{D(0)}}{1 + \left(\frac{R_0}{R_{DA}}\right)^6}. \quad (7)$$

## 2.2 Time-resolved fluorescence

In single-molecule FRET experiments with pulsed excitation, the detected photons are also characterized by their delay time with respect to the excitation pulse. The distribution of delay times $t$ of photons emitted by a donor with a fluorescence lifetime $\tau_{D(0)}$ that is quenched by a FRET rate constant, $k_{FRET}$, follows an exponential decay:

$$f_{D|D}^{(DA)}(t) = k_{F,D}\mathrm{e}^{-t/\tau_{D(A)}} \text{ with } \tau_{D(A)} = 1/(\tau_{D(0)}^{-1} + k_{FRET}). \quad (8)$$

where $\tau_{D(A)}$ is the fluorescence lifetime of the donor in the presence of FRET. If the radiative rate constant, $k_{F,D}$, is independent of the FRET rate, the fluorescence decay of a mixture of species with different fluorescence lifetimes with the lifetime distribution $p(\tau_{D(A)})$ is given by:

$$f_{D|D}^{(DA)}(t) = k_{F,D}\int p(\tau_{D(A)})\mathrm{e}^{-\frac{t}{\tau_{D(A)}}}d\tau_{D(A)}, \quad (9)$$

where $f_{D|D}^{(DA)}(t)$ corresponds to the Laplace transform of the distribution of fluorescence lifetimes. Equation (9) can also be expressed in terms of the interdye distance, $R_{DA}$, directly as:

$$f_{D|D}^{(DA)}(t) = k_{F,D}\int p(R_{DA})\mathrm{e}^{-\frac{t}{\tau_{D(0)}}\left[1+\left(\frac{R_0}{R_{DA}}\right)^6\right]}dR_{DA}. \quad (10)$$

These equations highlight the potential to resolve the conformational heterogeneity in terms of the distribution of interdye distance $p(R_{DA})$ from the FRET-sensitized donor fluorescence decay. The interpretation hereby depends on the choice of the model function for $p(R_{DA})$; thus, it is imperative to consider the broadening introduced by the flexible linkers, which will be discussed in detail in sections 4.1and 0 below.

In this work, it is assumed that the properties of the fluorophores do not vary for different conformational states of the molecule (homogenous approximation). In practice, this assumption does often not hold when the environment of the fluorophores changes, leading to local quenching by aromatic residues, spectral shifts, or sticking interactions with the biomolecular surface. For details on how to account for a correlation between photophysical and conformational states, the reader is referred to ref. [50].

## 2.3 Intensity-based observable: FRET efficiency

The FRET efficiency can be quantified either from the number of photons emitted by the acceptor dyes due to FRET or from the decrease of the number of photons emitted by donor dye due to the transfer of energy to the acceptor. Using the fluorescence intensities $F$ that is fully corrected for the quantum yields and detection efficiencies, the FRET efficiency is given by:

$$E = \frac{F_{A|D}^{(DA)}}{F_{A|D}^{(DA)} + F_{D|D}^{(DA)}} = \frac{F_{D|D}^{(D0)} - F_{D|D}^{(DA)}}{F_{D|D}^{(D0)}}. \quad (11)$$

The superscripts refer to the sample type ($DA$ is a FRET sample), and the subscripts refer to the excitation, (... $|X$), and detection, ($X|$ ...), channels. D and A refer to the donor and acceptor fluorophore, respectively. For instance, $F_{A|D}^{(DA)}$ is the fluorescence intensity of the acceptor ($A|$ ... ) of a FRET molecule ($DA$) given that the donor was excited (... $|D$). In practice, the detected raw signals in the donor, ($I_{D|D}$), and acceptor, ($I_{A|D}$), channels need to be corrected (for details, see [38]) to yield fluorescence intensities, $F$.

In a time-resolved experiment, the fluorescence intensity $F$ is determined by integrating the fluorescence intensity decay $f(t)$. For the distribution of fluorescence lifetimes, $p(\tau_{D(A)})$, the integrated donor fluorescence intensity is given by:

$$F_{D|D}^{(DA)} = k_{F,D} \int \int p(\tau_{D(A)}) \mathrm{e}^{-\frac{t}{\tau_{D(A)}}} d\tau_{D(A)} dt. \tag{12}$$

The integral over $t$ is equivalent to the fluorescence lifetime, $\int \mathrm{e}^{-t/\tau_{D(A)}} dt = \tau_{D(A)}$, and the fluorescence intensity is hence proportional to the species-averaged fluorescence lifetime, $\langle \tau_{D(A)} \rangle_x$:

$$F_{D|D}^{(DA)} = k_{F,D} \int \tau_{D(A)} \, p(\tau_{D(A)}) \, d\tau = k_{F,D} \langle \tau_{D(A)} \rangle_x. \tag{13}$$

Through the definition of the FRET efficiency from the photon counts of the donor fluorophore in the presence and absence of FRET, $F_{D|D}^{(DA)}$ and $F_{D|D}^{(D0)}$ (eq. (11)), we can relate the intensity-averaged FRET efficiency $E$ to the time-resolved fluorescence decays of the donor in the presence and absence of FRET, $f_{D|D}^{(DA)}(t)$ and $f_{D|D}^{(D0)}(t)$:

$$E = 1 - \frac{k_{F.D} \int f_{D|D}^{(DA)}(t) dt}{k_{F,D} \int f_{D|D}^{(D0)}(t) dt} = 1 - \frac{\langle \tau_{D(A)} \rangle_x}{\langle \tau_{D(0)} \rangle_x}. \tag{14}$$

For now, we consider the case of a single-exponential donor lifetime, that is $\langle \tau_{D(0)} \rangle_x = \tau_{D(0)}$, and consider the effect of multi-exponential donor fluorescence lifetimes in section 3.7.

## 2.4 Lifetime-based observable: Average delay time

In smFRET with pulsed excitation, the detected photons are characterized by their delay time with respect to the excitation pulse. Due to the limited number of photons available in a single-molecule experiment, it is impossible to recover the distribution of fluorescence lifetimes $p(\tau_{D(A)})$. However, an average delay time, $\langle t \rangle$, can be determined reliably.

The average delay time $\langle t \rangle$ is defined by:

$$\langle t \rangle = \int t \cdot p_{D|D}^{(DA)}(t) dt = \frac{\int t \cdot f_{D|D}^{(DA)}(t) dt}{\int f_{D|D}^{(DA)}(t) dt}, \tag{15}$$

where $p_{D|D}^{(DA)}(t)$ is the normalized fluorescence decay that describes the probability distribution of delay times. For a distribution of fluorescence lifetimes $p(\tau_{D(A)})$ the average $\langle t \rangle$ is then:

$$\langle t \rangle = \frac{\int t \cdot \left[ \int p(\tau_{D(A)}) \mathrm{e}^{-t/\tau_{D(A)}} d\tau_{D(A)} \right] dt}{\int \int p(\tau_{D(A)}) \mathrm{e}^{-t/\tau_{D(A)}} d\tau_{D(A)} dt} = \frac{\int p(\tau_{D(A)}) \left[ \int t \cdot \mathrm{e}^{-t/\tau_{D(A)}} dt \right] d\tau_{D(A)}}{\int p(\tau_{D(A)}) \left[ \int \mathrm{e}^{-t/\tau_{D(A)}} dt \right] d\tau_{D(A)}}. \tag{16}$$

The inner integrals are given by $\int \mathrm{e}^{-t/\tau_{D(A)}} dt = \tau_{D(A)}$ and $\int t \cdot \mathrm{e}^{-t/\tau_{D(A)}} dt = \tau_{D(A)}^2$, resulting in the following expression for the average delay time:

$$\langle t \rangle \rightarrow \frac{\int \tau_{D(A)}^2 \, p(\tau_{D(A)}) d\tau_{D(A)}}{\int \tau_{D(A)} \, p(\tau_{D(A)}) d\tau_{D(A)}} = \frac{\overline{\tau_{D(A)}^2}}{\overline{\tau_{D(A)}}}, \tag{17}$$

where $\overline{\tau_{D(A)}}$ and $\overline{\tau_{D(A)}^2}$ are the first and second moments of the lifetime distribution. Thus, in an ideal measurement (i.e., in the absence of shot-noise or other experimental imperfections), the average delay time $\langle t \rangle$ converges to the ratio of the second and first moments of the lifetime distribution. Importantly, the average delay time $\langle t \rangle$ informs on the second moment $\overline{\tau_{D(A)}^2}$ of the fluorescence lifetime distribution, which conveys information about its variance.

It is important to note that the average delay time $\langle t \rangle$ is equivalent to the intensity-weighted average fluorescence lifetime, which we denote by $\langle \tau_{D(A)} \rangle_F$ to distinguish it from the species-weighted average fluorescence lifetime, $\langle\tau_{D(A)}\rangle_x$, introduced above. Consider that the fluorescence intensity of a species with lifetime $\tau_{D(A)}$, $F(\tau_{D(A)})$, is proportional to its fluorescence lifetime:

$$F(\tau_{D(A)}) = p(\tau_{D(A)}) \int k_{F,D} e^{-\frac{t}{\tau_{D(A)}}} dt = k_{F,D}\, \tau_{D(A)}\, p(\tau_{D(A)}). \tag{18}$$

Then, the intensity-weighted average lifetime, $\langle\tau_{D(A)}\rangle_F$, is given by:

$$\langle\tau_{D(A)}\rangle_F = \frac{\int F(\tau_{D(A)})\tau_{D(A)} d\tau_{D(A)}}{\int F(\tau_{D(A)}) d\tau_{D(A)}} = \frac{\int_0^\infty \tau_{D(A)}^2\, p(\tau_{D(A)}) d\tau_{D(A)}}{\int_0^\infty \tau_{D(A)}\, p(\tau_{D(A)}) d\tau_{D(A)}} = \frac{\overline{\tau_{D(A)}^2}}{\overline{\tau_{D(A)}}}, \tag{19}$$

which is equivalent to the result for the average delay time above.

So far, we have assumed that the fluorescence is excited by an ideal $\delta$-pulse. Experimentally, the analysis is complicated due to the finite width of the laser excitation pulse and characteristics of the detection electronics, defining the instrument response function (IRF). In the analysis, the IRF is accounted for by convolution with the ideal decay model. For low photon numbers, accurate lifetimes are best extracted using a maximum likelihood estimator (MLE) that correctly accounts for the noise characteristics of the photon detection, anisotropy effects and the presence of background signal [73]. The fluorescence lifetime obtained by maximizing the likelihood function is equivalent to the intensity-averaged fluorescence lifetime, i.e., $\tau_{MLE} = \langle\tau_{D(A)}\rangle_F$ (see Supplementary Note 1).

# 3 Concepts

## 3.1 Comparison to intensity-based approaches

Combining fluorescence lifetimes and intensities is superior in detecting and visualizing fast conformational dynamics than approaches that rely on intensities alone. In purely intensity-based approaches, the average inter-photon time limits the ability to detect conformational dynamics. We demonstrate this limitation by simulations of smFRET experiments of molecules that undergo conformational dynamics between distinct states at increasing interconversion rates. We simulate typical smFRET experiments with a count rate per molecule of 100 kHz. The simulated smFRET data was processed using the popular burst variance analysis (BVA)[68] analysis procedure.

In BVA, the variance of the FRET efficiency is estimated for every detected single molecule to reveal conformational dynamics happening on a timescale of the single-molecule burst duration. The basic idea of BVA is to obtain an estimate of the distribution of FRET efficiencies within a single-molecule event by sampling the FRET efficiency with a higher rate than the structural dynamics. In BVA, the FRET efficiency trace of single-molecule bursts is subsampled (Figure 3, Figure 1A), and the standard deviation of the FRET efficiency within a single-molecule burst is estimated by

$$\sigma_E = \sqrt{\frac{1}{M}\sum_{i=1}^{M}(E_i - E)^2}. \tag{20}$$

Here, $E^{(i)}$ is the FRET efficiency of a sample, $M$ is the total number of samples, and $E$ is the average FRET efficiency of the single-molecule event obtained by equation (11)). The standard deviation $\sigma_E$ is then plotted against the average FRET efficiency $E$ (Figure 3B). The lower boundary for the standard deviation of the FRET efficiency is given by the theoretical shot-noise limit, determined by the number of photons per sample $N$ [76]:

$$\sigma_E = \sqrt{\frac{E(1-E)}{N}}. \tag{21}$$

This previous equation is the corresponding static FRET-line in BVA. Single-molecule events that exceed this limit are considered dynamic.

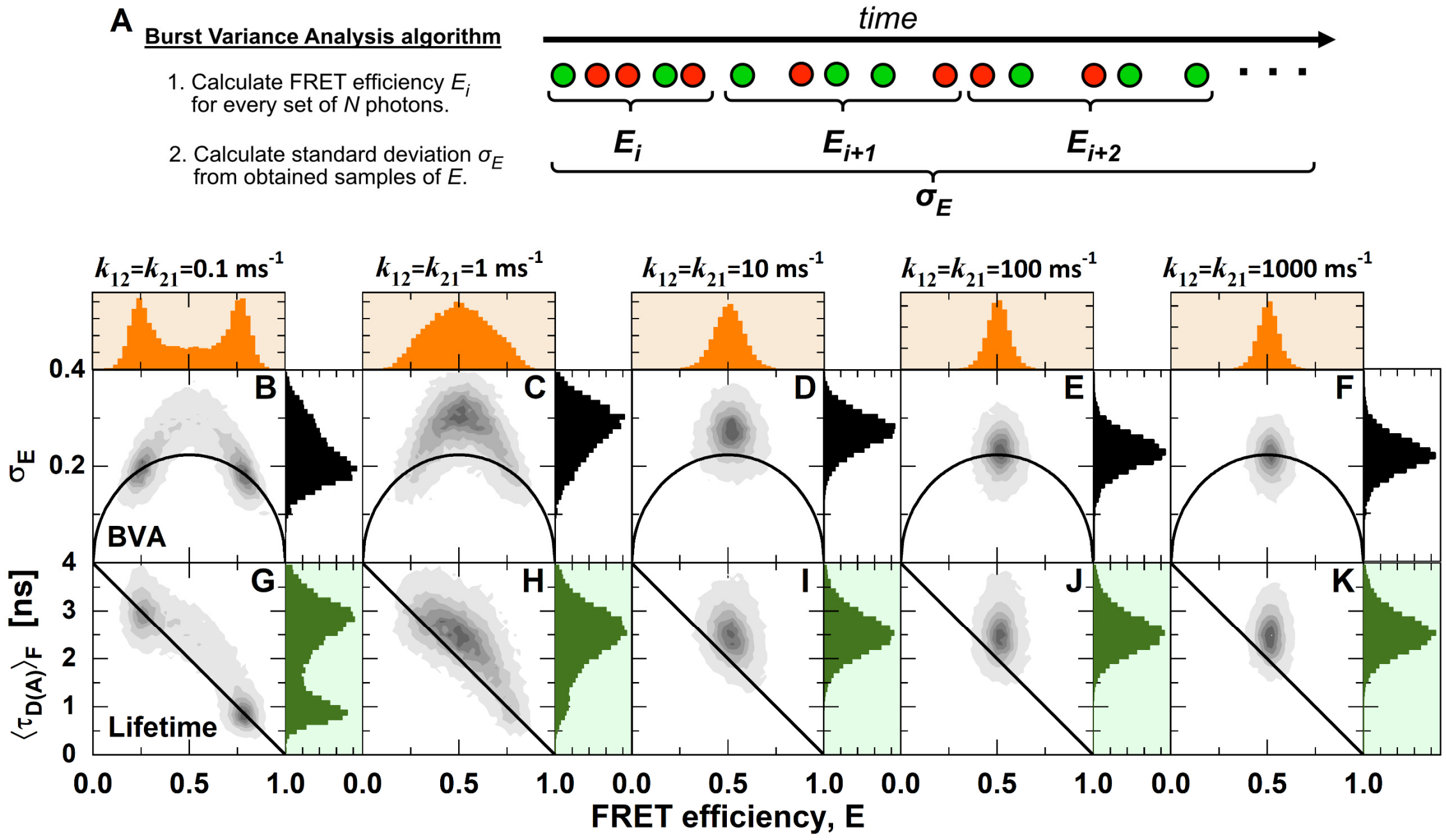

**Figure 3: Comparison of intensity-based and lifetime-based indicators of dynamics. A)** Illustration of the algorithm used in burst variance analysis (BVA) to estimate the standard deviation of the FRET efficiency for a set of fluorescence photons of a single-molecule event. The photon trace is sub-sampled by the number of photons, $N$ (typically $N = 5$). For every sub-sample, the FRET efficiency, $E$, is estimated. An estimate of the standard deviation of the FRET efficiency, $\sigma_E$, is obtained by the estimates of $E$. **B-K)** Molecule-wise histograms of simulated datasets with indicated interconversion rates between two states with FRET efficiencies of 0.25 and 0.8 and a diffusion time of 0.5 ms. **B-F)** In BVA, conformational dynamics increase the standard deviation of the FRET efficiency $\sigma_E$ beyond the expected shot-noise variance (black line) given by eq. (21)). **G-K)** The comparison of the two estimators of the FRET efficiency, $E$ and $\langle\tau_{D(A)}\rangle_F$, reveals conformational dynamics as a shift from the static FRET-line (diagonal line) given by $E = 1 - \frac{\langle\tau_{D(A)}\rangle_F}{\tau_{D(0)}}$. BVA is most sensitive to slow dynamics, while the standard deviation of the FRET efficiency is underestimated at fast dynamics. In contrast, the lifetime-based indicator detects conformational dynamics irrespective of their timescale. The dashed magenta line indicates the expected position of the dynamic population on the y-axis. For BVA, a photon window of $N$ = 5 was used.

The simulated data was processed by BVA with a photon window of $N$ = 5 (Figure 3B-F). A standard deviation $\sigma_E$ observed in BVA that exceeds the shot-noise limit decreases as the dynamics become faster (Figure 3B-F). In the simulations, the average inter-photon time was 10 µs, and the time resolution is further reduced due to the need to average over a given photon number (typically, $N$ = 5)[68]. All faster processes than this limit will be averaged over the sampling time and thus not detected as dynamic (Figure 3E-F). The dependency on the timescale of dynamics makes it difficult to predict the exact shape of the observed distributions, which requires taking into account the experimental photon count distribution[68]. Hence, they have mainly been used as qualitative indicators of conformational dynamics. It should be noted that dynamics on timescales faster than the inter-photon time can still be detected by fluorescence correlation spectroscopy (FCS), wherein the effective time resolution is determined mainly by the signal-to-noise ratio. However, in contrast to the single-molecule analysis, it is challenging to directly identify states or their connectivity from the FCS curves.

On the other hand, using the relation between the FRET efficiency $E$ and the intensity-averaged donor fluorescence lifetime $\langle\tau_{D(A)}\rangle_F$, conformational dynamics are identified even if they are fast (Figure 3G-K). This lifetime-based indicator is independent of the detection count rate because it relies only on the deviation of the fluorescence decay from the ideal single-exponential behavior. Hence, all dynamic processes that are slower than the fluorescence lifetime (> ns) are detected, and no decrease of the dynamic shift is observed at increasing timescales of the dynamics.

In practice, one has to consider some artifacts that potentially lead to a false-positive detection of dynamics. Examples include dark states of the acceptor (e.g., due triplet states). Acceptor dark states always affect intensity-based indicators of dynamics as they result in fluctuations of the apparent FRET efficiency. The effect of dark acceptor states on the donor fluorescence lifetime depends on the nature of the photophysical change. Triplet states often still act as FRET acceptors with a similar Förster radius as the single state; a similar situation is found for the *cis-trans* isomerization of cyanine dyes such as Cy5[77-79]. Radical or ionic dark states, on the other hand, often are not viable FRET acceptors. In this case, the donor lifetime will fluctuate as a function of the photophysical state of the acceptor[80].

## 3.2 FRET-lines of static and dynamic molecules

In addition to detecting the presence of dynamics, FRET-lines are a powerful tool to obtain information about the nature of the dynamic exchange and identify the limiting states and their connectivity. So far, we have introduced the static FRET-line that describes the ideal relationship between the fluorescence lifetime of the donor fluorophore and the FRET efficiency in the absence of dynamics. For the experimental observables $E$ and $\langle\tau_{D(A)}\rangle_F$, the static FRET-line is defined as:

$$\text{static FRET-line: } E = 1 - \frac{\langle\tau_{D(A)}\rangle_F}{\tau_{D(0)}}. \qquad (22)$$

Importantly, this relationship only holds for a fixed distance between the dyes resulting in a single FRET rate, in which case the intensity-weighted average fluorescence lifetime is equal to the species average, $\langle\tau_{D(A)}\rangle_F = \langle\tau_{D(A)}\rangle_x$. In the case of a distribution of distances (and hence lifetimes) that are sampled during the observation time, the intensity-averaged fluorescence lifetime is biased towards species with long fluorescence lifetimes and thus low FRET efficiencies, and $\langle\tau_{D(A)}\rangle_F > \langle\tau_{D(A)}\rangle_x$. This results in the shift of the populations from the static FRET-line (Figure 1C). We call this shift the "dynamic shift" and define it as the minimal distance to the static FRET-line for a given point in the $E - \langle\tau_{D(A)}\rangle_F$ histogram.

We now consider the simplest case of dynamics wherein the molecule switches between two defined conformations during the observation time:

$$E^{(1)},\ \tau_{D(A)}^{(1)} \underset{k_{12}}{\overset{k_{21}}{\rightleftarrows}} E^{(2)},\ \tau_{D(A)}^{(2)}, \qquad (23)$$

where $k_{12}$ and $k_{21}$ are the microscopic interconversion rates between the two states that define the probability that a molecule spends a fraction of time $x^{(i)}$ in state $i$ during the observation time. Fractions $x^{(i)}$ are stochastic quantities and change from one observation to another. For now, we are not interested in the exact distribution of the state occupancy $x^{(1)}$ and treat it as the independent parameter of the model. This is equivalent to the assumption of a uniform distribution for $x^{(1)}$. The effect of occupancies distribution is discussed in detail in Part II.

Assuming that each state is characterized by the same donor fluorescence lifetime, the species-weighted and fluorescence-weighted average lifetimes depend only on the state occupancy of the individual states, $x^{(1)}$ and $x^{(2)} = 1 - x^{(1)}$:

$$\langle\tau_{D(A)}\rangle_x = x^{(1)}\tau_{D(A)}^{(1)} + \left(1 - x^{(1)}\right)\tau_{D(A)}^{(2)}; \qquad (24)$$

$$\langle\tau_{D(A)}\rangle_F = \frac{\langle\tau^2_{D(A)}\rangle_x}{\langle\tau_{D(A)}\rangle_x} = \frac{x^{(1)}\left(\tau^{(1)}_{D(A)}\right)^2 + \left(1-x^{(1)}\right)\left(\tau^{(2)}_{D(A)}\right)^2}{x^{(1)}\tau^{(1)}_{D(A)} + (1-x^{(1)})\tau^{(2)}_{D(A)}}. \quad (25)$$

Here, we changed the notation from the continuous distribution of lifetimes to the discrete case, that is:

$$p\left(\tau_{D(A)}\right) = \begin{cases} x^{(1)} & \text{for } \tau_{D(A)} = \tau^{(1)}_{D(A)} \\ 1-x^{(1)} & \text{for } \tau_{D(A)} = \tau^{(2)}_{D(A)} \\ 0 & \text{otherwise} \end{cases} \quad (26)$$

To obtain a general relationship between the observables $E$ and $\langle\tau_{D(A)}\rangle_F$, we find the line that describes all values of $x^{(1)}$ by combining equations (24) and (25), relating the species-weighted average lifetime to the intensity-weighted average lifetime:

$$\text{dynamic FRET-line: } E = 1 - \frac{\langle\tau_{D(A)}\rangle_x}{\tau_{D(0)}} = 1 - \frac{1}{\tau_{D(0)}}\cdot\left[\frac{\tau^{(1)}_{D(A)}\cdot\tau^{(2)}_{D(A)}}{\tau^{(1)}_{D(A)}+\tau^{(2)}_{D(A)}-\langle\tau_{D(A)}\rangle_F}\right]. \quad (27)$$

This relationship is defined for $\langle\tau_{D(A)}\rangle_F$ in the interval $\left[\tau^{(1)}_{D(A)},\ \tau^{(2)}_{D(A)}\right]$, which is equivalent for $x^{(1)}$ being in the interval [0, 1]. Because eq. (27) describes the FRET-line for a binary system in dynamic exchange; we call it the *dynamic FRET-line*. Dynamic FRET-lines connect two static states. They were first introduced by Kalinin et al.[40], and Gopich and Szabo[81] later described analogous relations.

Figure 4 illustrates the concept of static and dynamic FRET-lines. Static FRET-lines describe pure states, which are described by sharp distributions ($\delta$-functions) in terms of the lifetime distribution $p(\tau_{D(A)})$ (Figure 4A). In contrast, dynamic FRET-lines describe the mixing of two pure states as a function of the state occupancy $x^{(1)}$. The corresponding donor fluorescence decays are single-exponential for pure states and bi-exponential in the case of mixing between pure states (Figure 4B). In the $E - \langle\tau_{D(A)}\rangle_F$ plot, the dynamic FRET-line connects the two points of the contributing pure states on the static FRET-line by a curved line (Figure 4D).

To quantify the sensitivity of the dynamic exchange, it is helpful to consider the maximum separation between the dynamic and static FRET-lines. We define this *dynamic shift*, ds, orthogonal to the static FRET-line (Figure 4D). Like the dynamic FRET-line, the value of the dynamic shift depends only on the FRET efficiencies of the limiting states $E_1$ and $E_2$, and is given by (Supplementary Note 2):

$$ds = \frac{1}{\sqrt{2}}\left(\sqrt{1-E_1} - \sqrt{1-E_2}\right)^2. \quad (28)$$

Note that this equation for the dynamic shift is valid for a plot of the FRET efficiency $E$ against the *normalized* intensity-averaged donor fluorescence lifetime, $\langle\tau_{D(A)}\rangle_F/\tau_{D(0)}$. Exemplary dynamic FRET-lines for different FRET efficiencies $E_1$ and $E_2$ are shown in Figure 4E with their corresponding dynamic shifts. By visualizing the dynamic shift as a function of the FRET efficiencies of the limiting states (Figure 4F), one can define sensitive and insensitive regions depending on a given detectability threshold for the dynamic shift. This threshold depends on how well the experimental setup is calibrated, the accuracy of the fluorescence lifetime estimation and the measurement statistics that typical threshold values for the detectability of dynamic shifts are on the order of 0.05 or less, potentially reaching a sensitivity of 0.01 for well-calibrated setups and carefully performed experiments. This places the purple dynamic FRET-line shown in Figure 4E on the border of the insensitive region, while the other two examples with a dynamic shift above 0.1 are clearly in the sensitive region.

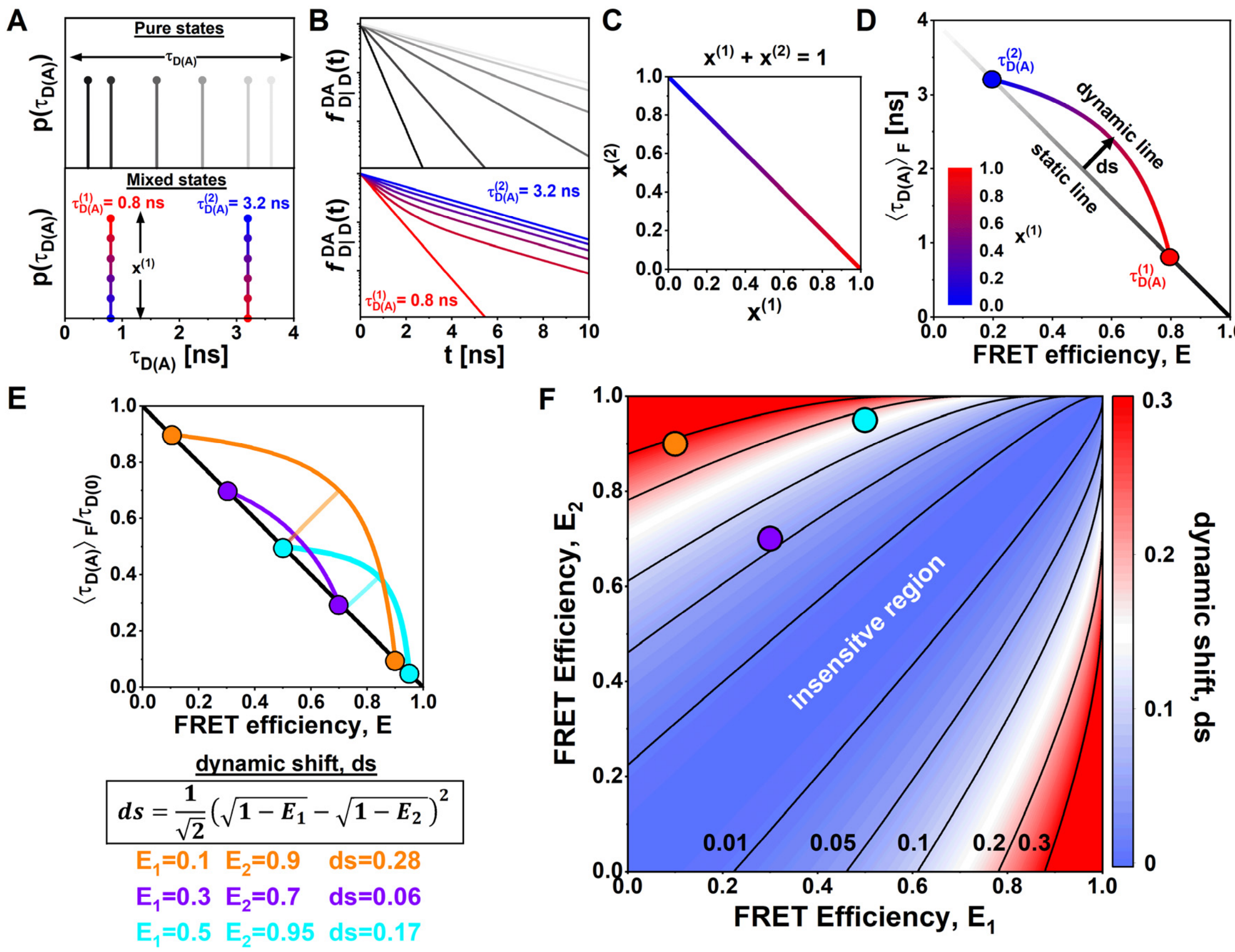

$$ds = \frac{1}{\sqrt{2}}\left(\sqrt{1-E_1} - \sqrt{1-E_2}\right)^2$$

**Figure 4:** FRET-lines of dynamic molecules. **A)** Pure states are characterized by a single lifetime, and the corresponding lifetime distributions show a single peak. In the presence of dynamics, pure states are mixed at different ratios. The lifetime distributions show two peaks weighted by the species fractions $\boldsymbol{x^{(1)}}$ and $\boldsymbol{x^{(2)}} = \boldsymbol{1 - x^{(1)}}$. The pure states are defined by lifetimes $\boldsymbol{\tau_{D(A)}^{(1)}}$ = 0.8 ns and $\boldsymbol{\tau_{D(A)}^{(1)}}$ = 3.2 ns. Species fractions are color coded from red ($\boldsymbol{x^{(1)} = 1}$) to blue ($\boldsymbol{x^{(1)} = 0}$). **B)** The corresponding fluorescence decays of the lifetime distributions shown in A. For pure states, the decays are single exponentials, while mixed states have two-lifetime components. C) The dependency between the species fractions $\boldsymbol{x^{(1)}}$ and $\boldsymbol{x^{(2)}}$ is given by $\boldsymbol{x^{(1)} + x^{(2)} = 1}$. **D)** In a plot of the FRET efficiency $\boldsymbol{E}$ versus the intensity-weighted average fluorescence lifetime $\boldsymbol{\langle\tau_{D(A)}\rangle_F}$, pure states define the static FRET-line (grayscale diagonal line). Mixed states are displaced from the static FRET-line and fall onto a curved line connecting the pure states, described by equation (27). The dynamic FRET-line is color-coded by the contribution of species 1. The arrow indicates the maximum possible dynamic shift ds from the static FRET-line. **E)** Exemplary dynamic FRET-lines for limiting states with FRET efficiencies $E_1$=0.1/$E_2$=0.9 (orange, ds=0.28), $E_1$=0.3/$E_2$=0.7 (purple, ds=0.06) and $E_1$=0.5/$E_2$=0.95 (cyan, ds=0.17) are shown in a plot of the FRET efficiency versus the normalized intensity-weighted average fluorescence lifetime $\boldsymbol{\langle\tau_{D(A)}\rangle_F/\tau_{D(0)}}$. **F)** Contour plot of the dynamic shift, ds, as a function of the FRET efficiencies of the limiting states, $E_1$ and $E_2$. The dynamic shift for the examples given in E are shown as circles.

## 3.3 General definition of FRET-lines

FRET-lines are idealized relations between the FRET-related experimental observables $E$ and $\langle\tau_{D(A)}\rangle_F$ for different physical models of the system. Before considering more specific scenarios, such as the effect of the flexible dye linkers or disordered systems, we first present a general definition of FRET-lines.

Consider that the experiment is described by a physical model defined by a set of parameters $\Lambda$. The model encompasses all parameters of the experimental system and fully defines the two-dimensional distribution of the experimental observables, $p(E, \langle\tau_{D(A)}\rangle_F|\Lambda)$. For a complete description of the experiment, we would require the joint distribution of the experimental observables over the different realizations of the system parameters $\Lambda$, weighted by their probability of occurrence $p(\Lambda)$:

$$p\big(E, \langle\tau_{D(A)}\rangle_F\big) = \int p\big(E, \langle\tau_{D(A)}\rangle_F|\Lambda\big)\, p(\Lambda) d\Lambda. \tag{29}$$

This distribution is generally challenging to address as it depends on the photon statistics of the experiment; however, a derivation of the distribution for a two-state system may be found in reference [81].

In the ideal case of zero photon shot noise, the distribution $p(E, \langle\tau_{D(A)}\rangle_F|\Lambda)$ would simplify to ideal curves on the $(E, \langle\tau_{D(A)}\rangle_F)$ plane, which define parametric relations between $E$ and $\langle\tau_{D(A)}\rangle_F$ as a function of the model parameters $\Lambda$. If we choose a fixed value for all model parameters, we obtain a single point on the $(E, \langle\tau_{D(A)}\rangle_F)$ plane. If instead, we vary a single parameter, a defined curve – the FRET-line - is obtained. Let the variable parameter be $\lambda$ and the fixed values for the remaining model parameters be $\Lambda_f$. Then, the parametric relation between $E$ and $\langle\tau_{D(A)}\rangle_F$ for a given model is obtained from the moments of the lifetime distribution by the following equations:

$$E = 1 - \frac{\overline{\tau_{D(A)}}\big(\lambda, \Lambda_f\big)}{\tau_{D(0)}}; \tag{30}$$

$$\langle\tau_{D(A)}\rangle_F = \frac{\overline{\tau_{D(A)}^2}(\lambda, \Lambda_f)}{\overline{\tau_{D(A)}}(\lambda, \Lambda_f)}. \tag{31}$$

To derive the FRET-line for a given physical model, one has to compute the moments of the lifetime distribution, $\overline{\tau_{D(A)}}$ and $\overline{\tau_{D(A)}^2}$, as functions of the model parameters. As an example, our physical model might define the dynamic exchange between two distinct conformations, as described in the previous section. In this case, the parameters of the model are the FRET efficiencies of the distinct conformations and the fractional occupancy of the states, i.e., $\Lambda = \{E^{(1)}, E^{(2)}, x^{(1)}, x^{(2)}\}$, whereby we only have to consider one fractional occupancy as $x^{(2)} = 1 - x^{(1)}$. From this set of parameters, we have chosen $x^{(1)}$ as the free parameters ($\lambda = x^{(1)}$) and kept the FRET efficiencies constant ($\Lambda_f = \{E^{(1)}, E^{(2)}\}$).

We can write a general expression for the first and second moments of the lifetime in equations (30) and (31) using the definition of the moments:

$$\overline{\tau_{D(A)}^\nu}\big(\lambda, \Lambda_f\big) = \int \tau_{D(A)}^\nu p\big(\tau_{D(A)}|\lambda, \Lambda_f\big) d\,\tau_{D(A)}, \qquad \nu = \{1,2\} \tag{32}$$

Thus, the problem reduces to find an expression of the lifetime distribution $p\big(\tau_{D(A)}|\lambda, \Lambda_f\big)$ for a given model. If such an expression is available, we can derive equations for $E$ and $\langle\tau_{D(A)}\rangle_F$ (or any related observable) as a function of the variable parameter $\lambda$. Finally, to obtain the explicit form of the FRET-line, the free parameter $\lambda$ can be eliminated by substitution, and the resulting expression defines a direct relation between the observables $E$ and $\langle\tau_{D(A)}\rangle_F$. A detailed description of this general formalism is given in Supplementary Note 3.

**Table 1: Overview of experimental parameters and corresponding model parameters. A dash indicates that there is no corresponding parameter.**

| Model | | Experiment |
|---|---|---|
| Probability distribution | ↔ | Random realization |
| Expected value | ↔ | Experimental observable |
| Probability density function | ↔ | Histogram |
| FRET-lines | ↔ | Distribution of FRET efficiency, fluorescence lifetime, or related quantities |
| Expectation value of FRET efficiency | ↔ | Species-averaged FRET efficiency |
| $E = 1 - \frac{\overline{\tau_{D(A)}}}{\tau_{D(0)}}$ | ≙ | $E = \frac{F_{A\|D}}{F_{A\|D} + F_{D\|D}} = 1 - \frac{\langle \tau_{D(A)} \rangle_x}{\tau_{D(0)}}$ |
| First moment of the lifetime distribution | ↔ | Species-averaged lifetime |
| $\overline{\tau_{D(A)}} = \int_0^\infty \tau_{D(A)}\, p(\tau_{D(A)}) d\tau_{D(A)}$ | ≙ | $\langle \tau_{D(A)} \rangle_x = (1 - E)\tau_{D(0)}$ |
| Second moment of the lifetime distribution | ↔ | Species-averaged *squared* lifetime |
| $\overline{\tau^2_{D(A)}} = \int_0^\infty \tau^2_{D(A)}\, p(\tau_{D(A)}) d\tau_{D(A)}$ | ≙ | $\langle \tau^2_{D(A)} \rangle_x = \langle \tau_{D(A)} \rangle_F (1 - E)\tau_{D(0)}$ |
| Ratio of the second and first moment of the lifetime distribution | ↔ | Intensity-weighted average fluorescence lifetime, average delay time |
| $\frac{\overline{\tau^2_{D(A)}}}{\overline{\tau_{D(A)}}}$ | ≙ | $\langle \tau_{D(A)} \rangle_F$ or $\langle t \rangle$ |
| - | ↔ | Intensity-weighted average FRET efficiency |
| $1 - \frac{1}{\tau_{D(0)}} \frac{\overline{\tau^2_{D(A)}}}{\overline{\tau_{D(A)}}}$ | ≙ | $E_\tau = 1 - \frac{\langle \tau_{D(A)} \rangle_F}{\tau_{D(0)}}$ |
| Variance of the lifetime distribution | ↔ | - |
| $\mathrm{Var}(\tau_{D(A)}) = \overline{\tau^2_{D(A)}} - \overline{\tau_{D(A)}}^2 = \mathrm{Var}(E)\tau^2_{D(0)}$ | ≙ | $(1 - E)(E - E_\tau)\tau^2_{D(0)}$ |
| Difference between the normalized first and second moment of the lifetime distribution | ↔ | - |
| $\Gamma = \frac{\overline{\tau_{D(A)}}}{\tau_{D(0)}} - \frac{\overline{\tau^2_{D(A)}}}{\tau_{D(0)}^2}$ | ≙ | $(1 - E)E_\tau$ |

## 3.4 Experimental observables and moments of the lifetime distribution

The theoretical description of the average delay time $\langle t \rangle$ and the FRET efficiency in sections 2.3 and 2.4 had naturally led us to the moments of the lifetime distribution (eq. (32)). The first moment of $p(\tau_{D(A)})$ is equal to the expected value of the fluorescence lifetime. The second moment is given by the expected value of the square of the fluorescence lifetime. The variance $\mathrm{Var}(\tau_{D(A)})$ is the second *central* moment, defined as the average squared deviation from the mean, which is related to the first and second moments by:

$$\mathrm{Var}(\tau_{D(A)}) = \langle (\tau_{D(A)} - \overline{\tau_{D(A)}})^2 \rangle = \overline{\tau^2_{D(A)}} - \overline{\tau_{D(A)}}^2. \quad (33)$$

Thus, the second moment, and consequentially $\langle \tau_{D(A)} \rangle_F$, relates to the *variance* of the lifetime distribution. Using the relations between the experimental observables $E$ and $\langle \tau_{D(A)} \rangle_F$ and the moments of the lifetime distribution, we obtain:

$$\mathrm{Var}(\tau_{D(A)}) = \langle \tau^2_{D(A)} \rangle_x - \langle \tau_{D(A)} \rangle_x^2 = (1 - E)(E - E_\tau)\tau^2_{D(0)}, \quad (34)$$

where we have introduced the quantity $E_\tau$ defined as:

$$E_\tau = 1 - \frac{\langle\tau_{D(A)}\rangle_F}{\tau_{D(0)}}. \tag{35}$$

Due to the linear relation between the FRET efficiency and the fluorescence lifetime, the variance of the lifetime distribution is directly proportional to the variance of the FRET efficiency distribution by:

$$\mathrm{Var}(E) = \frac{\mathrm{Var}(\tau_{D(A)})}{\tau_{D(0)}^2}. \tag{36}$$

This provides an alternate approach to BVA to estimate the variance of the FRET efficiency distribution from the observables $E$ and $\langle\tau_{D(A)}\rangle_F$. The result is identical to the expression obtained in reference [81], relating $\langle\tau_{D(A)}\rangle_F$ to the variance of the FRET efficiency distribution:

$$\langle\tau_{D(A)}\rangle_F = \tau_{D(0)}\left[1 - E + \frac{\mathrm{Var}(E)}{1-E}\right]. \tag{37}$$

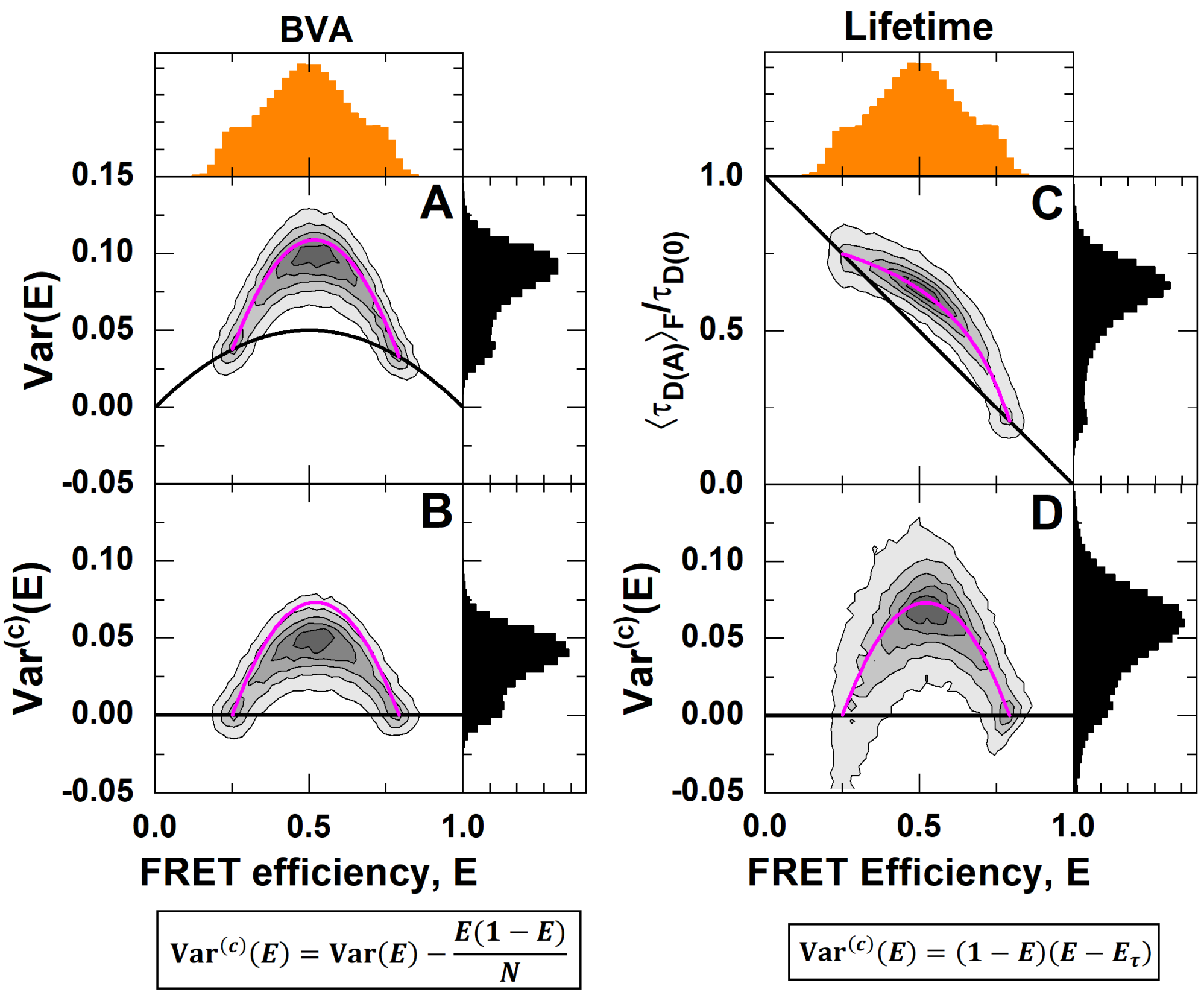

**Figure 5:** Estimating the variance of the FRET efficiency distribution. Shown is a simulation of molecules interconverting between two distinct states with $E^{(1)}$=0.25 and $E^{(2)}$=0.8, interconversion rates of $k_{12} = k_{21}$ = 1 ms$^{-1}$ and a diffusion time of 0.5 ms. **A)** Burst variance analysis (BVA) quantifies the total variance of the FRET efficiency through analysis of the photon time trace (compare Figure 3A), which contains contributions from photon shot noise and conformational dynamics (magenta line). The shot-noise variance is given as a black line. **B)** A simple subtraction of the photon shot noise reveals the variance due to conformational dynamics, $\mathbf{Var^{(c)}}(E)$. **C)** The plot of the two observables $E$ and $\langle\tau_{D(A)}\rangle_F$ reveals the dynamics as a right-ward shift from the static FRET-line (black). **D)** The estimated variance of the FRET efficiency from the observables follows the expected line as given by equation (41).

For a single lifetime component, the distribution of lifetimes is given by a Dirac delta function $\delta$:

$$p(\tau_{D(A)}) = x^{(i)}\delta\left(\tau - \tau_{D(A)}^{(i)}\right) = \begin{cases} x^{(i)} & \tau_{D(A)} = \tau_{D(A)}^{(i)} \\ 0 & \text{else} \end{cases} \tag{38}$$

The $\nu$-th moment is then given by $\tau_{D(A)}^{\nu}$ and the variance of the distribution, as given by equation (33), is zero, defining the equivalent static FRET-line. Thus, the static FRET-line (eq. (22)) corresponds to the particular case of lifetime distributions with vanishing variance. For two-component lifetime distributions, the distribution of lifetimes is given by the weighted sum of two $\delta$-functions, leading to the following expression for the moments of the lifetime distribution:

$$\begin{aligned} \langle\tau_{D(A)}\rangle_x &= x^{(1)}\tau_{D(A)}^{(1)} + \left(1 - x^{(1)}\right)\tau_{D(A)}^{(2)}; \\ \langle\tau_{D(A)}^2\rangle_x &= x^{(1)}\left(\tau_{D(A)}^{(1)}\right)^2 + \left(1 - x^{(1)}\right)\left(\tau_{D(A)}^{(2)}\right)^2. \end{aligned} \tag{39}$$

Note that the moments of the lifetime distribution are linear functions of the species fraction $x^{(1)}$. For the mixing between two states (eq. (39)), the variance is then given by:

$$\mathrm{Var}(\tau_{D(A)}) = \overline{\tau_{D(A)}^2} - \overline{\tau_{D(A)}}^2 = x^{(1)}\left(1 - x^{(1)}\right)\left(\tau_{D(A)}^{(1)} - \tau_{D(A)}^{(2)}\right)^2. \tag{40}$$

We can eliminate the variable $x^{(1)}$ to obtain the relation between $\mathrm{Var}(\tau_{D(A)})$ and $\langle\tau_{D(A)}\rangle_x$:

$$\mathrm{Var}(\tau_{D(A)}) = \left[\langle\tau_{D(A)}\rangle_x - \tau_{D(A)}^{(1)}\right]\left[\tau_{D(A)}{}^{(2)} - \langle\tau_{D(A)}\rangle_x\right], \tag{41}$$

from which the variance of the FRET efficiency distribution is obtained by equation (36). Equation (41) defines the dynamic FRET-line for data displayed in the mean-variance representation.

To illustrate that we can indeed estimate the variance of the FRET efficiency distribution from the two experimental observables $E$ and $\langle\tau_{D(A)}\rangle_F$, we compare the variance estimate with that obtained from burst variance analysis (BVA) for a simulated dataset (Figure 5). BVA correctly identifies the presence of conformational dynamics between the two states at FRET efficiencies of 0.25 and 0.8 (Figure 5A). The variance estimate obtained from BVA, however, includes the contribution of photon shot noise (eq. (21), black line in Figure 5A), and the dynamics is shown as excess variance beyond the shot-noise limit. To obtain the contribution to the variance due to conformational dynamics ($\mathrm{Var}^{(c)}(E)$), we subtract the shot-noise contribution given by $\sigma_{SN}^2 = E(1 - E)/N$, where $N = 5$ is the photon window used for the analysis (Figure 5B). Compared to the expected variance given by equation (41) (pink line), BVA underestimates the variance of the FRET efficiency, caused by the averaging over the photon window used in the calculation. It must also be considered that BVA measures the combined variance of the FRET efficiency caused by the contributions of shot-noise and dynamics. However, these contributions are not strictly additive. The simple subtraction of the shot-noise contribution performed here is thus only approximative. In the ($E$, $\langle\tau_{D(A)}\rangle_F$) representation, the same dataset shows a dynamic shift from the diagonal line that is described by the dynamic FRET-line (Figure 5C). From the experimental observables, we calculate the variance of the FRET efficiency distribution. Unlike the variance obtained by BVA, this variance estimate represents the pure contribution of the conformational dynamics and follows the expected dynamic FRET-line. Note, however, that the molecule-wise distribution of the variance estimated from the observables $E$ and $\langle\tau_{D(A)}\rangle_F$ generally shows a broader distribution compared to BVA. Conceptual static and two-state dynamic FRET-lines for the mean-variance representation of the data are shown in Figure 6B.

### 3.5 Alternative representation of dynamic lines

For the dynamic mixing between pure species, i.e., species whose lifetime distributions are described by $\delta$-functions, the moments of the lifetime distribution are simply given by the linear combination of the moments of the pure components (compare equation (39)):

$$p^{(i)}(\tau_{D(A)}) = \sum x^{(i)}\delta\left(\tau - \tau_{D(A)}^{(i)}\right) \quad \Rightarrow \quad \begin{aligned}\langle\tau_{D(A)}\rangle_x &= \sum x^{(i)}\tau_{D(A)}^{(i)} \\ \langle\tau_{D(A)}^2\rangle_x &= \sum x^{(i)}\left(\tau_{D(A)}^{(i)}\right)^2\end{aligned}. \tag{42}$$

Any linear combination of the quantities $\langle\tau_{D(A)}\rangle_x$ and $\langle\tau_{D(A)}^2\rangle_x$ will thus retain this property. This implies that we can further simplify the expression for the dynamic FRET-line by choosing the first and second moments as the parameters, which results in a linear expression for the dynamic FRET-line.

In the parameter space of the first two moments, $(\langle\tau_{D(A)}\rangle_x, \langle\tau_{D(A)}^2\rangle_x)$, the static FRET-line is given by $\langle\tau_{D(A)}^2\rangle_x = \left(\langle\tau_{D(A)}\rangle_x\right)^2$, which is the equation for an ordinary parabola. In other words, while the dynamic FRET-line is linearized, we now have a quadratic relation for the static FRET-line. Using the parameters $(\langle\tau_{D(A)}\rangle_x, \langle\tau_{D(A)}^2\rangle_x)$, static and dynamic FRET-lines are however not well separated, making it challenging to distinguish static from dynamic molecules. To overcome this problem, we replace the second moment with the difference between the normalized first and second moments:

$$\Gamma = \frac{\langle\tau_{D(A)}\rangle_x}{\tau_{D(0)}} - \frac{\langle\tau_{D(A)}^2\rangle_x}{\tau_{D(0)}^2}. \tag{43}$$

This *moment difference* $\Gamma$ is related to the experimental observables $E$ and $\langle\tau_{D(A)}\rangle_F$ by:

$$\Gamma = (1-E)\left(1 - \frac{\langle\tau_{D(A)}\rangle_F}{\tau_{D(0)}}\right) = (1-E)E_\tau, \tag{44}$$

where we defined $E_\tau = 1 - \frac{\langle\tau_{D(A)}\rangle_F}{\tau_{D(0)}}$.

In this representation, the static FRET-line transforms to:

$$\Gamma_{\text{static}} = \frac{\langle\tau_{D(A)}\rangle_x}{\tau_{D(0)}} - \frac{\left(\langle\tau_{D(A)}\rangle_x\right)^2}{\tau_{D(0)}^2} = \frac{\langle\tau_{D(A)}\rangle_x}{\tau_{D(0)}}\left(1 - \frac{\langle\tau_{D(A)}\rangle_x}{\tau_{D(0)}}\right) = (1-E)E. \tag{45}$$

Equation (45) describes a parabola that crosses the FRET efficiency axis at points (0, 0) and (1, 0) and has a maximum at (1/2, ¼) (Figure 6C). In the case of dynamics, the difference of the normalized lifetime moments is given by:

$$\Gamma_{\text{dynamic}} = x^{(1)}\frac{\tau^{(1)}}{\tau_{D(0)}}\left(1 - \frac{\tau^{(1)}}{\tau_{D(0)}}\right) + \left(1 - x^{(1)}\right)\frac{\tau^{(2)}}{\tau_{D(0)}}\left(1 - \frac{\tau^{(2)}}{\tau_{D(0)}}\right). \tag{46}$$

From this, we obtain the simple form of the dynamic FRET-line:

$$\Gamma_{\text{dynamic}} = \left(1 - \frac{\tau^{(1)}}{\tau_{D(0)}} - \frac{\tau^{(2)}}{\tau_{D(0)}}\right)\frac{\langle\tau_{D(A)}\rangle_x}{\tau_{D(0)}} + \frac{\tau^{(1)}\tau^{(2)}}{\tau_{D(0)}^2} = (1 - E^{(1)} - E^{(2)})E + E^{(1)}E^{(2)}. \tag{47}$$

The expression for the dynamic FRET-line is linear to the FRET efficiency $E$, directly connecting the two points belonging to the pure states (Figure 6C).

In the difference between the first and second normalized moments $\Gamma$, we found a parameter that linearizes the dynamic mixing while retaining a simple relation for the static FRET-line. The linearization of dynamics in this *moment representation* dramatically simplifies the graphical analysis of kinetic networks by providing direct visualization of the kinetic connectivity. To highlight its usefulness, we show the moment representation together with the histogram of the experimental observables, $E$ and $\langle\tau_{D(A)}\rangle_F$, in the following discussions of more complex scenarios. The moment representation resembles the analysis of fluorescence lifetimes in the phasor approach to fluorescence lifetime imaging (Phasor-FLIM)[82]. In both approaches, single-exponential fluorescence decays are found on a curve, a parabola in the moment representation, and a circle in Phasor-FLIM. Moreover, bi-exponential decays are shifted inwards from the curve and lie on the line connecting the coordinates of the pure components. The phasor calculation only requires fluorescence decays; thus, also applicable to study quenching without FRET. In principle, the moment representation could thus be combined with the phasor information to add another dimension to the analysis. The different transformations of the observables $E$ and $\langle\tau_{D(A)}\rangle_F$ and their theoretical equivalents are summarized in Table 1.

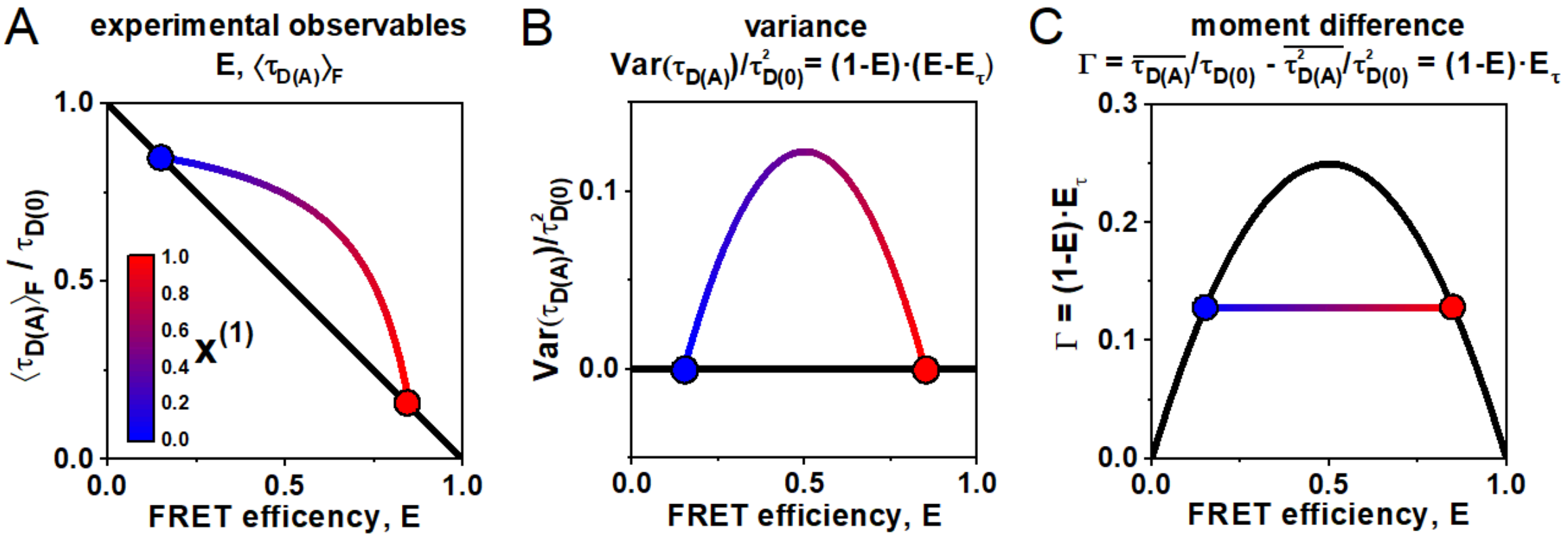

**Figure 6:** Different representations of the FRET estimators $\boldsymbol{E}$ and $\langle\boldsymbol{\tau_{D(A)}}\rangle_{\boldsymbol{F}}$ and derived quantities for a two-state system with $\boldsymbol{E}^{(1)} = 0.25$ and $\boldsymbol{E}^{(2)} = 0.75$. **A)** In a plot of the two observables $\boldsymbol{E}$ and $\langle\boldsymbol{\tau_{D(A)}}\rangle_{\boldsymbol{F}}$, dynamics show as a curved line (color-coded by the contribution of the low-FRET species) that deviates from the diagonal static FRET-line (black). **B)** In the mean-variance representation (bottom), the static FRET-line is given by zero variance, while the dynamic line curves upwards. **C)** Using the difference between the first and second moments **Γ**, the static FRET-line transforms into a parabola, while the dynamic FRET-line is given by a line. All lifetimes and moments are normalized to the donor-only lifetime $\boldsymbol{\tau_{D(0)}}$ to simplify the illustration.

### 3.6 Multistate systems

The concept of FRET-lines is beneficial to characterize complex kinetic schemes with more than two states. Consider a kinetic network involving three conformational states:

$$\begin{array}{ccc} & x^{(2)} & \\ \swarrow\!\!\nearrow & & \nwarrow\!\!\searrow \\ x^{(1)} & \leftrightharpoons & x^{(3)} \end{array} \tag{48}$$

with the fraction of the three states $x^{(1)}$, $x^{(2)}$, and $x^{(3)}$, where $x^{(3)} = 1 - x^{(1)} - x^{(2)}$. The equations for the moments of the lifetime distribution are easily extended for the three-state system, but we can only eliminate one of the two free parameters $x^{(1)}$ and $x^{(2)}$. Consequently, the conversion function between species-averaged and fluorescence-averaged fluorescence lifetime additionally depends on one of the three species fractions, and we can only define the equivalent of FRET-lines by fixing this species fraction at a specific value. Because there are thus two degrees of freedom, multi-state systems are described by an area instead of a line (Figure 7A-C). This area is enclosed by limiting binary dynamic FRET-lines, which describe the direct exchange among two of the three states and are obtained by fixing one of the species fractions to zero. To define multi-state FRET-lines analogous to the two-state system, it is necessary to include an additional boundary condition. For example, the lines crossing the area in Figure 7A-C are obtained by varying one of the fractions while requiring the other two to be equal.

When more than two states are involved, the equilibrium population potentially lies enclosed by the limiting binary dynamic FRET-lines. The dynamic FRET area then serves as a reference to reveal the spatial and temporal heterogeneities of the sample. The color of the area in Figure 7A-C represents the population fraction of each of the states. The position of a single-molecule event on the plane is related to the occupancy fractions $x^{(i)}$ of the measured molecule, which in the limit of fast dynamics (or long observation time) tend to the equilibrium fractions. Using the representation of the moment difference (Figure 7C), it is possible to determine the state occupancies of the different states graphically from the two-dimensional plot (Figure 7D). As an example, we consider the high-FRET state 1 (red circle in Figure 7D) The red line connecting state 1 to the mixed population (orange) intersects the binary exchange line between the states 2 (green) and 3 (blue) at a given point (turquoise). Then, the state occupancy $x^{(1)}$ is obtained from the length of the segments of the red line, $a^{(1)}$ and $b^{(1)}$, defined by the position of the mixed population along the line, by:

$$x^{(1)} = \frac{a^{(1)}}{a^{(1)} + b^{(1)}}. \tag{49}$$

The state occupancies $x^{(2)}$ and $x^{(3)}$ are obtained analogously as described for $x^{(1)}$ above, as indicated by the dashed lines in Figure 7C. A detailed derivation of this expression is given in Supplementary Note 4.

In multi-state systems, FRET-lines are especially helpful in identifying the minimal set of states and their kinetic connectivity. This information can reduce the complexity of the kinetic model by eliminating exchange pathways, providing crucial information for further quantitative analysis of the dynamic network by dynamic photon distribution analysis or fluorescence correlation spectroscopy. This aspect of FRET-lines is illustrated in detail in the second part of the paper.

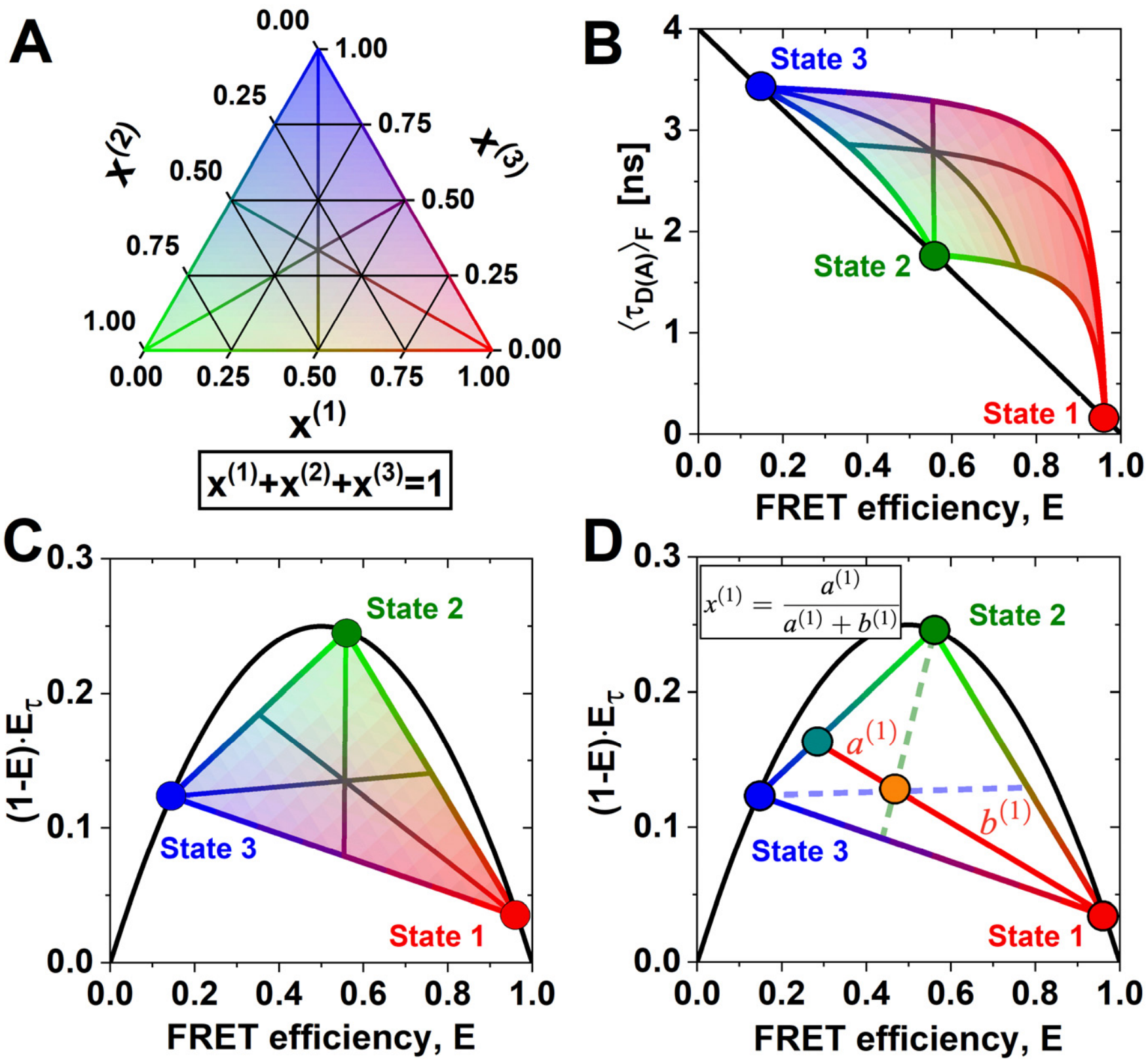

**Figure 7:** FRET-lines in three-state systems. **A)** Ternary plot of the fractions of the three species. The area is colored according to the contribution of the species (red: $\boldsymbol{x}^{(1)}$, green: $\boldsymbol{x}^{(2)}$, blue: $\boldsymbol{x}^{(3)}$). **B-C)** In the $(\boldsymbol{E}, \langle\boldsymbol{\tau}_{D(A)}\rangle_{F})$ parameter space (B), three-state mixing is described by an area that is confined by the two-state dynamic FRET-lines. In the moment representation (C), the dynamic mixing is simplified to a triangle with straight lines that describe the dynamic interconversion. Additionally, specific examples of limiting FRET-lines are given in A-C. For these lines, one species fraction is varied while the other two fractions are kept equal, e.g., $\boldsymbol{x}^{(1)} \in [\mathbf{0}, \mathbf{1}]$ and $\boldsymbol{x}^{(2)} = \boldsymbol{x}^{(3)} = \mathbf{0.5}\left(\mathbf{1} - \boldsymbol{x}^{(1)}\right)$. These lines intersect at $\boldsymbol{x}^{(1)} = \boldsymbol{x}^{(2)} = \boldsymbol{x}^{(3)} = \mathbf{1/3}$. **D)** In the moment representation, the species fractions can be determined by graphical analysis from the sections $\boldsymbol{a}^{(1)}$ and $\boldsymbol{b}^{(1)}$ of the connecting line between the position of the population (orange) and the pure state. The solid line indicates the procedure to determine the fraction of species 1, while the corresponding lines for species 2 and 3 are given as dashed lines.

## 3.7 Multi-exponential donor decays

Up to now, we have assumed that the fluorescence decay of the donor dye in the absence of the acceptor is single exponential. Experimentally, however, this condition is often violated due to the effect of the local environment on the tethered dyes. The most common mechanisms that affect the quantum yield of tethered dyes are the quenching of rhodamine or xanthene based dyes by electron-rich amino acids such as tryptophane through photoinduced electron transfer (PET)[83-85], and the enhancement of the fluorescence of cyanine-based dyes due to steric restriction and dye-surface interactions that modulate the cis-trans isomerization[86-88]. Also, the used organic dyes may consist of a mixture of isomers with distinct fluorescence properties. The effect of multi-exponential fluorescence decays of the donor fluorophore on the static and dynamic FRET-lines depends on the timescale of the dynamic exchange between the different donor states. This exchange may be fast (e.g., in the case of dynamic quenching by PET), on a similar timescale as the observation time of a few milliseconds (e.g., for sticking of the fluorophore to the biomolecular surface), or non-existent (e.g., in the case of an isomer mixture).

Here, we consider two limiting cases of donor dyes with multi-exponential fluorescence decays in the absence of FRET: a static mixture and fast exchange with complete averaging during the observation time. As before, we assume the homogenous approximation wherein the fluorescence properties are identical in different conformational states of the host molecule, i.e., the FRET rate $k_{RET}$ does not depend on the donor-only lifetime $\tau_{D(0)}$. In this case, the donor fluorescence decay in the absence of FRET is described by a distribution of fluorescence lifetimes $p(\tau_{D(0)})$:

$$f_{D|D}^{(D0)}(t) = k_{F,D} \int p(\tau_{D(0)}) \mathrm{e}^{-t/\tau_{D(0)}} d\tau_{D(0)}. \tag{50}$$

For the donor fluorescence decay in the presence of the acceptor, we now have to consider a distribution of donor fluorescence lifetimes and FRET rates:

$$f_{D|D}^{(DA)}(t) = k_{F,D} \int \int p(\tau_{D(0)}) p(k_{RET}) \mathrm{e}^{-t/\tau_{D(A)}} dk_{RET} d\tau_{D(0)}, \tag{51}$$

where the donor fluorescence lifetime in the presence of the acceptor is given by $\tau_{D(A)} = \left(\tau_{D(0)}^{-1} + k_{RET}\right)^{-1}$, and $p(\tau_{D(0)})$ and $p(k_{RET})$ correspond to the donor-only lifetimes and FRET rates distributions, respectively. Note that due to the homogenous approximation, we have factored the joint distribution of donor and FRET states, that is $p\left(\tau_{D(0)}, k_{RET}\right) = p(\tau_{D(0)}) p(k_{RET})$.

The moments of the fluorescence lifetime distribution then evaluate to:

$$\overline{\tau_{D(A)}^{v}} = \int_0^\infty \int_0^\infty p\left(\tau_{D(0)}\right) p(k_{RET}) \tau_{D(A)}^{v} dk_{RET} d\tau_{D(0)}, \tag{52}$$

or in the discrete case of distinct donor-only and FRET states:

$$\langle \tau_{D(A)}^{v} \rangle_x = \sum_{i,j} x_{D(0)}^{(j)} x_{RET}^{(i)} \left(\tau_{D(A)}^{(i,j)}\right)^{v}, \tag{53}$$

where $x_{D(0)}^{(j)}$ and $x_{RET}^{(i)}$ are the fractions of the donor and FRET states, respectively. From the moments, the observable $\langle \tau_{D(A)} \rangle_F$ is then readily calculated.

A more complex situation arises for the intensity-based FRET efficiency $E$ because the fluorescence intensities obtained for the different donor states are weighted by their respective quantum yields. Consequentially, it becomes impossible to define a single distance-related FRET efficiency. Instead, we define the proximity ratio $E_{PR}$ in analogy to eq. (11) based on the average fluorescence intensities detected in the donor and acceptor channel $\overline{F_{D|D}^{(DA)}}$ and $\overline{F_{A|D}^{(DA)}}$ by:

$$E_{PR} = \frac{\overline{F_{A|D}^{(DA)}}}{\overline{F_{D|D}^{(DA)}} + \overline{F_{A|D}^{(DA)}}} = 1 - \frac{\langle \tau_{D(A)} \rangle_x}{\langle \tau'_{D(0)} \rangle_x}, \tag{54}$$

where the species-averaged lifetimes are calculated over all donor and FRET states. The effective donor-only lifetime $\tau'_{D(0)}$ in the presence of quenching is defined as:

$$\tau'_{D(0)} = \tau_{D(A)} + \gamma'\left(\tau_{D(0)} - \tau_{D(A)}\right) \tag{55}$$

where the factor $\gamma'$ is given by the ratio of the quantum yields of the acceptor and donor fluorophores, $\gamma' = \frac{\Phi_{F,A}}{\Phi_{F,D}}$. See Supplementary Note 5 for a derivation of eq. (55). For the moment representation, the moment difference $\Gamma$ in the case of a mixture of donor states is then defined as:

$$\Gamma = (1 - E_{PR})\left(1 - \frac{\langle\tau_{D(A)}\rangle_F}{\langle\tau_{D(0)}\rangle_F}\right) = (1 - E_{PR})E_{PR,\tau}, \tag{56}$$

where $\langle\tau_{D(0)}\rangle_F$ is the intensity-weighted average donor fluorescence lifetime and $E_{PR,\tau} = 1 - \frac{\langle\tau_{D(A)}\rangle_F}{\langle\tau_{D(0)}\rangle_F}$. The effect of a mixture of two distinct photophysical states of the donor is illustrated in Figure 8 for the ($E$-$\langle\tau_{D(A)}\rangle_F$) parameter space (A-C) and in the moment representation (D-F). We consider two different donor lifetimes of $\tau_{D(0)}^{(1)}$ = 4 ns and $\tau_{D(0)}^{(2)}$ = 1 ns that correspond to distinct donor quantum yields of $\Phi_{F,D}^{(1)}$ = 0.8 and $\Phi_{F,D}^{(2)}$ = 0.2. When separate measurements are performed (Figure 8A,D), accurate FRET efficiencies $E$ can be calculated for each measurement, and the ideal static and dynamic FRET-lines are obtained. For the dynamic exchange, we assume equilibrium fractions of $x_{D(0)}^{(1)}$ = 0.25 and $x_{D(0)}^{(2)}$ = 0.75 for the two donor states. In the case of exchange on a timescale much slower than the observation time (Figure 8B,E), an individual correction of the different populations is not possible. For the proximity ratio $E_{PR}$, curved static FRET-lines are obtained for the two species as an effect of the averaging in eq. (54) . In the moment representation, this effect shows as an increased (for the species with $\tau_{D(0)}^{(2)}$ = 4 ns) or decreased (for the species with $\tau_{D(0)}^{(2)}$ = 1 ns) curvature of the static FRET-lines, while the linearity of the dynamic FRET-lines is retained. The effect of fast exchange between the different donor states (i.e., complete averaging during the observation time) is illustrated in Figure 8C,F. For the ($E$-$\langle\tau_{D(A)}\rangle_F$) parameter space, a single convex static FRET-line is obtained. This line falls between the curved static FRET-lines obtained for the slow exchange and intersects the $\langle\tau_{D(A)}\rangle_F$ axis at the intensity-weighted average donor fluorescence lifetime $\langle\tau_{D(0)}\rangle_F$ = 2.71 ns. In the moment representation (Figure 8F), the static FRET-line shows a higher curvature than the ideal static FRET-line (dashed gray line). Notably, even in the case of fast exchange between different donor states, the dynamic FRET-line in the moment representation remains linear (see Supplementary Note 5). Here, we have not considered the calculation of accurate FRET efficiencies for distributions of donor and acceptor states and instead introduced the proximity ratio. Using the general formalism introduced here, however, reference static and dynamic FRET-lines can still be defined even for uncorrected data if the corrections are instead accounted for in the FRET-lines.

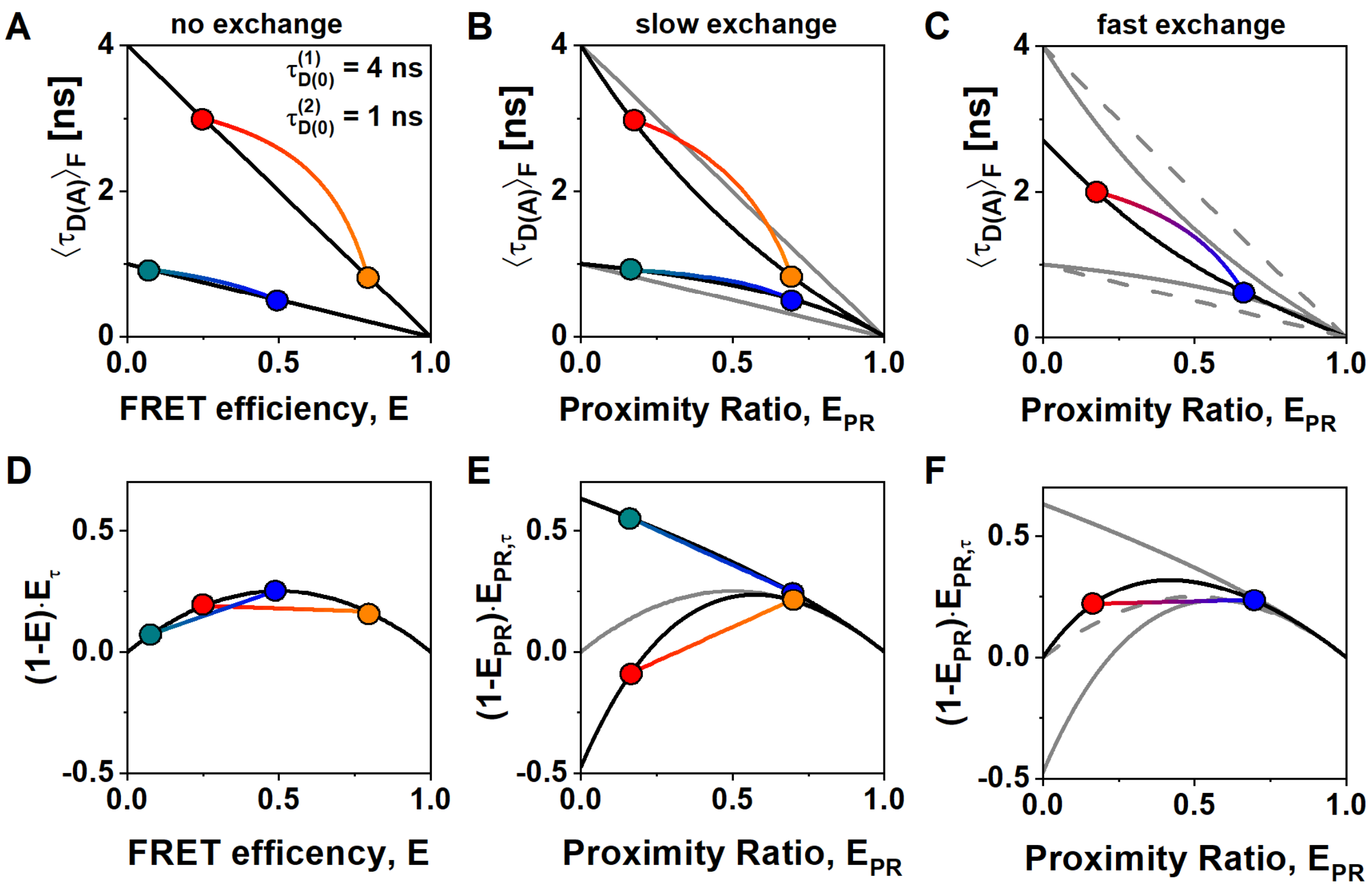

**Figure 8:** Static and dynamic FRET-lines for mixtures of distinct photophysical states of the donor in the ($E$-$\langle\tau_{D(A)}\rangle_F$) parameter space (A-C) and in the moment representation (D-F). **A,D)** Static and binary dynamic FRET lines for a superposition of two measurements with distinct donor-only lifetimes of $\tau_{D(0)}^{(1)}$ = 4 ns and $\tau_{D(0)}^{(2)}$ = 1 ns, corresponding to donor quantum yields of $\Phi_{F,D}^{(1)}$ = 0.8 and $\Phi_{F,D}^{(2)}$ = 0.2. Static FRET-lines are shown in black. The inter-dye distances of the two FRET species are $R_{DA}^{(1)}$ = 40 Å (blue, orange) and $R_{DA}^{(2)}$ = 60 Å (teal, red). The Förster radius of the donor state with $\tau_{D(0)}^{(1)}$ = 4 ns is $R_0$ = 50 Å. The acceptor quantum yield is chosen as $\Phi_{F,A}$ = 0.8. **B,E)** Static and binary dynamic FRET lines for the mixture of the two species shown in A and D in slow exchange, i.e., on a timescale slower than the observation time, with equilibrium fractions of $x_{D(0)}^{(1)}$ = 0.25 and $x_{D(0)}^{(2)}$ = 0.75. For the proximity ratio $E_{PR}$, a curvature of the static FRET-lines arises even in the absence of dynamics. Gray lines correspond to the ideal static FRET-lines shown in A. Note that in E, the moment difference $(1 - E_{PR})E_{PR,\tau}$ can assume negative values. **C,F)** Static and binary dynamic FRET-lines for the mixture of the two species shown in A in fast exchange, i.e., for complete averaging during the observation time, with equilibrium fractions of $x_{D(0)}^{(1)}$ = 0.25 and $x_{D(0)}^{(2)}$ = 0.75. Solid gray lines correspond to the static FRET-lines for slow exchange as shown in B and E. Dashed gray lines correspond to the ideal static FRET-lines of the two species as shown in A and D. Note that the static FRET-line is convex in this case.

# 4 Practical aspects and application

## 4.1 Dye-linker dynamics

So far, we have assumed that a conformational state of the molecule is described by a single donor fluorescence lifetime and will be represented by a point lying on the ideal, diagonal static FRET-line. A heterogeneous mixture of molecules with different FRET efficiencies, i.e., different donor-acceptor distance, would then follow this static FRET-line. This line does, however, not describe experimental data accurately. It is consistently observed that the population mean of static molecules deviates from the ideal static-FRET-line, exhibiting a bias towards longer fluorescence-weighted donor lifetimes, $\langle\tau_{D(A)}\rangle_F$. The deviation from the ideal static FRET-line is caused by the use of long, flexible linkers of 10-20 Å length that tether the fluorophore to the biomolecules[40, 74, 89]. Fast variations of the donor-acceptor distance $R_{DA}$ during the observation time result in a distribution of donor lifetimes $p(\tau_{D(A)})$ that are sampled in each single-molecule event (Figure 9A-B). Due to the finite width of the distribution, the population is thus shifted towards longer donor fluorescence lifetimes, whereby the deviation from the ideal static FRET-line increases with increasing linker length and thus distribution width $\sigma_{DA}$ (Figure 9C). Recently, we estimated that the translational diffusion coefficient of dyes tethered to proteins is on the order of 5 to 10 Å$^2$/ns[50]. Assuming free three-dimensional diffusion, this estimate of 10 Å$^2$/ns would translate to an expected root-mean-square displacement $\langle x\rangle = \sqrt{6Dt}$ of ~10 Å per 2 ns, resulting in significant changes of the inter-dye distance during the excited state lifetime[90, 91]. However, it is to be expected that the effective displacement is reduced due to the restriction of the dye's movement by the linker. Under the assumption that the diffusion of the fluorophore is slow compared to the fluorescence lifetime, the fluorescence decays may be approximated by a static distribution of distances[50, 74, 92]. The observation time for every single molecule on the order of milliseconds is long compared to the diffusional motion of the dyes, resulting in complete averaging of the spatial distribution of the dyes around their attachment during the observation time. Under these assumptions, we can calculate the averaged quantities and moments of the lifetime distribution based on the equilibrium distance distribution.

Different approaches for modeling the spatial distribution of tethered fluorophores have been developed[74, 93-95]. In the accessible volume (AV) approach[89], possible positions of the fluorophore in the three-dimensional space are identified through a geometric search algorithm. By considering all possible combinations of donor-acceptor distances, the inter-dye distance distribution can be obtained from the accessible volumes of the donor and acceptor dyes. Extensions of the AV approach have incorporated surface trapping of fluorophores[3, 50] or accounted for the energetic contributions of linker conformation[96, 97]. More accurate models of the spatial distribution of tethered dyes are obtained by coarse-grained[50, 95, 98] or all-atom[99, 100] molecular dynamics simulations, from which explicit inter-dye distance distributions may be obtained. However, as will be discussed below, the contribution of the linker flexibility is mainly defined by the width of the inter-dye distance distribution, $\sigma_{DA}$, and shows only a weak dependence on the explicit shape of the distribution. Experimentally, the width of the linker distribution may be obtained from the fluorescence decay of the donor by modeling the fluorescence decays with a model function that includes a distribution of distances[50, 97]. Alternatively, by using the two-dimensional histogram of $\langle\tau_{D(A)}\rangle_F$ vs. $E$, one can vary the width parameter of the static FRET-line such that it intersects with the population of static molecules. Typically, we consider a fixed standard deviation of $\sigma_{DA}$~6 Å that satisfies benchmarking experiments on rigid DNA molecules[40, 74].

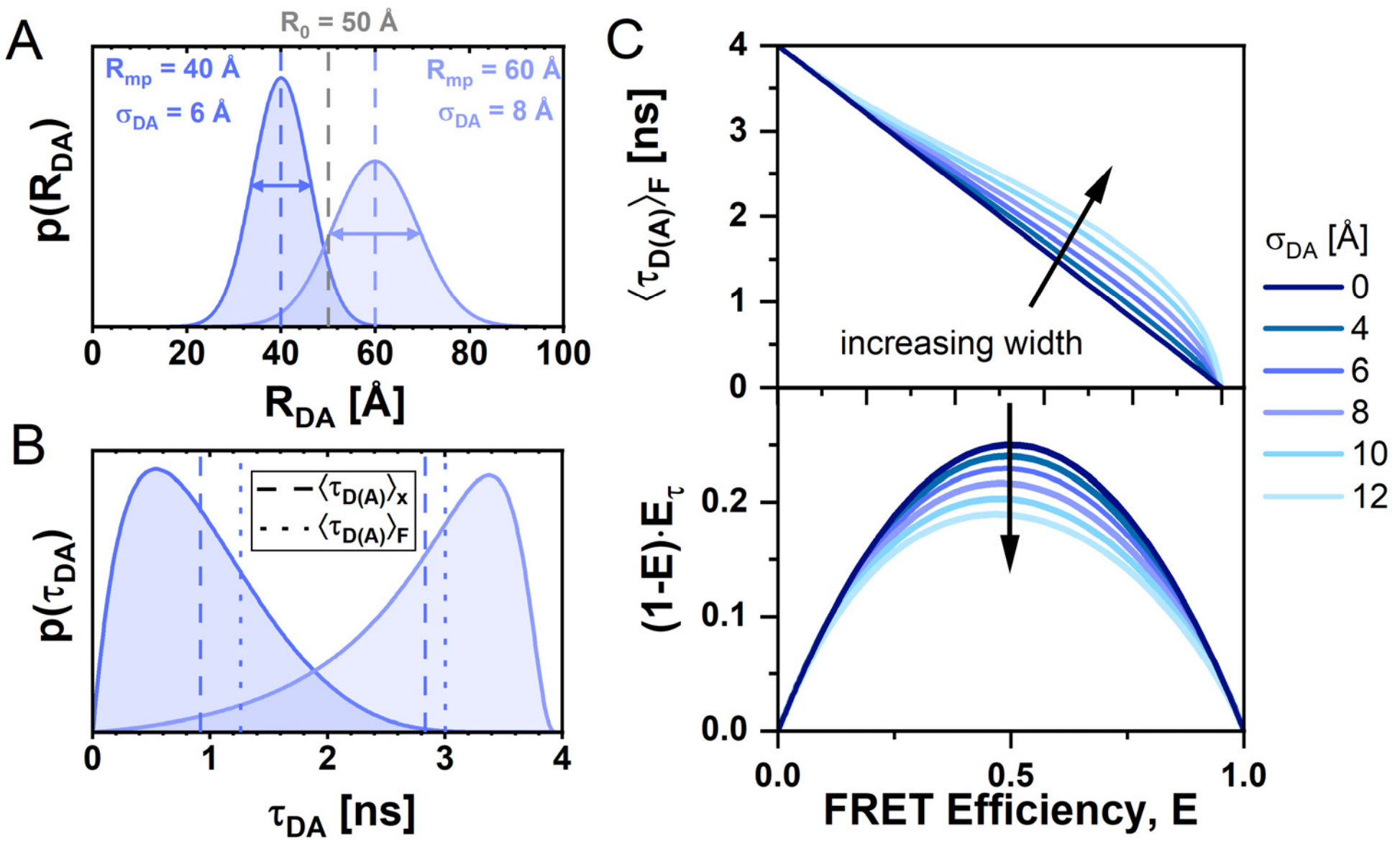

**Figure 9:** FRET-lines in the presence of fast distance fluctuations (linker dynamics). **A-B)** The distribution of inter-dye distances $\boldsymbol{p(R_{DA})}$ (A) is transformed into the corresponding distribution of donor fluorescence lifetimes $\boldsymbol{p(\tau_{D(A)})}$ (B) by the Förster relation $\boldsymbol{\tau_{D(A)} = \tau_{D(0)}\left[1+\left(\frac{R_0}{R}\right)^6\right]^{-1}}$. The Förster radius, $\boldsymbol{R_0}$, is 50 Å. **C)** The broadening of the lifetime distribution causes a deviation of the static FRET-line in the $(\boldsymbol{E}, \langle\boldsymbol{\tau_{D(A)}}\rangle_{\boldsymbol{F}})$ representation towards higher values of the intensity-weighted average lifetime $\langle\boldsymbol{\tau_{D(A)}}\rangle_{\boldsymbol{F}}$. Higher values for the width of the distance distribution result in stronger deviation. In the moment representation, the contribution of the distribution width causes an inward shift of the static FRET-line.

### 4.1.1 General description in the presence of distance fluctuations

Before we discuss different models of the equilibrium distribution of inter-dye distances, we describe how the moments of the lifetime distribution can generally be calculated in the presence of distance heterogeneity. It is assumed that the fluctuations of the inter-dye distance due to the dynamics of the linkers ($\tau_{\text{linker}}$) are (i) slow compared to the fluorescence lifetime $\tau_{D(A)}$, leading to a distribution of lifetimes, but (ii) fast compared to the observation time $T_{\text{obs}}$ (limited by the diffusion time $t_{\text{diff}}$), allowing us to treat the distance distribution as stationary:

$$\tau_{D(A)} \ll \tau_{\text{linker}} \ll T_{\text{obs}} \approx t_{\text{diff}} \tag{57}$$

The fluorescence lifetime of the donor, $\tau_{D(A)}$, is related to the inter-dye distance, $R_{DA}$, by:

$$\tau_{D(A)}(R_{DA}) = \tau_{D(0)}\left[1+\left(\frac{R_0}{R_{DA}}\right)^6\right]^{-1}. \tag{58}$$

Then, the moments of the fluorescence lifetime distribution, in terms of a distribution of inter-dye distances $p(R_{\text{DA}})$, are obtained from equation (32) by a change of variables $\tau_{D(A)} \rightarrow \tau_{D(A)}(R_{DA})$ given by eq. (58):

$$\overline{\tau_{D(A)}} = \int_0^\infty p(R_{DA})\,\tau_{D(A)}(R_{DA})dR_{DA}$$
$$\overline{\tau_{D(A)}^2} = \int_0^\infty p(R_{DA})\left(\tau_{D(A)}(R_{DA})\right)^2 dR_{DA} \quad (59)$$

which allow us to calculate the different quantities used for the representations above as a function of these two moments. In general, the distance distribution will depend on the donor-acceptor separation $R_{mp}$ (here defined as the distance between the mean positions) and a set of parameters $\Lambda$ that describe the shape of the distribution (e.g., its width). To construct the static FRET-line for a given distance distribution model $p(R_{DA}|R_{mp},\Lambda)$, we vary the mean donor-acceptor distance, $R_{mp}$, and compute the fluorescence averaged lifetime and the FRET efficiency from the moments of the lifetime distribution given by equation (59). The integrals in equation (59) are difficult to solve analytically, even for the simple case of a normal distribution of distances, due to the sixth-power dependence between $\tau_{D(A)}$ and $R_{D(A)}$. However, they can be calculated numerically for arbitrary models of the distribution. The shape of the distribution may potentially also depend on the conformation of the biomolecule and thus the donor-acceptor distance $R_{mp}$, in which case the shape parameters would depend on the conformation, $\Lambda \rightarrow \Lambda(R_{mp})$. From the experimental observables $E$ and $\langle\tau_{D(A)}\rangle_F$, we can only access the first and second moment of the lifetime distribution. Consequentially, it is not possible to address the shape of the lifetime distribution $p(\tau_{D(A)})$ explicitly. The same dynamic shift can thus be observed for different distributions, as long as their mean and variance (or equivalently, their first and second moments) are identical.

### 4.1.2 FRET-lines of flexibly linked dyes

In practice, it is desirable to have access to simple reference static FRET-lines that can be used for graphical analysis of the measured data and the comparison of different models. To this end, an analytical model for the distance distribution is required. We first consider the simple case where the distribution of the dye positions in space follows an isotropic normal distribution (Figure **10**A). This model can be interpreted as two ideal (Gaussian) chain polymer linkers which are separated by a distance $R_{mp}$ and show no interaction with the biomolecule. In this case, the inter-dye distance vector, $\boldsymbol{R}_{DA}$, is also normally distributed with width $\sigma_{DA}=\sqrt{\sigma_D^2+\sigma_A^2}$ where $\sigma_D$ and $\sigma_A$ are the width of the spatial distributions of the donor and acceptor dyes, respectively. The distribution of inter-dye distances, $R_{DA}$, is then given by the non-central $\chi$-distribution with the distance between the mean positions of the dyes, $R_{mp}$, as the non-centrality parameter and $\sigma_{DA}$ as the width parameter:

$$\chi(R_{DA}|R_{mp},\sigma_{DA}) = \frac{R_{DA}}{R_{mp}}\left[\mathcal{N}^{(+)}(R_{DA}|R_{mp},\sigma_{DA}) - \mathcal{N}^{(+)}(R_{DA}|-R_{mp},\sigma_{DA})\right];$$
$$\mathcal{N}^{(+)}(R_{DA}|R_\mu,\sigma_{DA}) = \frac{1}{\sigma_{DA}\sqrt{2\pi}}e^{-\frac{1}{2}\left(\frac{R_{DA}-R_\mu}{\sigma_{DA}}\right)^2} \text{ with } x \geq 0. \quad (60)$$

Here, $\mathcal{N}^{(+)}(R_{DA}|R_\mu,\sigma_{DA})$ is a part of a normal distribution with a mean $\mu$ and a width $\sigma$ taken at non-negative values $x \geq 0$ (positive truncation) to avoid non-sensical distance values below zero. At small variance-to-mean ratios (i.e., at large distances), the $\chi$-distribution tends to the normal distribution. Therefore, the distribution $p(R_{DA})$ may be approximated by a normal distribution with mean inter-dye distance $R_{mp}$:

$$p(R_{DA}|R_{mp},\sigma_{DA}) \approx \lim_{\frac{\sigma_{DA}}{R_{mp}}\rightarrow 0}\chi(R_{DA}|R_{mp},\sigma_{DA}) = \mathcal{N}^{(+)}(R_{DA}|R_{mp},\sigma_{DA}). \quad (61)$$

As experimental distances are usually larger than 35 Å and the apparent distribution widths of the inter-dye distance are on the order of 5-10 Å, this approximation is often valid. However, for broader distributions, the truncation of the normal distribution with $R_{DA} \geq 0$ results in a significant deviation from the $\chi$-distribution at small inter-dye distances (see Figure 10B). Compared to the $\chi$-distribution, the truncated Gaussian distance model overestimates the contribution of small distances (corresponding to high FRET efficiencies). Overall, this results only in minor deviations of the generated static FRET-lines compared to the $\chi$-distribution, which are most pronounced at large distribution widths and high FRET efficiencies (Figure 10C). However, the two models show significant deviations in terms of the average FRET efficiency at identical center distances $R_{mp}$. To illustrate this effect, we plot the change of the average FRET efficiency at constant $R_{mp}$ and increasing $\sigma_{DA}$ for the Gaussian and $\chi$ distance distributions in Figure 10C (see vertical blue and red lines, respectively). The deviation of the average FRET efficiencies between the two models increases with increasing width $\sigma_{DA}$. Notably, the interpretation of average FRET efficiencies in terms of the distance between the mean positions of the dyes $R_{mp}$ is thus biased by choice of the model function for the linker distribution.

In summary, the choice of the distance distribution model function has only a minor effect on the shape of the static FRET-lines, which is mainly determined by the width parameter. However, we propose that the $\chi$-distribution should be preferred for the interpretation of linker-averaged FRET efficiencies in terms of physical distances when broad linker distributions are expected.

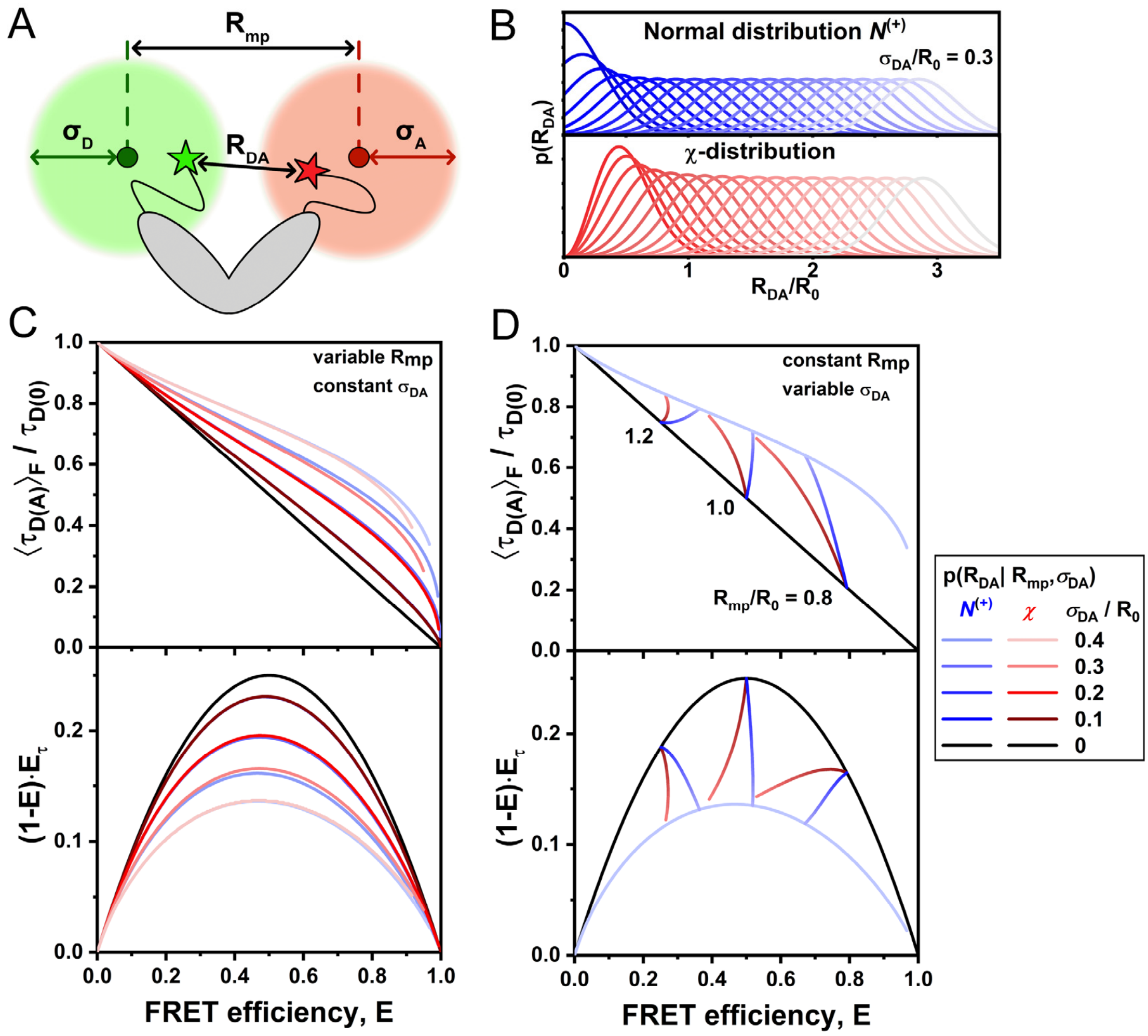

**Figure 10: Calculation of static FRET-lines with dye linker diffusion: Difference between normal and $\chi$ distribution. A)** The donor and acceptor dyes are tethered to the biomolecule by flexible linkers. As a result, they can occupy an accessible volume described by the dyes' mean position and a width parameter $(\sigma_D, \sigma_A)$ that describes the linker flexibility. The distance $R_{mp}$ describes the distance between the mean positions of the dyes, while $R_{DA}$ is given by the instantaneous distance between the two dyes. **B)** Normally distributed (top) and $\chi$-distributed (bottom) inter-dye distance distribution with constant width parameter $\sigma_{DA}^{(l)} = 15$ Å at $R_0 = 50$ Å. Notice that the normal distributions are truncated at small inter-dye distances. **C)** Linker-corrected static FRET-lines for a normal distribution (dashed lines) and $\chi$-distribution (solid lines) at $\sigma_{DA}^{(l)}$ =5, 10, 15 and 20 Å (from dark to light) and $R_0 = 50$ Å in the $(E, \langle\tau_{D(A)}\rangle_F)$ (top) and moment representation (bottom). The left panel shows the FRET-lines for variable center distance at constant distribution width and the right panel shows the FRET-lines for constant center distance and variable distribution width. While the shape of the resulting FRET-lines for the normal and $\chi$-distribution are similar (left panel), a systematic deviation is observed for the linker-averaged FRET-efficiency at increasing width parameter (right panel).

## 4.2 Conformational dynamics in the presence of linker broadening

So far, we have only considered the effects of linker broadening for static conformations of molecules. In the presence of conformational dynamics, the total distance heterogeneity will be given by the combination of both contributions. If the timescale of the dynamics of the linkers is comparable to the timescale of conformational dynamics (e.g., for intrinsically disordered proteins), one would require a joint probability distribution of the conformational dynamics and the linker configuration. Generally, however, the dynamics of tethered dyes are much faster than the dynamics of the host molecule. It can then be assumed that the linker distribution is entirely sampled for every single molecule, allowing it to be treated as a stationary distribution for each conformational state. Consider that the conformational states are characterized by different mean donor-acceptor distances $R_{mp}^{(c)}$ which are populated with probability $p(R_{mp}^{(c)}|\Lambda^{(\mathrm{dyn})})$, where $\Lambda^{(\mathrm{dyn})}$ is the set of parameters describing the conformational dynamics, i.e., the transition rate matrix. The linker distributions in the different conformational states are given by $p(R_{DA}|R_{mp}^{(c)}, \Lambda_l^{(c)})$, whereby the parameters of the linker distance distribution, $\Lambda_l^{(c)}$, may potentially be different for the conformational states. The combined distance distribution $p(R_{DA})$ then takes the general form:

$$p(R_{DA}) = \int p\left(R_{DA}|R_{mp}^{(c)}, \Lambda_l^{(c)}\right) p(R_{mp}^{(c)}|\Lambda^{(\mathrm{dyn})}) dR_{mp}^{(c)}, \tag{62}$$

where the integration is performed over all possible conformational states.

We first turn to the specific case wherein we describe the linker distribution in each conformational state by a $\chi$-distribution characterized by the mean inter-dye distance $R_{mp}^{(c)}$ and its corresponding width, $\sigma_{D(A)}^{(c)}$. For the case of two conformational states, the combined distribution of inter-dye distances integral in equation (62) then simplifies to the discrete sum:

$$p(R_{DA}) = \sum_{c=1}^{2} x^{(c)} \chi\left(R_{DA}\middle|R_{mp}^{(c)}, \sigma_{DA}^{(c)}\right). \tag{63}$$

The dynamic FRET-line in the presence of flexible linkers is obtained by varying the species fraction $x^{(1)}$ and numerically calculating the moments, as described above in eq. (59).

### 4.2.1 Separating the contributions of linkers and conformational dynamics

The presented approach is applicable if an analytical model is available to describe the contributions of linker dynamics to the broadening of the distance distribution. In the experiment, however, we might not know the exact distribution but are able to measure the moments of the lifetime distribution in the distinct (static) conformational states experimentally. Without having to model the linker distribution explicitly, we thus have access to the linker-averaged moments of each conformational state $c$, $\langle\tau_{D(A)}\rangle_l^{(c)}$ and $\langle\tau_{D(A)}^2\rangle_l^{(c)}$, defined as:

$$\langle\tau_{D(A)}^v\rangle_l^{(c)} = \int \tau_{D(A)}^v(R_{DA}) p\left(R_{DA}|R_{mp}^{(c)}, \Lambda_l^{(c)}\right) dR_{DA}. \tag{64}$$

For the general description of the distance distribution given in equation (62), the moments of the lifetime distribution in the presence of conformational dynamics are given by the double integral:

$$\langle\tau_{D(A)}^v\rangle_x = \int\int \tau_{D(A)}^v(R_{DA}) p\left(R_{DA}|R_{mp}^{(c)}, \Lambda_l^{(c)}\right) p(R_{mp}^{(c)}|\Lambda^{(dyn)}) dR_{mp}^{(c)} dR_{DA}. \tag{65}$$

To separate the contributions of the conformational dynamics and the linker fluctuations, we rearrange the integral to first integrate over the linker distribution, which is possible due to the separation of the timescales of the linker and conformational dynamics:

$$\langle \tau_{D(A)}^v \rangle_x = \int \left[ \int \tau_{D(A)}^v(R_{DA}) p\left(R_{DA} | R_{mp}^{(c)}, \Lambda_l^{(c)}\right) dR_{DA} \right] p(R_{mp}^{(c)} | \Lambda^{(dyn)}) dR_{mp}^{(c)}$$
$$= \int \langle \tau_{D(A)}^v \rangle_l^{(c)} p\left(R_{mp}^{(c)} \middle| \Lambda^{(dyn)}\right) dR_{mp}^{(c)}. \quad (66)$$

Thus, in the calculation of the moments, we can separate the contributions of the linker distribution by first evaluating the moments of the linker distribution in each conformational state, $\langle \tau_{D(A)}^v \rangle_l^{(c)}$, which is then used to evaluate the moments in the presence of conformational dynamics. From equation (66), it can be shown that the variances of the linker distributions and the conformational dynamics are additive, that is:

$$\mathrm{Var}(\tau_{D(A)}) = \mathrm{Var}^{(c)}\left(\langle \tau_{D(A)} \rangle_l^{(c)}\right) + \overline{\mathrm{Var}^{(l)}(\tau_{D(A)})}^{(c)}, \quad (67)$$

where $\mathrm{Var}^{(c)}\left(\langle \tau_{D(A)} \rangle_l^{(c)}\right)$ is the variance of the linker-averaged lifetime for all conformational states and $\overline{\mathrm{Var}^{(l)}(\tau_{D(A)})}^{(c)}$ is the average of the linker-variances over the different states (see Supplementary Note 3.5 for a derivation of eq. (67)).

The importance of these equations is that the contributions of the linkers can be treated separately from the conformational dynamics. We only require to know the linker-averaged moments, $\langle \tau_{D(A)}^v \rangle_l^{(c)}$, of the lifetime distribution of the different conformational states, which may be calculated for a particular model of the linker distance distribution (eq. (64)) or be obtained from the observables $E$ and $\langle \tau_{D(A)} \rangle_F$ of the pure states. The linker-averaged moments then replace the corresponding powers of the pure state lifetimes in the calculation of dynamic FRET-lines (eq. (39)). Thus, the moments of the lifetime distribution for two-state dynamic exchange, i.e., $c \in \{1,2\}$, are given by:

$$\langle \tau_{D(A)} \rangle_x = x^{(1)} \langle \tau_{D(A)} \rangle_l^{(1)} + \left(1 - x^{(1)}\right) \langle \tau_{D(A)} \rangle_l^{(2)}$$
$$\langle \tau_{D(A)}^2 \rangle_x = x^{(1)} \langle \tau_{D(A)}^2 \rangle_l^{(1)} + \left(1 - x^{(1)}\right) \langle \tau_{D(A)}^2 \rangle_l^{(2)}, \quad (68)$$

from which the dynamic FRET-lines in the different representations are obtained by varying the species fraction $x^{(1)}$ as described before. Therefore, the linearity of the dynamic mixing of the moments for conformational dynamics is still valid in the presence of linker fluctuations. Dynamic FRET-lines thus stay linear in the moment representation.

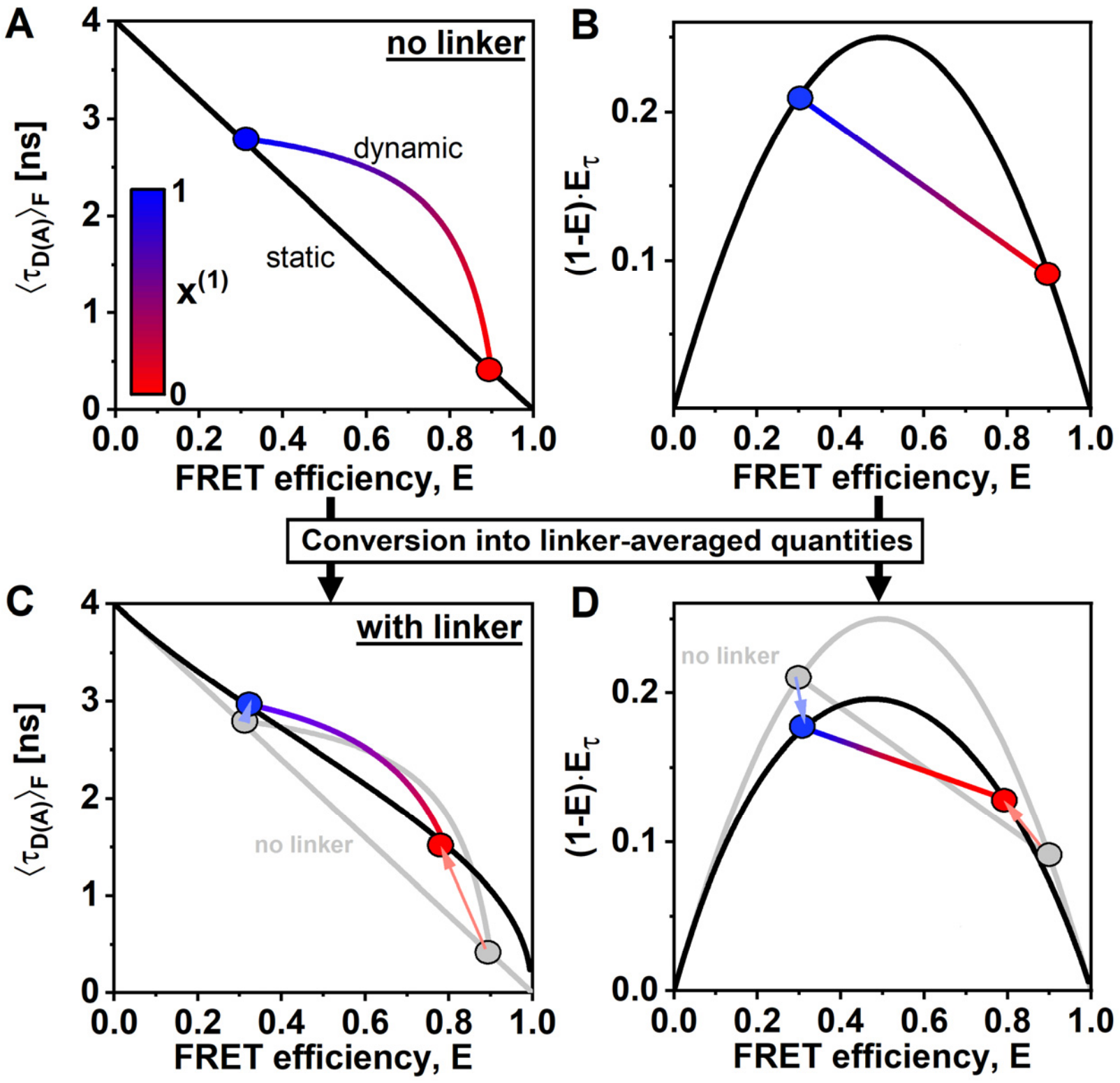

**Figure 11** Dynamic FRET-lines in the presence of flexible linkers. **A-B)** Static and dynamic FRET-lines in the $(\boldsymbol{E}, \langle\boldsymbol{\tau_{D(A)}}\rangle_{\boldsymbol{F}})$ parameter space (A) and in the moment representation (B) in the absence of flexible linkers. The static FRET-line is given in black, and the dynamic FRET-line is colored according to the relative contribution of the two species. **C-D)** Static and dynamic FRET-lines in the presence of flexible linkers (black and colored lines) are shown in the $(\boldsymbol{E}, \langle\boldsymbol{\tau_{D(A)}}\rangle_{\boldsymbol{F}})$ parameter space (C) and in the moment representation (D). The FRET-lines in the absence of flexible linkers, as shown in A-B, are displayed in gray. Arrows indicate the shift of the pure states after averaging over the linker distance distribution. No simple relation exists between the dynamic FRET-line in the presence and absence of flexible linkers for the $(\boldsymbol{E}, \langle\boldsymbol{\tau_{D(A)}}\rangle_{\boldsymbol{F}})$ representation (C). In the moment representation (D), the linear relationship for the dynamic exchange is retained in the presence of flexible linkers. The dynamic FRET-line is simply obtained by connecting the shifted coordinates of the pure states in the presence of flexible linkers. The curves are obtained for a donor lifetime of $\boldsymbol{\tau_{D(0)}} = \mathbf{4}$ ns, a Förster radius of $\boldsymbol{R_0} = \mathbf{50}$ Å and interdye distances $\boldsymbol{R_{mp}^{(1)}} = \mathbf{57.5}$ Å and $\boldsymbol{R_{mp}^{(2)}} = \mathbf{34.5}$ Å. The distribution width for the linker broadening was $\boldsymbol{\sigma_{DA}} = \mathbf{7.5}$ Å.

Dynamic FRET-lines in the presence of flexible linkers are illustrated in Figure 11. In the $(E, \langle\tau_{D(A)}\rangle_F)$ representation, it is not possible to perform a simple graphical construction of the dynamic FRET-line for flexible linkers. In the moment representation, however, the dynamic FRET-line for flexible linkers is simply obtained by connecting the linker-averaged coordinates of the two states. This simplification has important implications for the accurate description of dynamic FRET-lines in complex experimental systems, where no model for the linker distribution is available. Consider, for example, the case that the width of the linker distribution depends on the inter-dye distance in an unknown manner. In this case, it is not possible to obtain a general static FRET-line. However, with the presented formalism, we only require knowing the positions of the limiting static states, which are sufficient to fully describe the corresponding dynamic FRET-line. For the $(E, \langle\tau_{D(A)}\rangle_F)$ representation, the linker-averaged first and second moment of the limiting states, $\langle\tau_{D(A)}\rangle_l^{(i)}$ and $\langle\tau_{D(A)}^2\rangle_l^{(i)}$, can be determined from the averaged FRET observables $E$ and $\langle\tau_{D(A)}\rangle_F$ of the static populations, from which the dynamic FRET-line is obtained by a linear combination of the moments (eq. (68)) and conversion back into the $(E, \langle\tau_{D(A)}\rangle_F)$ parameter space. In the moment representation, the dynamic FRET-line is simply obtained graphically by connecting the conformational states with a straight line. Thus, for the construction of the dynamic FRET-line, it is generally not required to know the linker distance distribution in analytical form. If structural information is available, the linker distribution may also be obtained from the accessible volumes of the dyes in distinct conformations. In a three-state system, the dynamic FRET-lines in the presence of flexible linkers are shifted towards the center of the area enclosed by the limiting lines (Figure 12).

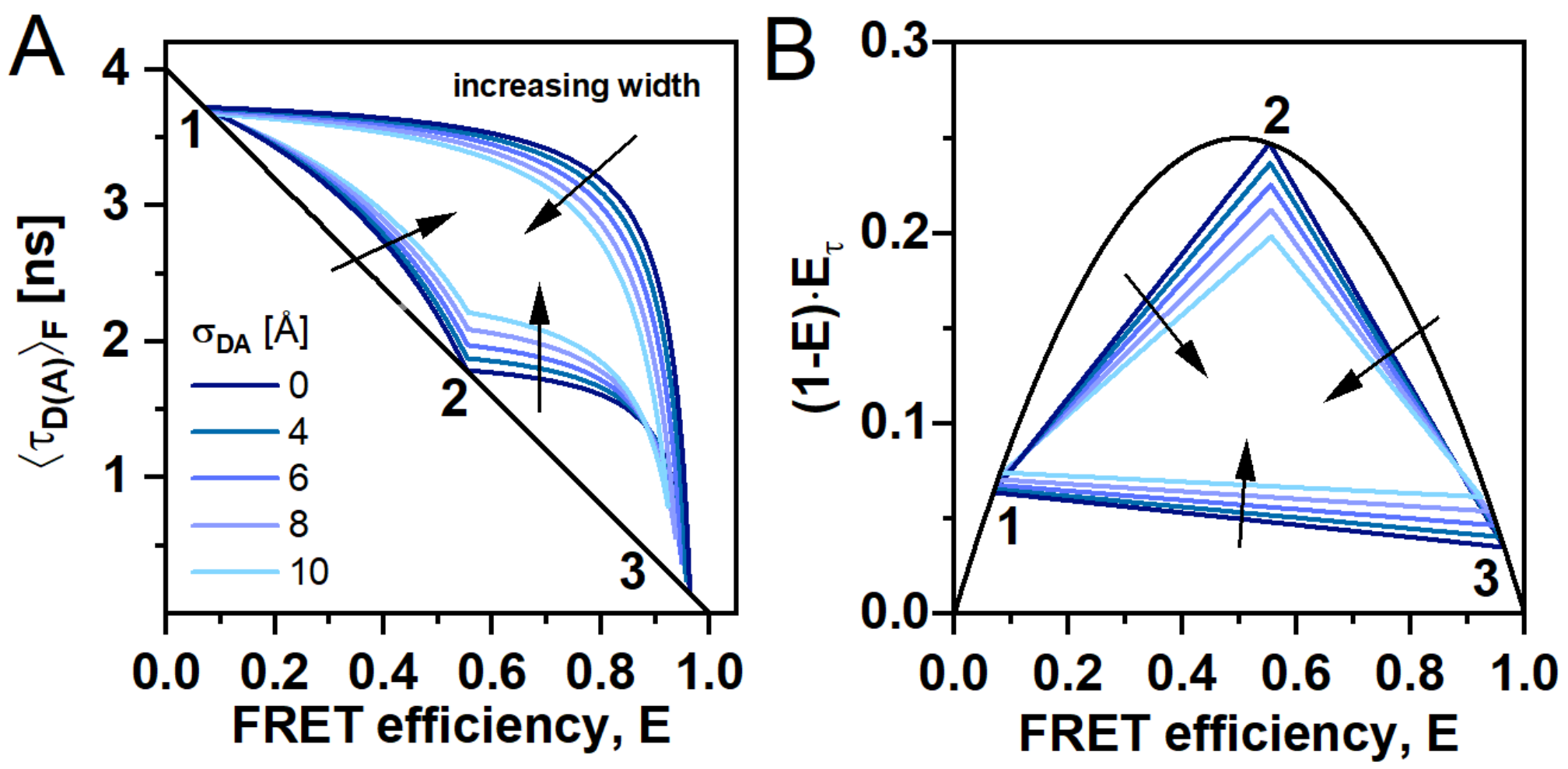

**Figure 12:** Dynamic FRET-lines in the presence of flexible linkers in three-state systems for the $(\boldsymbol{E}, \langle\boldsymbol{\tau_{D(A)}}\rangle_{\boldsymbol{F}})$ parameter space (**A**) and the moment representation (**B**). With increasing linker width, the dynamic FRET-lines are shifted inwards for both representations. In the moment representation, the linearity of the dynamic FRET-lines is retained in the presence of flexible linkers. The distances between the mean positions of the dyes, $\boldsymbol{R_{mp}}$, for the three states are 30, 50, and 80 Å with a Förster radius $\boldsymbol{R_0}$ = 52 Å.

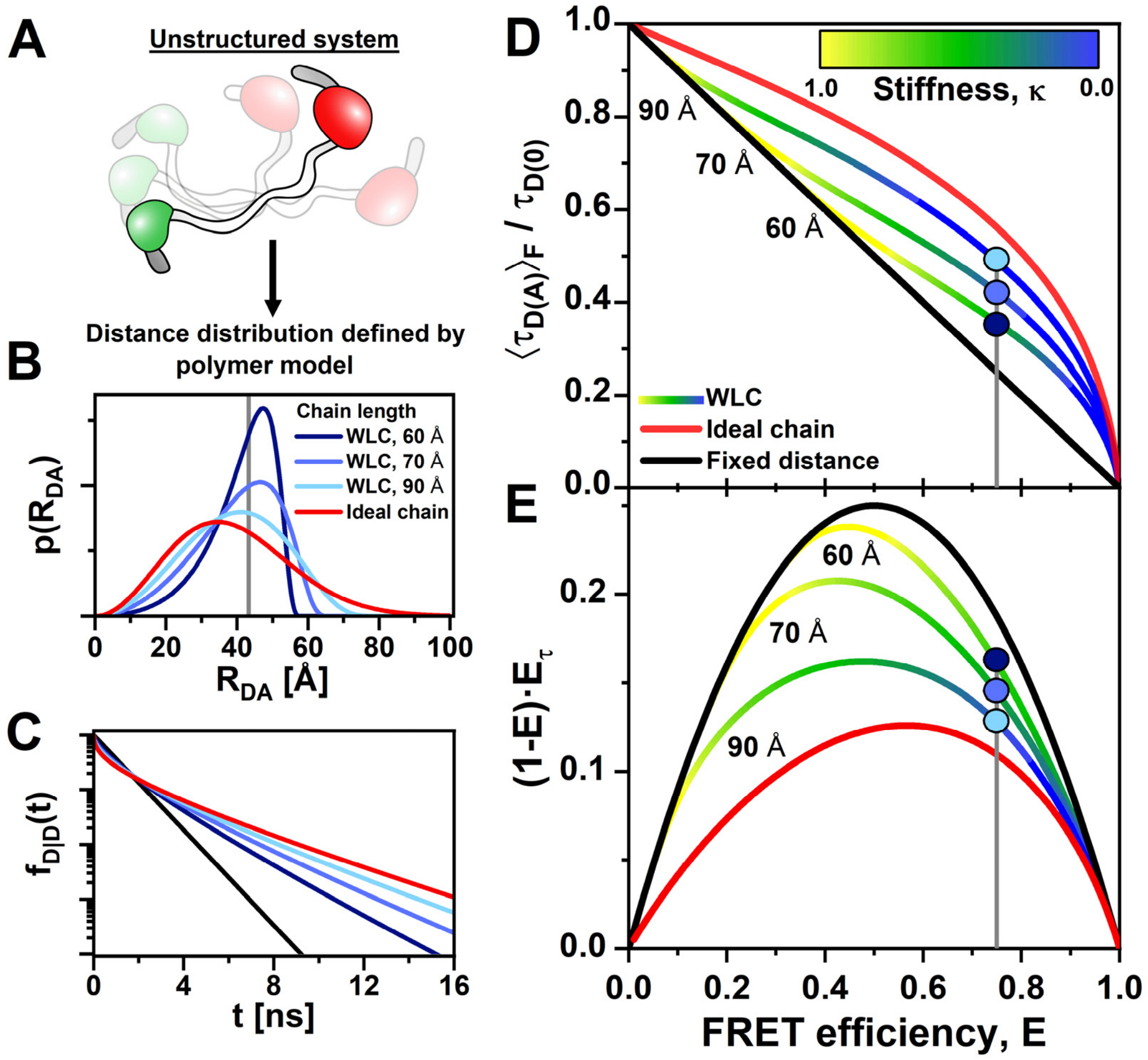

**Figure 13:** FRET-lines for disordered states. **A)** Unstructured biomolecules, such as intrinsically disordered proteins, rapidly interconvert between an ensemble of structures. **B)** The interdye distance distributions $\boldsymbol{p(R_{DA})}$ of an unstructured system may be described by polymer models, here given by worm-like chain (WLC) of different lengths of 60, 70 and 90 Å. **C)** The distance distributions define the corresponding fluorescence decays $\boldsymbol{f_{D|D}(t)}$. **D)** FRET-lines for the WLC model at different polymer lengths and stiffness $\boldsymbol{\kappa}$. The examples shown in A-B are given as colored dots. The static FRET-line for fixed dyes is given as a black line, and the FRET-line for the Gaussian chain model is given as a red line. **E)** The same data as shown in D in the moment representation. FRET-lines were calculated for a donor-lifetime $\boldsymbol{\tau_{D(0)}} = \mathbf{4}$ ns, and a Förster distance $\boldsymbol{R_0} = 52$ Å.

### 4.3 FRET-lines of flexible polymers

In the previous section, we have described the contributions of the flexible linkers to the static and dynamic FRET-lines. Through the stationary distance distribution, the effects of the fast linker dynamics could be accounted for. In principle, the linkers are equivalent to short, flexible polymers, which may be treated analogously to the procedure described above when a model for the equilibrium distance distribution is available. In the following, we present FRET-lines for different polymer models in the context of the potential application to the study of flexible biological polymers such as unfolded or intrinsically disordered proteins.

#### 4.3.1 Disordered states

Single-molecule FRET measurements are particularly suited to characterize biomolecules with partial or lack of stable tertiary structure, such as unfolded proteins, intrinsically disordered proteins (IDPs), and proteins with intrinsically disordered regions (IDRs)[101-106]. In the one-dimensional analysis of FRET efficiency histograms, the information about the fast dynamics of these systems is hidden, and complementary methods such as small-angle X-ray scattering (SAXS) have to be employed to assert the presence of disorder[102-104, 107]. In contrast, the knowledge of the fluorescence weighted average lifetime $\langle\tau_{D(A)}\rangle_F$ in addition to the FRET efficiency, $E$ allows dynamics to be identified directly from the single-molecule FRET experiment. As described above, these quantities allow one to address the mean and variance of the distribution of fluorescence lifetimes and thus contain information about the mean and variance of the distribution of inter-dye distances (Figure 13A-C). Here, we outline how to exploit this information to characterize IDPs and proteins with IDRs by means of FRET-lines of polymer models.
The conformational dynamics of IDPs or proteins with IDRs are usually fast compared to the observation time, with relaxation times on the order of 100 ns to several µs [108-110]. In the measurement, a single population is then observed at a position that corresponds to the average over the continuous distribution of conformations. We first consider that the disordered system is described by a Gaussian chain model (GC). This model approximates the conformational space by a quasi-continuum of states and has previously been applied to the description of experimental single-molecule FRET histograms of $E$ and $\langle\tau_{D(A)}\rangle_F$ of IDPs[20]. The distribution of interdye distances is given by the central $\chi$-distribution and depends only on the variance of the interdye distance, $\sigma_{DA}^2$:

$$p_{\mathrm{GC}}(R_{DA}|\sigma_{DA}) = \chi(R_{DA}|0,\sigma_{DA}) = 2\left(\frac{R_{DA}}{\sigma_{DA}}\right)^2 N^+(R_{DA}|0,\sigma_{DA})\,. \tag{69}$$

Often, this distribution is written in terms of the mean squared distance, $\overline{R_{DA}^2}$, which is related to the variance by $\overline{R_{DA}^2} = 3\sigma_{DA}^2$. As this model has only one variable parameter ($\sigma_{DA}$), only a single Gaussian chain FRET-line may be constructed. This FRET-line describes all polymers that behave like an ideal Gaussian chain (red line in Figure 13D-E). It can be thought of as a reference line for polymers that describes how ideal the studied system behaves, analogous to the static FRET-line for structured systems. More realistically, a disordered peptide chain may be described by the worm-like chain (WLC) model (see Supplementary Note 6)[111, 112]. The parameters defining the inter-dye distance distribution of the WLC model are the total chain length $L$ and the persistence length $l_p$ that define the stiffness of the chain by $\kappa = \frac{l_p}{L}$. In principle, the total length of the chain is known a priori from the protein sequence. From the experimentally observed position of the population in the two-dimensional histogram, the stiffness of the chain can then be estimated. FRET-lines for the WLC model are shown for different combinations of the parameters $\kappa$ and $L$ in Figure 13C-D. Notice that different combinations of $\kappa$ and $L$ can result in identical FRET efficiencies, as indicated by the horizontal line in the plot. To determine both parameters, in addition $\langle\tau_{D(A)}\rangle_F$ needs to be known.

### 4.3.2 Order/disorder transitions

Another scenario that can be identified and described by FRET-lines is the spontaneous transition between folded and unfolded states. Suppose that the distance distribution in the folded and the unfolded states are given by $p^{(f)}\left(R_{DA}|R_{mp}^{(f)}\right)$ and $p^{(u)}\left(R_{DA}|\Lambda^{(u)}\right)$, respectively. Then, the combined distance distribution is given by:

$$p(R_{DA}) = x^{(f)}\,p^{(f)}\left(R_{DA}|R_{mp}^{(f)}\right) + \left(1 - x^{(f)}\right)p^{(u)}\left(R_{DA}|\Lambda^{(u)}\right), \quad (70)$$

where $x^{(f)}$ is the species fraction of the molecules in the folded state, $p^{(f)}\left(R_{DA}|R_{mp}^{(f)}\right)$ describes the linker distribution in the folded state around the average distance $R_{mp}^{(f)}$ (Figure 14B), and $p^{(u)}\left(R_{DA}|\Lambda^{(u)}\right)$ describes the distance distribution in the unfolded state, dependent on the parameters of the polymer model, $\Lambda^{(u)}$ (Figure 14A-C). By varying $x^{(f)}$ while keeping the parameters of the distance distributions ($R_{mp}^{(f)}$ and $\Lambda^{(u)}$) constant, the dynamic FRET-line is obtained. These FRET-lines are conceptually identical to dynamic FRET-lines describing the exchange between two folded states, under the assumption that the sampling of the distance distribution in the unfolded state is fast compared to the transition rate to the folded state (see section 4.2.1). The broad distance distribution of the unfolded state shifts the endpoint of the resulting folding FRET-line far from the static FRET-line (Figure 14D-E). Dynamic transitions between a single folded state, characterized by $R_{mp}^{(f)}$, and different unfolded states, each described by the WLC model with varying stiffness at constant length ($\Lambda^{(u)} = \{\kappa, L\}$), are illustrated in Figure 14C-D in the $(E,\ \langle\tau_{D(A)}\rangle_F)$ and moment representations. Notice how all unfolded states are described by a single curve defined by the total chain length. Even though both folded and unfolded states are described by a distribution of distances, the folding FRET-line in the moment representation remains linear (Figure 14D). Dynamic unfolding FRET-lines describe folding/unfolding transitions of proteins similar to binary dynamic FRET-lines[113, 114]. The position of the population on the folding/unfolding FRET-lines informs on kinetic rate constants of the folding/unfolding events (see part II). For fast- folding/unfolding transitions on the microsecond timescale, the position of the population along the folding/unfolding FRET-line may thus be used to determine the equilibrium constant of the folding process.

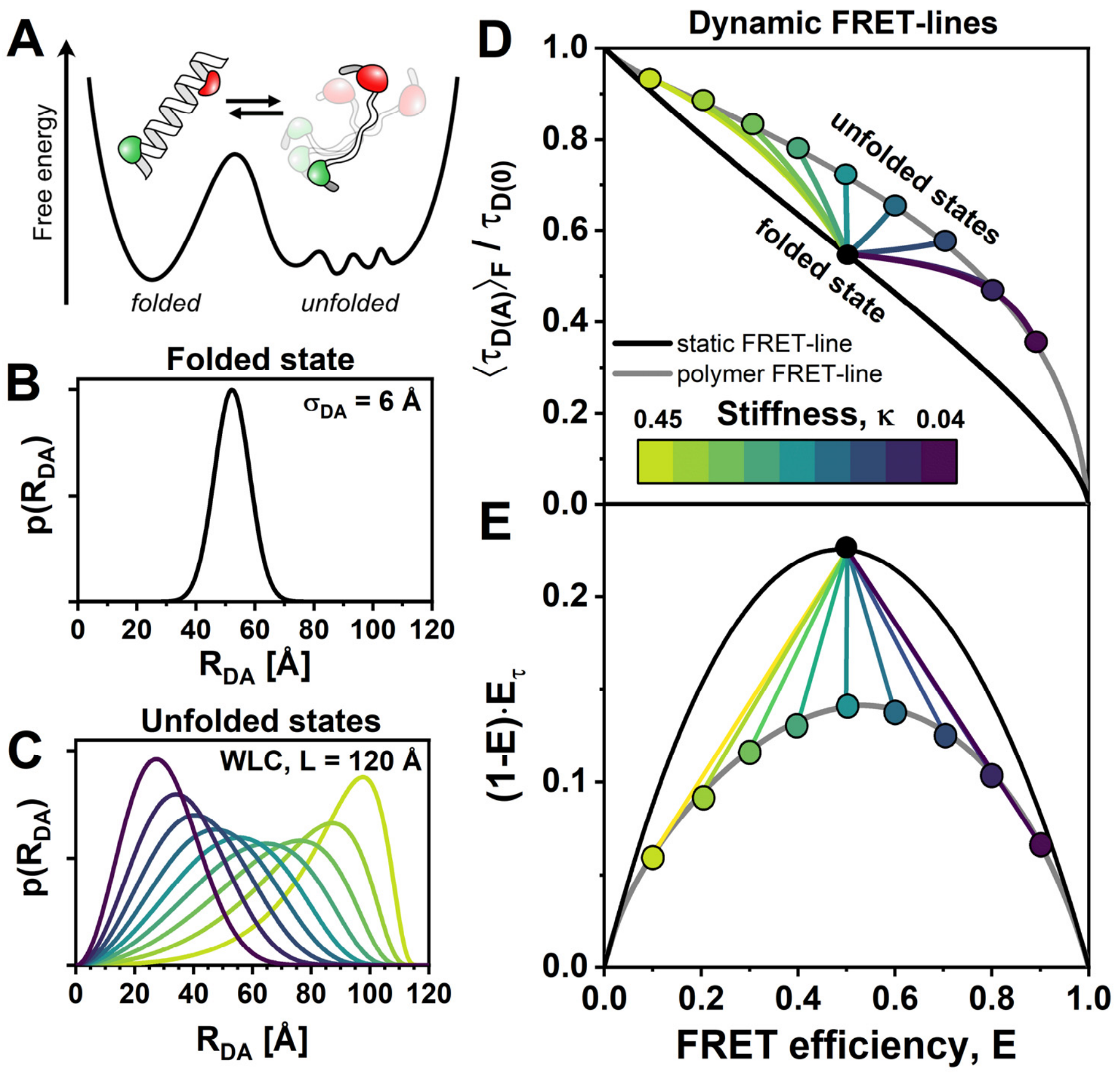

**Figure 14:** FRET-lines for order-disorder transitions **A)** Free energy landscape of a folding/unfolding transition. **B)** Distance distribution of the folded state given by a non-central $\chi$-distribution centered at 52 Å with a width parameter of 6 Å. **C)** Distance distributions $p(R_{DA})$ for unfolded states described by the worm-like chain (WLC) model for a polymer of 120 Å length with varying stiffness $\kappa$ (see color scale in D). **D-E)** Dynamic FRET-lines for the exchange between the folded and unfolded states in the parameter space of the experimental observables $E$ and $\langle \tau_{D(A)} \rangle_F$. (D) and in the moment representation (E). The folded state lies on the static FRET-line (black) for fixed distances, while all unfolded states are positioned on the dashed gray line defined by the WLC model with different stiffness $\kappa$. The WLC distance distributions of the unfolded state were calculated according to reference [111] and as described in Supplementary Note 6. The Förster radius is 52 Å and the donor lifetime in the absence of the acceptor is $\tau_{D(0)} = 4$ ns.

## 5 Conclusions

FRET-lines are guides that are superimposed on the two-dimensional histograms of the FRET observables $E$ and $\langle\tau_{D(A)}\rangle_F$ and provide a graphical analysis of complex kinetic networks in smFRET experiments. Here, we described a theoretical framework for FRET-lines based on a rigorous mathematical treatment and derived expressions for FRET-lines of static and dynamic molecules. In this framework, the mobility of the flexible dye linkers can be decoupled from the motion of the biomolecule, and it is readily applicable to disordered and unstructured systems. Based on the theoretical description of the experimental observables $E$ and $\langle\tau_{D(A)}\rangle_F$, we propose an alternative representation based on the moments of the underlying distribution of the donor fluorescence lifetime that simplifies the data representation. In this moment representation, the static FRET-line is transformed into a parabola, while dynamic FRET-lines are linearized. This enables a graphical analysis of complex kinetic networks, which can be performed "by hand" without having to apply complex equations and provides direct visualization of the kinetic exchange. This simplification of dynamic FRET-lines in the moment representation remains even for complex dynamics occurring in unstructured systems such as unfolded proteins.

In the second part of the paper, we focus on a quantitative analysis of the kinetics in multi-state systems by fluorescence correlation spectroscopy and fluorescence decay analysis. While FRET-lines do not consider the timescales of the dynamics explicitly, they provide important information on the connectivity of the states, which turns out to be the missing key towards finding unique solutions for the kinetics of multi-state systems.

## 6 Code availability

Computational tools for the calculation of FRET-lines discussed in this work are available at https://github.com/Fluorescence-Tools/FRETlines. The repository includes example Jupyter notebooks for the interactive exploration of static, dynamic, and the different polymer FRET-lines, as well as a python library for the generation of the FRET-lines discussed in this work. In addition, we provide a graphical user interface in the program "FRET-lines explorer" that is available with the software package for multiparameter fluorescence spectroscopy available at https://www.mpc.hhu.de/software/3-software-package-for-mfd-fcs-and-mfis.

**Table 3.** Used symbols and definitions

| | |
|---|---|
| **Theory - Experimental observables and their relations** | |
| $E$ | FRET efficiency, calculated from the integrated photon counts |
| $\tau_{D(A)}$, $\tau_{D(0)}$ | donor fluorescence lifetime in the presence and absence of the acceptor |
| $\langle\tau_{D(A)}\rangle_F$ | intensity-averaged donor fluorescence lifetime |
| $\langle t\rangle$ | average TCSPC delay time |
| $\langle\tau_{D(A)}\rangle_x$, $\langle\tau_{D(0)}\rangle_x$ | species-averaged donor fluorescence lifetime in the presence and absence of the acceptor |
| $k_{\mathrm{RET}}$ | rate constant of energy transfer from D to A |
| $k_{F,\mathrm{D}}$ | radiative rate constant of the donor fluorescence |
| $\Phi_{\mathrm{F,D}}$ | fluorescence quantum yield of the donor, D |
| $R_{\mathrm{DA}}$ | donor-acceptor separation distance |
| $R_0$ | characteristic distance referred to as Förster radius |
| $\kappa^2$ | orientation factor for the transition dipoles of the FRET dyes |
| $J(\lambda)$ | spectral overlap integral of the donor fluorescence and acceptor absorption spectrum |
| $n$ | refractive index of the medium |
| $k_Q^{(j)}$ | quenching rate constant of process $j$ |
| $t$ | TCSPC delay time |
| $f_{D\|D}^{(DA)}(t)$, $f_{D\|D}^{(D0)}(t)$ | time-dependent fluorescence intensity or fluorescence decay of the donor after donor excitation in the presence or absence of the acceptor |
| $p(\tau_{D(A)})$ | distribution of fluorescence lifetimes of the donor fluorophore |
| $p(R_{DA})$ | interdye distance distribution |
| $F_{D\|D}^{(DA)}$, $F_{D\|A}^{(DA)}$, $F_{D\|D}^{(D0)}$ | corrected (ideal) fluorescence intensities after excitation of the donor fluorophore of the acceptor (A\|D) and donor (D\|D) in presence of the acceptor (DA) or for a donor in absence of FRET (D0) |
| $p_{D\|D}^{(DA)}(t)$ | probability distribution of delay times for the donor after donor excitation in the presence of the acceptor |
| $\overline{\tau_{D(A)}}$, $\overline{\tau_{D(A)}^2}$ | first and second moment of the distribution of donor fluorescence lifetimes |
| $F(\tau_{D(A)})$ | fluorescence intensity of the species with lifetime $\tau_{D(A)}$ |
| $\tau_{MLE}$ | lifetime estimate obtained from maximum likelihood estimation |
| **Comparison between FRET-lines and intensity-based approaches** | |
| $\sigma_E$ | BVA standard deviation of the FRET efficiency within a single-molecule event |
| $N$ | Number of photons used for the sampling window to estimate $\sigma_E$ in BVA |
| $E_i$ | Sample obtained for the FRET efficiency within a single-molecule event in BVA |
| **FRET-lines of static and dynamic molecules** | |
| $\tau_{D(A)}^{(i)}$ | pure-state donor lifetime of species $i$ |
| $E^{(i)}$ | FRET efficiency of species $i$ |
| $k_{ij}$ | microscopic interconversion rate constant from state $j$ to state $i$ |
| $\langle\tau_{D(A)}^2\rangle_x$ | species-averaged squared donor fluorescence lifetime |
| $x^{(i)}$ | species-fraction of species $i$ |
| $\tau_{D(A)}^{(i)}$ | pure-state donor lifetime of species $i$ |
| $\delta(x)$ | Dirac delta function |
| ds | dynamic shift, defined as the maximum deviation of the dynamic FRET-line orthogonal to the static FRET-line |
| **General definition of FRET-lines** | |
| $\Lambda$, $p(\Lambda)$ | set of parameters describing the experiment and model and their probability |
| $\lambda$ | variable parameter used for the generation of FRET-line |
| $\Lambda_f$ | parameters that are fixed for the FRET-line |
| $p(E, \langle\tau_{D(A)}\rangle_F \| \lambda, \Lambda_f)$ | conditional distribution of the experimental observables $E$ and $\langle\tau_{D(A)}\rangle_F$ given $\lambda$ and $\Lambda_f$ |
| $p(\tau_{D(A)} \| \lambda, \Lambda_f)$ | conditional distribution of donor fluorescence lifetimes given $\lambda$ and $\Lambda_f$ |
| $\overline{\tau_{D(A)}^{v}}(\lambda, \Lambda_f)$ | $v$-th moment of the lifetime distribution given $\lambda$ and $\Lambda_f$ ($v = \{1,2\}$) |
| **Moments of the lifetime distribution and alternative representations** | |
| $\mathrm{Var}(\tau_{D(A)})$, $\mathrm{Var}(E)$ | variance of the donor fluorescence lifetime or FRET efficiency |
| $\mathrm{Var}^{(c)}(E)$ | contribution of conformational dynamics to the variance of the FRET efficiency |
| $\sigma_{SN}^2$ | contribution of shot noise to the variance of the FRET efficiency |
| $\Gamma$ | difference between the normalized first and second moments of the lifetime distribution |
| **FRET-lines for multi-exponential donor decays** | |
| $E_{PR}$ | proximity ratio, i.e. the uncorrected FRET efficiency |
| $\tau'_{D(0)}$ | effective donor-only lifetime in the presence of quenching |
| $\gamma'$ | ratio of the acceptor to donor quantum yield |
| **FRET-lines in the presence of linker dynamics** | |
| $\tau_{\text{linker}}$ | characteristic timescale of linker fluctuations |
| $T_{\text{obs}}$ | observation time of a single-molecule events |
| $\sigma_{DA}$ | width parameter of the inter-dye distance distribution |
| $\sigma_D, \sigma_A$ | Width of the positional distributions of the donor or acceptor fluorophore |
| $R_{mp}$ | distance between mean dye positions |
| $\chi(R_{DA}\|R_{mp}, \sigma_{DA})$ | $\chi$-distribution of inter-dye distance $R_{DA}$ |
| $\mathcal{N}^{(+)}(R_{DA}\|R_{mp}, \sigma_{DA})$ | positive-truncated normal distribution of inter-dye distance $R_{DA}$ |
| $R_{DA}^{(c)}, R_{mp}^{(c)}$ | interdye distance and distance between mean dye positions in conformational state *c* |
| $\Lambda_l^{(c)}$ | linker parameters describing the inter-dye distance distribution in conformational state *c* |
| $\Lambda^{(\text{dyn})}$ | parameters describing the conformational dynamics (transition rate matrix) |
| $\sigma_{DA}^{(c)}$ | linker distribution width in conformation *c* |
| $\langle\tau_{D(A)}^{v}\rangle_l^{(c)}$ | $v$-th linker-averaged moment of the fluorescence lifetime of conformational state *c* |
| **FRET-lines for flexible polymers** | |
| $\overline{R_{DA}^2}$ | mean squared interdye distance used in Gaussian chain polymer model |
| $l_p$ | persistence length of the chain |
| $\kappa$ | stiffness of the chain |
| $L$ | length of the chain |

# Acknowledgements

We are thankful to Don C. Lamb and Mark Bowen for their comments. HS acknowledges support from the Alexander von Humboldt foundation, Clemson University start-up funds, and NSF (CAREER MCB-1749778), and NIH (R01MH081923 and P20GM121342). CS acknowledges support by the European Research Council through the Advanced Grant 2014 hybridFRET (number 671208). TP thanks the International Helmholtz Research School of Biophysics and Soft Matter (IHRS BioSoft). This research was supported by the ERC grant "Hybrid-FRET".

# Supporting Information:

# Unraveling multi-state molecular dynamics in single-molecule FRET experiments

# Part I: Theory of FRET-Lines

Anders Barth[1,a,*], Oleg Opanasyuk[1,*], Thomas-Otavio Peulen[1,b,*], Suren Felekyan[1], Stanislav Kalinin[1], Hugo Sanabria[2‡], Claus A.M. Seidel[1,‡]

[1] Institut für Physikalische Chemie, Lehrstuhl für Molekulare Physikalische Chemie, Heinrich-Heine-Universität, Düsseldorf, Germany

[2] Department of Physics and Astronomy, Clemson University, Clemson, S.C., USA

[a] Present address: Department of Bionanoscience, Kavli Institute of Nanoscience, Delft University of Technology, Delft, The Netherlands

[b] Present address: Department of Bioengineering and Therapeutic Sciences, University of California, San Francisco, California, USA

*Contributed equally

‡Corresponding authors: cseidel@hhu.de, hsanabr@clemson.edu

## Table of Contents

## Supplementary Note 1: Estimating fluorescence lifetimes from experimental data

Experimentally, it is challenging to determine accurate fluorescence lifetimes for single-molecule events due to the limited number of photons available. For the case of single-exponential decays, a maximum-likelihood estimator (MLE) performs best in retrieving an unbiased lifetime from the sparse dataset [1,2]. More traditional least-squares fitting based on a reduced $\chi^2$ estimator, on the other hand, systematically underestimates the lifetime at low photon numbers (< 1000).

In the main text of the manuscript, we have assumed that the experimentally determined lifetime by the MLE, using a single-exponential model function, corresponds to the fluorescence-weighted average lifetime defined by:

$$\langle\tau\rangle_F = \frac{\int t\, f(t)dt}{\int f(t)dt} \tag{1}$$

where $f(t)$ is the fluorescence decay. Here, we prove that the lifetime estimated from the MLE is indeed equivalent to $\langle\tau\rangle_F$ even in the general case of an arbitrary distribution of lifetimes.

The MLE is based on the minimization of the $2I^*$ function, defined as:

$$2I^* = -2\ln L \tag{2}$$

where $\ln L$ is the logarithm of the likelihood. For normally distributed errors, $2I^*$ corresponds to the $\chi^2$ goodness of fit estimator. In the analysis of normalized fluorescence decays, the counting statistics are instead modelled by a multinomial distribution, and the corresponding $2I^*$ is given by (Maus, 2001):

$$2I^* = 2\sum_{i=1} D_i \ln\left(\frac{D_i}{M_i}\right) \tag{3}$$

where the index $i$ runs over all bins in the TCSPC histogram, $D_i$ is the normalized measured fluorescence decay of arbitrary shape, and $M_i$ is the value of the model function in TCSPC bin $i$. To simplify the equation, we replace the discrete sum with the corresponding integral and treat the measured decay and model function as continuous:

$$2I^* = 2\int D(t)\, \ln\left(\frac{D(t)}{M(t)}\right)dt = 2\int D(t)\, \ln D(t)dt - 2\int D(t)\, \ln M(t)dt \tag{4}$$

The model function is given by an exponential distribution determined by the single estimated lifetime $\tau_{MLE}$:

$$M(t) = \frac{1}{\tau_{MLE}}\exp(-t/\tau_{MLE}) \tag{5}$$

To find the estimated lifetime $\tau_{MLE}$, we need to find the minimum of the $2I^*$, which should satisfy the condition:

$$\frac{\partial(2I^*)}{\partial\tau_{MLE}} = 0 \tag{6}$$

The first term in equation (4) vanishes, as it is only determined by the data $D(t)$ and not a function of $\tau_{MLE}$. Then, for the second term, we can move the derivative into the integral to obtain:

$$\frac{\partial(2I^*)}{\partial\tau_{MLE}} = -2\int D(t)\, \frac{\partial \ln M(t)}{\partial\tau_{MLE}}dt = 0 \tag{7}$$

The logarithm of the model function is given by:

$$\ln M(t) = -\ln\tau_{MLE} - \frac{t}{\tau_{MLE}} \tag{8}$$

and the derivate with respect to $\tau_{MLE}$ is given by:

$$\frac{\partial \ln M(t)}{\partial\tau_{MLE}} = -\frac{1}{\tau_{MLE}} + \frac{t}{\tau_{MLE}^2} \tag{9}$$

Substituting equation (9) into equation (7), and dropping the factor 2, we obtain:

$$\frac{1}{\tau_{MLE}}\int D(t)\ dt - \frac{1}{\tau_{MLE}^2}\int t\,D(t)\,dt = 0 \qquad (10)$$

This expression simplifies to:

$$\tau_{MLE} = \frac{\int t\,D(t)\,dt}{\int D(t)\,dt} = \langle\tau\rangle_F \qquad (11)$$

which is equivalent to the definition of the fluorescence-weighted average lifetime $\langle\tau\rangle_F$ above.

When the instrument response function ($IRF$) is considered, the decay is given by the convolution ($*$):

$$D'(t) = (IRF * D)(t) \qquad (12)$$

Then the expression for the MLE estimated lifetime takes form:

$$\tau_{MLE} = \frac{\int t\,(IRF * D)(t)\,dt}{\int (IRF * D)(t)\,dt} = \frac{\int t \int IRF(t-t')D(t')dt'\,dt}{\iint IRF(t-t')D(t')dt'\,dt}$$

Integrals in the expression above can be separated by the change of variables $t \rightarrow t'' = t - t'$, which gives:

$$\tau_{MLE} = \frac{\left[\int t'D(t')dt'\right]\left[\int IRF(t'')dt''\right] + \left[\int t''IRF(t'')dt''\right]\left[\int D(t')dt'\right]}{\left[\int D(t')dt'\right]\left[\int IRF(t'')dt''\right]}$$

Which, finally, can be simplified to:

$$\tau_{MLE} = \frac{\int t\,D(t)\,dt}{\int D(t)\,dt} + \frac{\int t\,IRF(t)\,dt}{\int IRF(t)\,dt} = \langle\tau\rangle_F + \langle t\rangle_{IRF} \qquad (13)$$

As the two processes of excited state emission and instrument response are independent, the obtained lifetime $\tau_{MLE}$ is shifted by the arrival time averaged over the instrument response function. If the IRF is accounted for in the model function $M(t)$, the maximum likelihood estimate is again equivalent to the intensity-weighted average fluorescence lifetime.

## Supplementary Note 2: Definition of the dynamic shift for binary exchange

The ideal relationship between the FRET efficiency $E$ and the donor fluorescence lifetime $\langle\tau_{D(A)}\rangle_F$ for a static system is given by the static FRET-line:

$$E^{(\text{stat})} = 1 - \frac{\langle\tau_{D(A)}\rangle_F}{\tau_{D(0)}} \tag{14}$$

where $\langle\tau_{D(A)}\rangle_F$ and $\tau_{D(0)}$ are the intensity-weighted average donor fluorescence lifetimes in the presence and absence of the acceptor, respectively. As described in section 3.1 of the main text, modifications of eq. 1 are required if the effects of the flexible linkers are to be considered[3]. As the incorporation of linker fluctuations requires numerical approaches, they are not considered here for the derivation of an analytical solutions for the dynamic shift.

Dynamic exchange between two limiting conformational states with FRET efficiencies $E^{(1)}$ and $E^{(2)}$ and corresponding donor fluorescence lifetimes $\tau_{D(A)}^{(1)}$ and $\tau_{D(A)}^{(2)}$ is described by the binary dynamic FRET-line:

$$E^{(\text{dyn})} = 1 - \frac{\tau_{D(A)}^{(1)}\tau_{D(A)}^{(2)}}{\tau_{D(0)}\left(\tau_{D(A)}^{(1)} + \tau_{D(A)}^{(2)} - \langle\tau_{D(A)}\rangle_F\right)} \tag{15}$$

The dynamic FRET-line does not consider the timescales of the dynamics explicitly but defines the curve on which single-molecule events fall that interconverted between the two limiting states. Using the simple relations:

$$\tau_{D(A)}^{(i)} = \tau_{D(0)}\left(1 - E^{(i)}\right);\ \ i = 1,2 \tag{16}$$

we can then write eq. (15 as a function of the FRET efficiencies as:

$$E^{(\text{dyn})} = 1 - \frac{\left(1 - E^{(1)}\right)\left(1 - E^{(2)}\right)}{\left(2 - E^{(1)} - E^{(2)} - \frac{\langle\tau_{D(A)}\rangle_F}{\tau_{D(0)}}\right)} \tag{17}$$

Now can calculate the difference between the static and dynamic FRET-lines, $\Delta_E$ (i.e. the red line in Figure 21 A):

$$\Delta_E = E^{(\text{dyn})} - E^{(\text{stat})} \tag{18}$$

$$\Delta_E\left(\langle\tau_{D(A)}\rangle_F\right) = \frac{\langle\tau_{D(A)}\rangle_F}{\tau_{D(0)}} - \frac{\left(1 - E^{(1)}\right)\left(1 - E^{(2)}\right)}{\left(2 - E^{(1)} - E^{(2)} - \frac{\langle\tau_{D(A)}\rangle_F}{\tau_{D(0)}}\right)} \tag{19}$$

The function $\Delta_E\left(\langle\tau_{D(A)}\rangle_F\right)$ is unimodal with the maximum defined by efficiencies $E^{(1)}$ and $E^{(2)}$:

$$\Delta_{E,max} = \left(\sqrt{1 - E^{(1)}} - \sqrt{1 - E^{(2)}}\right)^2 \tag{20}$$

The FRET efficiency difference $\Delta_E$ for the given example is shown in Figure 21 A,B.

We define the *dynamic shift* ds as the maximally possible FRET efficiency difference $\Delta_E$ between the static and dynamic FRET-lines at the same $\langle\tau_{D(A)}\rangle_F$ normalized by $\sqrt{2}$:

$$\boxed{\text{ds} \stackrel{\text{def}}{=} \frac{\Delta_{E,max}}{\sqrt{2}} = \frac{1}{\sqrt{2}}\left(\sqrt{1 - E^{(1)}} - \sqrt{1 - E^{(2)}}\right)^2} \tag{21}$$

The reason for the $\sqrt{2}$ normalization is illustrated in the Figure 21 A. In the ideal case discussed here, the dynamic shift is then the largest separation between dynamic and static FRET-lines along the orthogonal to the static FRET-line in the $\left(E, \frac{\langle\tau_{D(A)}\rangle_F}{\tau_{D(0)}}\right)$-representation (blue line in Figure 21 A).

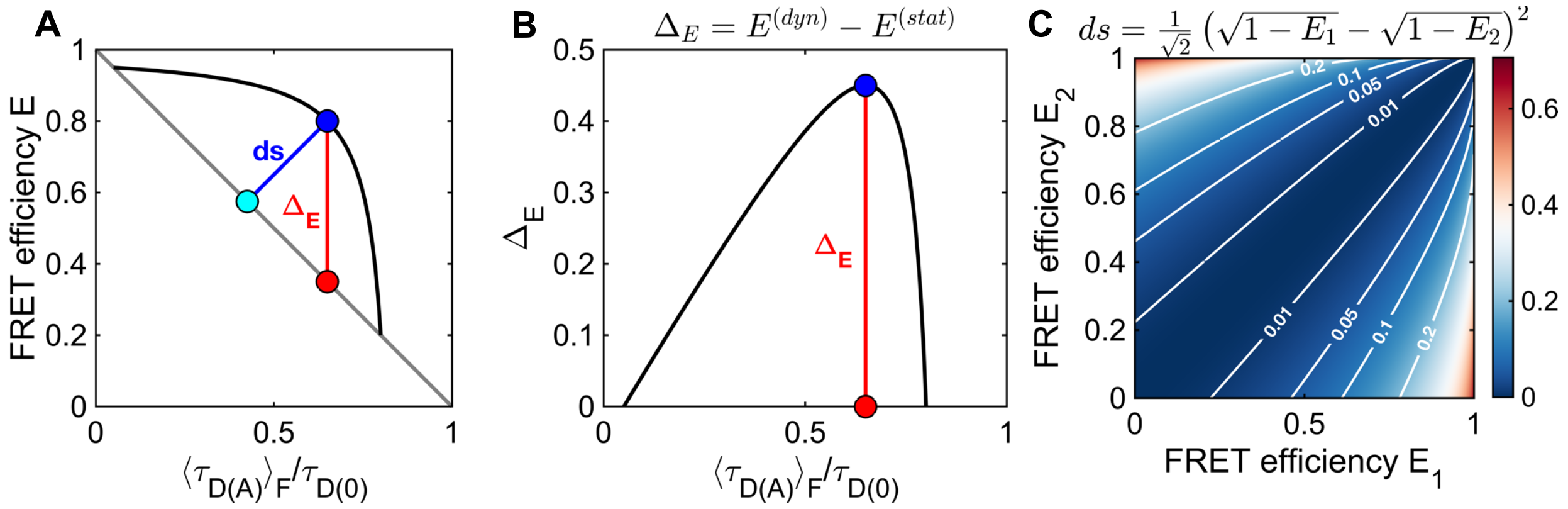

**Figure 21: The dynamic shift in the $(E, \langle\tau_{D(A)}\rangle_F)$ plot.** A) The dynamic shift ds between the static FRET-line (grey) and the dynamic FRET-line (black) is defined as the maximum distance orthogonal to the static FRET-line (blue line). The reference points for the dynamic shift on the static and dynamic FRET-lines are given in cyan and blue, respectively. The maximum difference between the static and dynamic FRET-lines in the FRET efficiency direction, $\Delta_E$, is given by the red line. **B)** The FRET efficiency difference $\Delta_E$ as a function of the normalized intensity-weighted average donor fluorescence lifetime $\langle\tau_{D(A)}\rangle_F/\tau_{D(0)}$. **C)** A contour plot of the dynamic shift as a function of the FRET efficiencies of the limiting states $E_1$ and $E_2$.

## Supplementary Note 3: A general model for FRET lines

### 3.1 General definition of FRET-lines

In this section, we describe a general theory for the description of "FRET-lines". FRET-lines are relations between the experimental observables $E$ and $\langle\tau_{D(A)}\rangle_F$ (or derived quantities) that depend on the physical model of the system. The ultimate aim of the model description would be to obtain the full distribution $p(E,\langle\tau_{D(A)}\rangle_F)$, which would enable a complete description of the experimental 2D histogram. The distribution $p(E,\langle\tau_{D(A)}\rangle_F)$, however, shows a complex dependence on the photon counting statistics. Instead, we focus on a description of $p(E,\langle\tau_{D(A)}\rangle_F)$ in the limiting case of the absence of photon shot noise. In this case, $p(E,\langle\tau_{D(A)}\rangle_F)$ collapses to defined curves on the $(E,\langle\tau_{D(A)}\rangle_F)$ plane, which we call FRET-lines. If the experimental data follows the calculated FRET-line, it is probable that the experimental system is described by the model of the theoretical line. As such, the graphical analysis provided by FRET-lines allows to check for consistency of the experimental data against different physical models and provides a starting point for further analysis.

To provide a general description, we switch from a discrete set of states ($i$) of the system (characterized by discrete distances and thus lifetimes and FRET efficiencies) to a continuous space of quantities ($\{\Lambda,\Lambda^m\}$) that define the observables:

$$\left\{E^{(i)},\tau_{D(A)}^{(i)},x_{D(A)}^{(i)}\right\}\rightarrow\{\Lambda,\Lambda^m\} \quad (22)$$

The set of quantities $\{\Lambda,\Lambda^m\}$ hereby combines characteristics of the studied molecular system and quantities characterizing the conditions of measurements. The parameters of the molecular system may be the donor/acceptor fluorescence lifetimes, FRET efficiencies, discrete or continuous distributions of donor-acceptor distances, the parameters of the linkers tethering the dyes to the molecule; the rates of transitions of species between different conformational states *etc*. The rate of the excitation, choice of emission filters or the dimensions of the focal volume are examples of measurement specific quantities. The total set $\{\Lambda,\Lambda^m\}$ consist of variables of two types. The variables of the first type ($\Lambda$) are not accessible during the experiment principally or intentionally. The variables of the second type ($\Lambda^m$) are model parameters that define the particular shape of the resulting distribution $p\left(E,\langle\tau_{D(A)}\rangle_F\right)=p(E,\langle\tau_{D(A)}\rangle_F;\Lambda^m)$ that describes the experimental histograms.

The division of variables to one or another type depends on the particular design of the experimental data. For example, usually the 2D histograms are built for single-molecule events of various durations $T$. Thus, each pixel of the 2D histogram contains contributions from SM event of different durations. In other words, the resulting histogram is integrated other all possible durations $T$. While all $T$ define the resulting histogram, the information about each of them is lost during the integration. In this case, the duration $T$ is an integrable parameter and belongs to the set $\Lambda$. However, we can also choose only SM events of a specific duration for the analysis and build a histogram only for this reduced data. In this case, the variable $T$ will become a parameter of the distribution $p(E,\langle\tau_{D(A)}\rangle_F)$ and will instead belong to the set $\Lambda^m$.

Another example is the fluorescence lifetime. If the system can be in two states (or there are two types of species), each characterized by specific fluorescence lifetimes ($\tau_{D(A)}^{(1)},\tau_{D(A)}^{(2)}$), this set of lifetimes will belong to the set of model parameters $\Lambda^m$ subject to determination during analysis. On the other hand, if the system is characterized by an infinite set of lifetimes quickly fluctuating in time (such as for a polymer chain), the information on the lifetimes of the specific microscopic conformational states will not be accessible during analysis. Therefore, the lifetime $\tau_{D(A)}$ will belong to the set $\Lambda$. Instead, if we are able to describe the distribution of lifetimes $p(\tau_{D(A)})$ we can get some characteristics of this distribution, such as the mean or variance, which will be part of the model parameters set $\Lambda^m$.

For a complete description of the experiment, we should know the joint distribution $p(E, \langle\tau_{D(A)}\rangle_F, \Lambda; \Lambda^m)$. The distribution of the experimental observables $p(E, \langle\tau_{D(A)}\rangle_F; \Lambda^m)$ for a set of model parameters $\Lambda^m$ is obtained by the marginalization (integration) of the joint distribution over the variables $\Lambda$:

$$p\big(E, \langle\tau_{D(A)}\rangle_F; \Lambda^m\big) = \int p\big(E, \langle\tau_{D(A)}\rangle_F, \Lambda; \Lambda^m\big)\, d\Lambda \tag{23}$$

The aim of the analysis is to determine the model parameters $\Lambda^m$ which best describe the experimental histogram.

The full distributions $p\big(E, \langle\tau_{D(A)}\rangle_F\big)$ is complicated even for simple systems. The example of detailed derivation of such distributions for two-state system can be found in reference [4]. The discussion of the full model is beyond the aim of the current work. Here, we would like to show how more simple, semi-quantitative models that define FRET-lines, can facilitate the interpretation of single-molecule FRET experiments.

### 3.2 Zero-shot noise approximation

Let the hidden variables $\Lambda$ be the actual numbers of photons emitted by donors and acceptors during the observation time: $\Lambda = \{N_A, N_D\}$ and let the set of model parameters be the average numbers of photons emitted by donors and acceptors: $\Lambda^m = \{n_A, n_D\}$. Then, the joint distribution describing the system is given by [Szabo, Gopich 2012][4]:

$$p\big(E, \langle\tau_{D(A)}\rangle_F, N_A, N_D; n_A, n_D\big) = \delta\left(E - \frac{N_A}{N_A + N_D}\right) p(N_A; n_A)\, p(N_D; n_D)\, p(\langle\tau_{D(A)}\rangle_F | N_D) \tag{24}$$

Where, distributions $p(N_A; n_A)\, p(N_D; n_D)$ are Poissonian and for the sufficiently large numbers of photons $N_A$ and $N_D$ can be approximated by normal distributions:

$$\begin{aligned} p(N_A; n_A) &\approx \mathcal{N}\big(N_A; n_A, \sqrt{n_A}\big); \\ p(N_D; n_D) &\approx \mathcal{N}\big(N_D; n_D, \sqrt{n_D}\big) \end{aligned} \tag{25}$$

The form of $p(\langle\tau_{D(A)}\rangle_F | N_D)$ is given in [Szabo, Gopich 2012] in general form and for the case of exchange between two conformational states. In the context of the current paper it is important that the mean and variance of this distribution are given by:

$$\overline{\tau_F} = \int t\, p_D(t) dt\,;\ \mathrm{Var}(\tau_F) = \frac{1}{N_D}\left(\int t^2\, p_D(t) dt - \overline{\tau_F}^2\right);\ p_D(t) = \frac{f_D(t)}{\int f_D(t) dt} \tag{26}$$

where $p_D(t)$ is the distribution of the arrival (delay) time $t$. In other words, $p_D(t)$ is normalized expectation of the donor fluorescence intensity decay $f_D(t)$.

As each pair of specific values of $(E, \tau_F)$ can be defined by various values of $N_A, N_D$, the distribution of observed variables $(E, \tau_F)$ is given by marginalization (integration) of the distribution of eq. (24) over the variables $\Lambda = \{N_A, N_D\}$:

$$p\big(E, \langle\tau_{D(A)}\rangle_F;\ n_A, n_D\big) = \iint \delta\left(E - \frac{N_A}{N_A + N_D}\right) p(N_A; n_A) p(N_D; n_D) p(\langle\tau_{D(A)}\rangle_F | N_D) dN_A dN_D \tag{27}$$

For very large numbers of photons, $N_A \to \infty$ and $N_D \to \infty$ (i.e. for zero shot-noise), the involved probability densities tend to delta functions:

$$\begin{aligned} p(N_A; n_A) &\to \delta(N_A - n_A) \\ p(N_D; n_D) &\to \delta(N_D - n_D) \\ p(\tau_F | N_D) &\to \delta\big(\langle\tau_{D(A)}\rangle_F - \overline{\tau_F}\big) \end{aligned} \tag{28}$$

And the integration over $N_A, N_D$ gives the distribution:

$$p(E, \tau_F; n_A, n_D) \to p'(E, \tau_F; \bar{E}, \overline{\tau_F}) = \delta(E - \bar{E})\delta(\tau_F - \overline{\tau_F});\quad \bar{E} = \frac{n_A}{n_A + n_D} \tag{29}$$

As shown in the main text, if there are multiple conformational states characterized by fluorescence lifetimes with distributions $p(\tau;\Lambda^m)$, the two quantities $\bar{E}$ and $\overline{\tau_F}$ can be expressed in terms of the first two moments of $p(\tau)$:

$$\begin{cases} \bar{E}(\Lambda^{m'}) = 1 - \dfrac{\overline{\tau_{DA}}(\Lambda^{m'})}{\tau_{DO}} \\ \overline{\tau_F}(\Lambda^{m'}) = \dfrac{\overline{\tau_{DA}^2}(\Lambda^{m'})}{\overline{\tau_{DA}}(\Lambda^{m'})} \end{cases}$$
$$\overline{\tau_{DA}^{\nu}}(\Lambda^{m'}) = \int \tau_{DA}^{\nu}\, p(\tau_{DA};\Lambda^{m'})d\tau_{DA}\,;\; \nu = \{1,2\}, \qquad (30)$$

where $\Lambda^{m'}$ is the set of (model) parameters of the distribution $p(\tau_{DA};\Lambda^{m'})$. Here, for simplicity, we omitted the distribution of donor-only lifetimes $\tau_{DO}$. Therefore, the zero shot-noise distribution of observables can be written in the form:

$$p'(E,\langle\tau_{D(A)}\rangle_F;\bar{E},\overline{\tau_F}) \to p'(E,\langle\tau_{D(A)}\rangle_F;\Lambda^{m'})$$
$$= \delta\left(E - \left(1 - \frac{\overline{\tau_{DA}}(\Lambda^{m'})}{\tau_{DO}}\right)\right)\delta\left(\tau_F - \left(\frac{\overline{\tau_{DA}^2}(\Lambda^{m'})}{\overline{\tau_{DA}}(\Lambda^{m})}\right)\right) \qquad (31)$$

The distribution $p'(E,\tau_F;\Lambda^{m'})$ is zero everywhere except the multidimensional volume in the space of variables $\{E,\tau_F;\Lambda^{m'}\}$ given by the set of equations:

$$\begin{cases} E = 1 - \dfrac{\overline{\tau_{DA}}(\Lambda^{m'})}{\tau_{DO}} \\ \langle\tau_{D(A)}\rangle_F = \dfrac{\overline{\tau_{DA}^2}(\Lambda^{m'})}{\overline{\tau_{DA}}(\Lambda^{m'})} \end{cases} \qquad (32)$$

Instead of using the distribution of fluorescence lifetimes $p(\tau_{DA};\Lambda^{m'})$ we can use the distribution of any other variable $\Lambda'$ that define $\tau_{DA}$ (an efficiency or a FRET distance for example). So generally, we can write for the fluorescence lifetimes moments:

$$\overline{\tau_{DA}^{\nu}}(\Lambda^{m'}) = \int \tau_{DA}^{\nu}(\Lambda')p(\Lambda';\Lambda^{m'})d\Lambda';\quad \nu = \{1,2\} \qquad (33)$$

Thus, in the case of zero shot-noise the description of the system is reduced to the system of equations (32) and some probability $p(\Lambda';\Lambda^{m'})$ which defines the moments of lifetimes distribution according to eq. (33).
In the following, we will consider only the zero-shot noise distribution and therefore will omit the prime superscripts for it and involved variables $(\Lambda';\Lambda^{m'})$.

### 3.3 General equation for the FRET-lines

Eq. (32) defines an $m$-dimensional volume in the in the ($m$+2)-dimensional space of quantities $\{E,\langle\tau_{D(A)}\rangle_F,\Lambda^m\}$, where $m$ is dimensionality of the space of parameters $\Lambda^m$. To facilitate a visual analysis, it is desirable to construct lines describing a defined set of properties of the system, which can be easily compared with the measured data in the $(E,\langle\tau_{D(A)}\rangle_F)$-plane. To obtain a line, we select one free parameter (λ) from the set $\Lambda^m$ and fix all other parameters. Let's designate all fixed values of parameters from the set $\Lambda^m$ by $\Lambda_f^m$. The line in the $(E,\langle\tau_{D(A)}\rangle_F)$-plane described by the parametric equations:

$$
\boxed{\begin{cases} E & = 1 - \dfrac{\overline{\tau_{DA}}(\lambda, \Lambda_f^m)}{\tau_{DO}} \\ \langle \tau_{D(A)} \rangle_F & = \dfrac{\overline{\tau_{DA}^2}(\lambda, \Lambda_f^m)}{\overline{\tau_{DA}}(\lambda, \Lambda_f^m)} \end{cases} \\ \overline{\tau_{DA}^{\nu}}(\lambda, \Lambda_f^m) = \int \tau_{DA}^{\nu}(\Lambda) p(\Lambda; \lambda, \Lambda_f^m) d\Lambda; \quad \nu = \{1,2\}} \quad , \tag{34}
$$

we call the *FRET-line* for the free parameter λ and fixed parameters $\Lambda_f^m$. FRET-lines define a relationship between the observables $E$ and $\langle \tau_{D(A)} \rangle_F$ for a sub-ensemble of single-molecule events characterized by the fixed parameters values $\Lambda_f^m$ and the free model parameter $\lambda$ in the limit of a zero shot-noise.

### 3.3.1 Example 1: Pure states FRET-line

Let consider the model for molecules with a single fluorescence lifetime $\tau_{D(A)}^{(1)}$. Applying the general scheme described above, we assign the parameters as follows:

$$
\begin{aligned} &\Lambda = \{\tau_{D(A)}\}; \quad \Lambda^m = \left\{\tau_{D(A)}^{(1)}\right\}; \quad \Lambda_f^m = \emptyset; \quad \lambda = \tau_{D(A)}^{(1)}; \\ &p(\Lambda; \lambda, \Lambda_f^m) = p\left(\tau_{D(A)}; \tau_{D(A)}^{(1)}\right) = \delta\left(\tau_{D(A)} - \tau_{D(A)}^{(1)}\right); \end{aligned} \tag{35}
$$

Now, we can calculate the moments of the lifetime distribution as functions of the parameter $\tau_{D(A)}^{(1)}$:

$$
\overline{\tau_{D(A)}^{\nu}}(\lambda, \Lambda^m) = \overline{\tau_{D(A)}^{\nu}}\left(\tau_{D(A)}^{(1)}\right) = \int \tau_{D(A)}^{\nu} \delta\left(\tau_{D(A)} - \tau_{D(A)}^{(1)}\right) d\tau_{D(A)} = \left(\tau_{D(A)}^{(1)}\right)^{\nu}, \quad \nu = \{1,2\}; \tag{36}
$$

The corresponding parametric relationship between $E$ and $\langle \tau_{D(A)} \rangle_F$ is given by:

$$
\begin{cases} E\left(\tau_{D(A)}^{(1)}\right) = 1 - \dfrac{\tau_{D(A)}^{(1)}}{\tau_{D(O)}} \\ \langle \tau_{D(A)} \rangle_F \left(\tau_{D(A)}^{(1)}\right) = \tau_{D(A)}^{(1)} \end{cases} \tag{37}
$$

Notice that in this simple example the set of fixed parameters is empty ($\Lambda_{\mathrm{f}}^m = \emptyset$) and the only parameter characterizing probability of lifetimes in the individual measurements is the mean lifetime $\tau_{D(A)}^{(1)}$ and it is naturally chosen as a free parameter ($\lambda = \tau_{D(A)}^{(1)}$).

The free parameter $\tau_{D(A)}^{(1)}$ can be eliminated and the explicit equation for the FRET efficiency is obtained:

$$
E(\langle \tau_{D(A)} \rangle_F) = 1 - \frac{\langle \tau_{D(A)} \rangle_F}{\tau_{D(O)}} \tag{38}
$$

We call this line the *pure states FRET-line.* Different points on this line correspond to different values of the free parameter $\tau_{D(A)}^{(1)}$, i.e. to different conformational states of the molecule.

### 3.3.2 Example 2: Binary mixed states FRET-line

Now, let us consider the FRET-line for molecules exchanging between two distinct conformational states, each characterized by a single fluorescence lifetime ($\tau_{D(A)}^{(1)}$ and $\tau_{D(A)}^{(2)}$). We additionally define the fraction of time $x^{(1)}$ that the molecule spends in state 1 during the observation time (then $x^{(2)} = 1 - x^{(1)}$). From the three model parameters $\tau_{D(A)}^{(1)}$, $\tau_{D(A)}^{(2)}$ and $x^{(1)}$, we pick the state occupancy $x^{(1)}$ as our free parameter. For this model, we assign the parameters as follows:

$$\Lambda = \{\tau_{D(A)}\};\ \Lambda^m = \left\{x^{(1)}, \tau_{D(A)}^{(1)}, \tau_{D(A)}^{(2)}\right\};\ \Lambda_f^m = \left\{\tau_{D(A)}^{(1)}, \tau_{D(A)}^{(2)}\right\};\ \lambda = x^{(1)}$$

$$p(\Lambda; \lambda, \Lambda_f^m) = \sum_{i=1}^{2} x^{(i)} \delta(\tau_{D(A)} - \tau_{D(A)}^{(i)}) \tag{39}$$

The moments of the fluorescence lifetime distribution as a function of the variables $\Lambda^m = \{x^{(1)}, \tau_{D(A)}^{(1)}, \tau_{D(A)}^{(2)}\}$ are:

$$\overline{\tau_{D(A)}^{\nu}}(\Lambda^m) = \sum_{i=1}^{2} x^{(i)}\ \tau_{D(A)}^{(i)\,\nu}, \qquad \nu = \{1,2\} \tag{40}$$

The expressions for the observed FRET efficiency and the fluorescence-weighted average lifetime take the form:

$$\begin{cases} E(x^{(1)}, \tau_{D(A)}^{(1)}, \tau_{D(A)}^{(2)}) & = 1 - \dfrac{x^{(1)}\tau_{D(A)}^{(1)} + (1 - x^{(1)})\tau_{D(A)}^{(2)}}{\tau_{D(O)}} \\ \langle \tau_{D(A)} \rangle_F (x^{(1)}, \tau_{D(A)}^{(1)}, \tau_{D(A)}^{(2)}) & = \dfrac{x^{(1)}\tau_{D(A)}^{(1)2} + (1 - x^{(1)})\tau_{D(A)}^{(2)2}}{x^{(1)}\tau_{D(A)}^{(1)} + (1 - x^{(1)})\tau_{D(A)}^{(2)}} \end{cases} \tag{41}$$

The explicit relationship between $E$ and $\langle \tau_{D(A)} \rangle_F$ can be obtained by elimination of the free parameter $x^{(1)}$:

$$E(\langle \tau_{D(A)} \rangle_F; \tau^{(1)}, \tau^{(2)}) = 1 - \frac{1}{\tau_{D(O)}} \frac{\tau^{(1)}\ \tau^{(2)}}{\tau^{(1)} + \tau^{(2)} - \langle \tau_{D(A)} \rangle_F} \tag{42}$$

We call this line the *dynamic FRET-line* for two states. Different points on this line correspond to different values of the variable $x^{(1)}$, i.e. to a different degree of mixing between the states 1 and 2. The expression for the pure states FRET-line (eq. (38)) is obtained from eq. (41) by setting the state occupancy of one of the states to zero (i.e. by fixing both occupancies) and varying the remaining fluorescence lifetime.

### 3.4 Expression for FRET-lines in terms of donor-acceptor distances

The examples above were written in terms of the fluorescence lifetimes as the defining characteristic of the different states. However, more often the models describing the studied system are formulated in terms of donor-acceptor distances (i.e. in the general scheme above $\Lambda^m = \{R_{DA}\}$) which are directly related to the structural properties of the molecules. If each state of the molecule is characterized by a single inter-dye distance, the expressions for the FRET-lines are obtained from the lifetime-based expression above by the trivial change of variables $\tau_{D(A)} \to R_{DA}$, whereby the dependence of the fluorescence lifetime on the inter-dye distance is defined by the Förster equation:

$$\tau_{D(A)}(R_{DA}) = \left[\frac{1}{\tau_{D(0)}} + k_{RET}\right]^{-1} = \tau_{D(0)} \left[1 + \left(\frac{R_0}{R_{DA}}\right)^6\right]^{-1} \tag{43}$$

In real experiments, however, the dyes are often tethered to molecules using flexible linkers. Each conformational state of the molecule is then characterized by a distribution of inter-dye distances. The inter-dye distance vector $\boldsymbol{R}_{DA}$ is given by the sum of the vector $\boldsymbol{R}_{DA}^{(c)}$, characterizing the conformation of molecule (*c*), and the vector $\boldsymbol{R}_{DA}^{(l)}$ that defines the linker conformation of donor and acceptor (*l*). The state of the system is fully characterized by the pair of vectors $\left\{\boldsymbol{R}_{DA}^{(c)}, \boldsymbol{R}_{DA}^{(l)}\right\}$, where $\boldsymbol{R}_{DA} = \boldsymbol{R}_{DA}^{(c)} + \boldsymbol{R}_{DA}^{(l)}$. As the lifetime depends on $\boldsymbol{R}_{DA}$ and we are interested in characterizing the conformations $\boldsymbol{R}_{DA}^{(c)}$, it is convenient to choose the model parameters as $\Lambda^m = \left\{R_{DA}, R_{DA}^{(c)}\right\}$. Here, we dropped the vector notation, as we are not considering the effect of the dye orientation on FRET and characterize conformations only by distances. Thus, the expression for the moments of the fluorescence lifetime distribution (eq. (33)) takes the form:

$$\overline{\tau^{\nu}_{D(A)}}(\Lambda^m) = \int\int \tau^{\nu}_{D(A)}(R_{DA})\, p\left(R^{(c)}_{DA}, R_{DA}; \Lambda^m\right)\, dR^{(c)}_{DA}\, dR_{DA}, \qquad \nu = \{1,2\} \tag{44}$$

If the timescale of the dynamics of the linkers is comparable with the timescale of conformational dynamics, the expression of a joined probability $p\left(R^{(c)}_{DA}, R_{DA}; \Lambda^m\right)$ is complicated. However, as the dyes are much smaller than the biomolecule, we assume that the dynamics of the tethered dyes are much faster than the conformational dynamics. In this case, the distribution of linker positions is averaged for every single-molecule event and takes its stationary form described by the parameters of the linkers ($\Lambda^{m,l}$). We also assume that the linker parameters are identical for all conformational states. Note that these assumptions do not make the conformational and linker distance distributions independent. They are still linked by the conformational distance $R^{(c)}_{DA}$, which is a free variable of the conformational distance distribution and a conditional parameter of the linker distance distribution. Therefore, we can rewrite eq. (44) in the form:

$$\overline{\tau^{\nu}_{D(A)}}(\Lambda^{m,c}, \Lambda^{m,l}) = \int \tau^{(l,\nu)}_{D(A)}\left(R^{(c)}_{DA}, \Lambda^{m,l}\right) p\left(R^{(c)}_{DA}|\Lambda^{m,c}\right) dR^{(c)}_{DA}, \qquad \nu = \{1,2\}, \tag{45}$$

$$\tau^{(l,\nu)}_{D(A)}\left(R^{(c)}_{DA}, \Lambda^{m,l}\right) = \int \tau^{\nu}_{D(A)}(R_{DA})\, p\left(R_{DA}|R^{(c)}_{DA}, \Lambda^{m,l}\right) dR_{DA}, \tag{46}$$

where $p\left(R^{(c)}_{DA}; \Lambda^{m,c}\right)$ is probability distribution function of the conformational distances with parameters $\Lambda^{m,c}$; $p\left(R_{DA}|R^{(c)}_{DA}; \Lambda^{m,l}\right)$ is the conditional probability density of linker positions for conformation ($c$) characterized by the distance $R^{(c)}_{DA}$, and $\tau^{(l,\nu)}_{D(A)}\left(R^{(c)}_{DA}, \Lambda^{m,l}\right)$ are the moments of the lifetime distribution for a given conformation ($c$). It should be noted that the re-parameterization of inter-dye distance in terms of the conformational ($R^{(c)}_{DA}$) and linker ($R^{(l)}_{DA}$) distances can be achieved in arbitrary ways. For example, we can choose $R^{(c)}_{DA}$ as the distance between the attachment points or the mean positions of the dyes or take it to be the mean distance between donor and acceptor in a given conformation.

### 3.5 FRET-lines in the mean-variance representation in the presence of linkers

The mean and variance of the lifetime distribution $\left(\langle\tau_{D(A)}\rangle, Var\left[\tau_{D(A)}\right]\right)$ are related to the moments of the lifetime distribution by:

$$\begin{cases} Var\left[\tau_{D(A)}\right] = \overline{\tau^2_{D(A)}} - \overline{\tau_{D(A)}}^2 \\ \langle\tau_{D(A)}\rangle \quad = \overline{\tau_{D(A)}} \end{cases} \tag{47}$$

This representation relates to the $\left(E, \langle\tau_{D(A)}\rangle_F\right)$ coordinates by:

$$\begin{cases} E \quad = 1 - \dfrac{\langle\tau_{D(A)}\rangle}{\overline{\tau_{D(0)}}} \\ \langle\tau_{D(A)}\rangle_F = \dfrac{Var\left[\tau_{D(A)}\right]}{\langle\tau_{D(A)}\rangle} + \langle\tau_{D(A)}\rangle \end{cases}; \tag{48}$$

And the reverse transformation is:

$$\begin{cases} Var\left[\tau_{D(A)}\right] = \langle\tau_{D(A)}\rangle_F \langle\tau_{D(A)}\rangle - \langle\tau_{D(A)}\rangle^2 \\ \langle\tau_{D(A)}\rangle \quad = \overline{\tau_{D(0)}}(1 - E) \end{cases} \tag{49}$$

The pure-states and mixed states FRET-lines in $\left(\langle\tau_{D(A)}\rangle, Var\left[\tau_{D(A)}\right]\right)$ representation for a two-state model are presented in the Fig. 6B of the main text.

Using the relationship of the variance and the first two moments of the distribution, we can rewrite eq. (47) in the form:

$$\begin{cases} Var\left[\tau_{D(A)}\right] = Var^{(c)}\left[\tau^{(l)}_{D(A)}\right] + \overline{Var^{(l)}\left[\tau_{D(A)}\right]}^{(c)} \\ \langle \tau_{D(A)} \rangle = \overline{\tau^{(l)}_{D(A)}}^{(c)} \end{cases} ; \tag{50}$$

$$\overline{Var^{(l)}\left[\tau_{D(A)}\right]}^{(c)} \left(\Lambda^{m,c}, \Lambda^{m,l}\right) = \int Var^{(l)}\left[\tau_{D(A)}\right]\left(R^{(c)}_{DA}, \Lambda^{m,l}\right) p\left(R^{(c)}_{DA}; \Lambda^{m,c}\right) dR^{(c)}_{DA}$$

Here, $\tau^{(l)}_{D(A)} = \tau^{(l,1)}_{D(A)}\left(R^{(c)}_{DA}, \Lambda^{m,l}\right)$ and $Var^{(l)}\left[\tau_{D(A)}\right]\left(R^{(c)}_{DA}, \Lambda^{m,l}\right)$ are the mean and the variance of fluorescence lifetimes within a given conformational state $c$ over the linker distribution with parameters $\Lambda^{m,l}$, $\overline{\tau^{(l)}_{D(A)}}^{(c)}$ and $Var^{(c)}\left[\tau^{(l)}_{D(A)}\right]$ are the mean and variance of the linker-averaged lifetime over all conformational states, and $\overline{Var^{(l)}\left[\tau_{D(A)}\right]}^{(c)}\left(\Lambda^{m,l}, \Lambda^{m,c}\right)$ is the mean of the linker variances over all conformational states.

To derive equation (50) above, consider that the moments of the lifetime distribution are given by the average of the linker moments, $\tau^{(l,\nu)}_{D(A)}$, over the conformational states:

$$\overline{\tau^{\nu}_{D(A)}} = \overline{\tau^{(l,\nu)}_{D(A)}}^{(c)}, \qquad \nu = \{1,2\}. \tag{51}$$

The variance is then given by:

$$Var\left[\tau_{D(A)}\right] = \overline{\tau^{2}_{D(A)}} - \overline{\tau_{D(A)}}^{2} = \overline{\tau^{(l,1)2}_{D(A)}}^{(c)} - \left(\overline{\tau^{(l,1)}_{D(A)}}^{(c)}\right)^2 \tag{52}$$

The square of the conformation-averaged first linker moment, $\left(\overline{\tau^{(l,1)}_{D(A)}}^{(c)}\right)^2$, can be expressed as:

$$\left(\overline{\tau^{(l,1)}_{D(A)}}^{(c)}\right)^2 = \overline{\tau^{(l,1)2}_{D(A)}}^{(c)} - Var^{(c)}\left[\tau^{(l,1)}_{D(A)}\right] \tag{53}$$

where $Var^{(c)}\left[\tau^{(l,1)}_{D(A)}\right]$ is variance of linker mean lifetimes over all conformational states. Note, that $\overline{\tau^{(l,1)2}_{D(A)}}^{(c)} \neq \tau^{(l,2)}_{D(A)}$. To find $\overline{\tau^{(l,1)2}_{D(A)}}^{(c)}$, we calculate the mean of the linker variances over the conformational states:

$$\begin{aligned} \overline{Var^{(l)}\left[\tau_{D(A)}\right]}^{(c)} &= \overline{\left(\tau^{(l,2)}_{D(A)} - \tau^{(l,1)2}_{D(A)}\right)}^{(c)} = \overline{\tau^{(l,2)}_{D(A)}}^{(c)} - \overline{\tau^{(l,1)2}_{D(A)}}^{(c)} \\ \Rightarrow \quad \overline{\tau^{(l,1)2}_{D(A)}}^{(c)} &= \overline{\tau^{(l,2)}_{D(A)}}^{(c)} - \overline{Var^{(l)}\left[\tau_{D(A)}\right]}^{(c)} \end{aligned} \tag{54}$$

Combining the expressions above we obtain:

$$\begin{aligned} Var\left[\tau_{D(A)}\right] &= \overline{\tau^{(l,2)}_{D(A)}}^{(c)} - \overline{\tau^{(l,1)}_{D(A)}}^{(c)2} \\ &= \overline{\tau^{(l,2)}_{D(A)}}^{(c)} - \overline{\tau^{(l,1)2}_{D(A)}}^{(c)} + Var^{(c)}\left[\tau^{(l,1)}_{D(A)}\right] \\ &= \overline{\tau^{(l,2)}_{D(A)}}^{(c)} - \overline{\tau^{(l,2)}_{D(A)}}^{(c)} + \overline{Var^{(l)}\left[\tau_{D(A)}\right]}^{(c)} + Var^{(c)}\left[\tau^{(l,1)}_{D(A)}\right] \end{aligned} \tag{55}$$

Thus, we obtain the important result:

$$\boxed{Var\left[\tau_{D(A)}\right] = Var^{(c)}\left[\tau^{(l,1)}_{D(A)}\right] + \overline{Var^{(l)}\left[\tau_{D(A)}\right]}^{(c)}} \tag{56}$$

The total variance of the lifetime is the sum of the contributions of the conformational dynamics and linker fluctuations. Note that the conformational variance is hereby evaluated using the linker-averaged lifetimes, while the linker-variance of the lifetime is given by the average over all conformational states. Let $\sigma^{(l)} \in \Lambda^{m,l}$ and $\sigma^{(c)} \in \Lambda^{m,c}$ be the parameters that define the width of the linker and conformational distance distributions. In the limiting case of vanishing width, i.e. $\sigma \to 0$, the conformational and linker distributions turn to Dirac delta distributions:

$$p\left(R_{DA}|R_{DA}^{(c)};\Lambda^{b,l}\right)\Big|_{\sigma^{(l)}\to 0}=\delta\left(R_{DA}-R_{DA}^{(c)}\right)$$
$$p\left(R_{DA}^{(c)};\Lambda^{b,c}\right)\Big|_{\sigma^{(c)}\to 0}=\delta\left(R_{DA}^{(c)}-R_{DA}^{(m,c)}\right);\; R_{DA}^{(m,c)}\in\Lambda^{m,c} \quad (57)$$

Using eq. (50) we can build two parametric FRET-line expressions for the limiting cases. If the distribution of linkers is very narrow, then $\sigma^{(l)}\to 0$, $\tau_{D(A)}^{(l)}\to\tau_{D(A)}$ and $\overline{Var^{(l)}\left[\tau_{D(A)}\right]}^{(c)}\to 0$. The resulting FRET-line we call the *fixed-dyes FRET-line*:

**Fixed-dyes FRET-lines, $\boldsymbol{\sigma^{(l)}\to 0}$ :**

$$\begin{cases} Var\left[\tau_{D(A)}\right]=Var^{(c)}\left[\tau_{D(A)}\right]\left(\Lambda^{m,c}\right) \\ \langle\tau_{D(A)}\rangle=\overline{\tau_{D(A)}}^{(c)}\left(\Lambda^{m,c}\right)\end{cases} \quad (58)$$

If the distribution of conformational states is very narrow, that is $\sigma^{(c)}\to 0$, then $\overline{\tau_{D(A)}^{(l)}}^{(c)}\to\tau_{D(A)}^{(l)}$, $\overline{Var^{(l)}\left[\tau_{D(A)}\right]}^{(c)}\to Var^{(l)}\left[\tau_{D(A)}\right]$ and $Var^{(c)}\left[\tau_{D(A)}^{(l)}\right]\to 0$. We call the corresponding FRET-line the *static FRET-line*:

**Static FRET-lines, $\boldsymbol{\sigma^{(c)}\to 0}$ :**

$$\begin{cases} Var\left[\tau_{D(A)}\right]=Var^{(l)}\left[\tau_{D(A)}\right]\left(R_{DA}^{(m,c)},\Lambda^{m,l}\right) \\ \langle\tau_{D(A)}\rangle=\tau_{D(A)}^{(l)}\left(R_{DA}^{(m,c)},\Lambda^{m,l}\right)\end{cases} \quad (59)$$

The *fixed-dyes FRET-line* is closely related to the $\tau_{D(A)}$–based FRET-line considered earlier in this section. The only difference between them is the choice of the variable $\Lambda$ defining the conformational states. In previous section, we chose $\Lambda=\tau_{D(A)}$, while in this section $\Lambda=R_{D(A)}$. We can convert between the two representations by the trivial change of the variables $\tau_{D(A)}$ and $R_{D(A)}$ through eq. (43). The *static FRET-line* is a generalization of the *pure states* line introduced in example – the pure state line is obtained from eq. (59) for the case of dyes tethered to a molecule by static linkers ($\sigma^{(l)}\to 0$). Contrary to the single pure states line, which we constructed by varying the pure state lifetime $\tau_{D(A)}^{(1)}$, we have now defined a set of lines for arbitrary linker distributions characterized by parameters $\Lambda^{m,l}$, each built by varying the conformational state parameter $R_{DA}^{(m,c)}$.

### 3.6 Binary dynamic FRET-lines in the presence of linkers

Let us now apply the procedure described above to the two-state example from the section 3.3.2. We assume that the inter-dye distances, $R_{DA}$, follow some distribution $p\left(R_{DA}|R_{DA}^{(c)},\sigma^{(c)}\right)$ defined by the non-centrality parameter $R_{DA}^{(c)}$, and the width $\sigma^{(c)}$. As the non-centrality parameter we choose the distance between the average dye position, $R_{DA}^{(c)}=R_{MP}^{(c)}$, and assume that this distance can take two discrete values, $R_{MP}^{(1)}$ and $R_{MP}^{(2)}$, with the fraction of the first state $x^{(1)}$. The full set of variables and distributions for this model is given by:

$$\begin{aligned}
\Lambda^{o} &=\{\Lambda^{o,c},\Lambda^{o,l}\}=\left\{R_{DA}^{(c)}=R_{MP}^{(c)},R_{DA}\right\};\\
\Lambda^{m,c} &=\left\{x^{(1)},R_{MP}^{(1)},R_{MP}^{(2)}\right\};\\
\Lambda^{m,l} &=\left\{R_{MP}^{(c)},\sigma^{(l)}\right\}\\
p(\Lambda^{o,c}|\Lambda^{m,c}) &=p\left(R_{MP}^{(c)}|x^{(1)},R_{MP}^{(1)},R_{MP}^{(2)}\right)=\sum_{i=1}^{2}x^{(i)}\delta\left(R_{MP}^{(c)}-R_{MP}^{(i)}\right)\\
p(\Lambda^{o,l}|\Lambda^{m,l}) &=p\left(R_{DA}|R_{MP}^{(c)},\sigma^{(l)}\right)
\end{aligned} \quad (60)$$

*The binary fixed-dyes FRET-line*

The fixed-dyes FRET-line is obtained by setting the width parameter of the linker distribution $\sigma^{(l)} = 0$. Then, the conformational states are described by a single lifetime $\tau^{(i)} = \tau_{D(A)}\left(R_{MP}^{(i)}\right); i = 1{,}2$, and the form of the fixed-dyes FRET-line follows the expression from example 1 (eq. (37)). In the ($\langle\tau_{D(A)}\rangle$, $Var\left[\tau_{D(A)}\right]$) representation, the FRET-line transforms to:

**Binary fixed-dyes FRET-lines:**

$$\begin{cases} Var\left[\tau_{D(A)}\right]\left(x^{(1)}, \tau^{(1)}, \tau^{(2)}\right) = x^{(1)}\left(1 - x^{(1)}\right)\left(\tau^{(1)} - \tau^{(2)}\right)^2 \\ \langle\tau_{D(A)}\rangle\left(x^{(1)}, \tau^{(1)}, \tau^{(2)}\right) = x^{(1)}\tau^{(1)} + (1 - x^{(1)})\tau^{(2)} \end{cases} \quad (61)$$

If the fraction $x^{(1)}$ is chosen as the free parameter of the FRET-line and $\tau^{(1)}$ and $\tau^{(2)}$ are fixed, the explicit expression of the FRET-line is given by:

**Binary fixed-dyes FRET-lines (explicit for free $x^{(1)}$) :**

$$Var\left[\tau_{D(A)}\right]\left(\langle\tau_{D(A)}\rangle, \tau^{(1)}, \tau^{(2)}\right) = \left[\langle\tau_{D(A)}\rangle - \tau^{(1)}\right]\left[\tau^{(2)} - \langle\tau_{D(A)}\rangle\right] \quad (62)$$

Notice that in the coordinates $\left(\langle\tau_{D(A)}\rangle, \sigma\left[\tau_{D(A)}\right]\right)$, where $\sigma\left[\tau_{D(A)}\right] = \sqrt{Var\left[\tau_{D(A)}\right]}$ is the standard deviation of fluorescence lifetime, this expression take the simple form of positive semi-circle with the center point $\left(\frac{\tau^{(2)}+\tau^{(1)}}{2}, 0\right)$ and radius $\left|\frac{\tau^{(2)}-\tau^{(1)}}{2}\right|$:

**Binary fixed-dyes FRET-lines (explicit for free $x^{(1)}$) :**

$$Var\left[\tau_{D(A)}\right]\left(\langle\tau_{D(A)}\rangle\right) = \Delta_\tau^2 - \left[\langle\tau_{D(A)}\rangle - \tau_c\right]^2; \quad \tau_c = \frac{\tau^{(2)} + \tau^{(1)}}{2}; \quad \Delta_\tau = \frac{\tau^{(2)} - \tau^{(1)}}{2} \quad (63)$$

*The static FRET-line*

The calculation of static FRET-line for the distribution $p\left(R_{DA}|R_{MP}^{(c)}; \sigma^{(l)}\right)$ with finite width $\sigma^{(l)}$ gives:

**Static FRET-lines:**

$$\tau_{D(A)}^{(l,\nu)}\left(R_{MP}^{(c)}, \sigma^{(l)}\right) = \int \tau_{D(A)}^{\nu}(R_{DA})\, p\left(R_{DA}|R_{MP}^{(c)}; \sigma^{(l)}\right) dR_{DA}, \; \nu = \{1,2\}$$

$$\begin{cases} Var^{(l)}\left[\tau_{D(A)}\right]\left(R_{MP}^{(c)}, \sigma^{(l)}\right) = \tau_{D(A)}^{(l,2)}\left(R_{MP}^{(c)}, \sigma^{(l)}\right) - \left[\tau_{D(A)}^{(l,1)}\left(R_{MP}^{(c)}, \sigma^{(l)}\right)\right]^2 \\ \langle\tau_{D(A)}\rangle^{(l)}\left(R_{MP}^{(c)}, \sigma^{(l)}\right) = \tau_{D(A)}^{(l,1)}\left(R_{MP}^{(c)}, \sigma^{(l)}\right) \end{cases} \quad (64)$$

To continue with the calculation of the binary dynamic FRET line for discrete conformational states with $R_{MP}^{(c)} = \{R_{MP}^{(1)}, R_{MP}^{(2)}\}$ and linker parameter $\sigma^{(l)}$, we only require the two points $(\tau_1^{(l,1)}, v^{(l,1)})$ and $(\tau_1^{(l,2)}, v^{(l,2)})$ of the static line, corresponding to the positions of the two conformational states, where $v^{(l,i)}$ is the linker variance of state $i$:

**Static FRET-points for two conformational states:**

$$\begin{cases} v^{(l,1)} = Var^{(l)}\left[\tau_{D(A)}\right]\left(R_{MP}^{(1)}, \sigma^{(l)}\right); \\ \tau^{(l,1)} = \langle\tau_{D(A)}\rangle^{(l)}\left(R_{MP}^{(1)}, \sigma^{(l)}\right); \end{cases} \quad \begin{cases} v^{(l,2)} = Var^{(l)}\left[\tau_{D(A)}\right]\left(R_{MP}^{(2)}, \sigma^{(l)}\right); \\ \tau^{(l,2)} = \langle\tau_{D(A)}\rangle^{(l)}\left(R_{MP}^{(2)}, \sigma^{(l)}\right); \end{cases} \quad (65)$$

Then, the expression for the mean of the linker-variances over the conformational states takes the form:

$$\overline{Var^{(l)}\left[\tau_{D(A)}\right]}^{(c)}\left(x^{(1)}, R_{MP}^{(1)}, R_{MP}^{(2)}, \sigma^{(l)}\right) = x^{(1)}v^{(l,1)} + (1 - x^{(1)})v^{(l,2)} \quad (66)$$

***Dynamic FRET-lines***

By substituting the values of $\tau_i^{(l)}$ from eq. (65) into the expression for fixed dyes (eq. (61)) and by adding the average variance (eq. (66)) according to eq.(56), we obtain the final expressions for the binary dynamic FRET-line:

**Binary dynamic FRET-lines in presence of linkers:**

$$\begin{cases} Var[\tau_{D(A)}](x^{(1)}) = x^{(1)}(1-x^{(1)})(\tau^{(l,2)}-\tau^{(l,1)})^2 + \\ \qquad + x^{(1)}v^{(l,1)} + (1-x^{(1)})v^{(l,2)} \\ \langle\tau_{D(A)}\rangle(x^{(1)}) = x^{(1)}\tau^{(l,1)} + (1-x^{(1)})\tau^{(l,2)} \end{cases} \quad (67)$$

The explicit FRET-line equation for the free fraction parameter $x^{(1)}$ takes the form:

**Binary dynamic FRET-lines in presence of linkers (explicit for free x⁽¹⁾):**

$$Var[\tau_{D(A)}](\langle\tau_{D(A)}\rangle) = \\ = [\tau^{(l,2)} - \langle\tau_{D(A)}\rangle]\left[\langle\tau_{D(A)}\rangle - \tau^{(l,1)} + \frac{v^{(l,1)} - v^{(l,2)}}{\tau^{(l,2)} - \tau^{(l,1)}}\right] + v^{(l,2)} \quad (68)$$

Eq. (68) in the $(\langle\tau_{D(A)}\rangle, Var[\tau_{D(A)}])$-coordinates describes the segment of a circle connecting the points $(\tau^{(l,1)}, v^{(l,1)})$ and $(\tau^{(l,2)}, v^{(l,2)})$. To highlight its shape this equation can be rewritten in the form:

**Binary dynamic FRET-lines in presence of linkers (explicit for free x⁽¹⁾):**

$$Var[\tau_{D(A)}](\langle\tau_{D(A)}\rangle) = \Delta_\tau^2 - [\langle\tau_{D(A)}\rangle - \tau_c]^2;$$

$$\tau_c = \tau_c^{(l)} + \frac{1}{2}\frac{\Delta_v^{(l)}}{\Delta_\tau^{(l)}} \; ; \; \Delta_\tau^2 = \Delta_\tau^{(l)2} + v_c^{(l)} + \left(\frac{1}{2}\frac{\Delta_v^{(l)}}{\Delta_\tau^{(l)}}\right)^2, \quad (69)$$

where:

$$\tau_c^{(l)} = \frac{\tau^{(l,2)} + \tau^{(l,1)}}{2}; \; \Delta_\tau^{(l)} = \frac{\tau^{(l,2)} - \tau^{(l,1)}}{2};$$

$$v_c^{(l)} = \frac{v^{(l,2)} + v^{(l,1)}}{2}; \; \Delta_v^{(l)} = \frac{v^{(l,2)} - v^{(l,1)}}{2}$$

Eq. (69) is the generalization of eq. (62). If the dyes are fixed, the static line reduces to the pure states line, as the variances of the fluorescence lifetime distributions in the two conformational states turn to zero and the linker-mean lifetimes take the corresponding fixed values. In other words, the two static line points $(\tau^{(l,1)}, v^{(l,1)})$ and $(\tau^{(l,2)}, v^{(l,2)})$ turn into points $(\tau^{(1)}, 0)$ and $(\tau^{(2)}, 0)$ and eq. (69) takes the form of eq. (62).

It should be noted that the procedure of linker-correction of dynamical FRET-lines outlined here does not require the knowledge of the whole static line. The linker means and variances of the fluorescence lifetime distribution need to be known only for the conformational states involved in the exchange. Thus, in principle, this procedure can be applied even for the cases when the linker distance distribution is not known in the analytical form. For example, the distributions of the dyes positions can be modeled as accessible volumes (AV) or estimated based on the knowledge about the local environment of the dyes on the molecule, in which case eq. (69) is still valid.

## Supplementary Note 4: Geometric determination of species fractions in multi-state systems

Here, we show that the fraction of a given species in three-state system can be obtained from the sections of the line connection the pure state and the mixture. In the example shown in Figure SN4.1, the species fraction of state 2 is obtained from the ratio of the magnitude of the vectors $\boldsymbol{p}_{13\to123}$ and $\boldsymbol{p}_{13\to2}$ as:

$$x^{(2)} = \frac{|\boldsymbol{p}_{13\to123}|}{|\boldsymbol{p}_{13\to2}|} = \frac{|\boldsymbol{p}_{123} - \boldsymbol{p}_{13}|}{|\boldsymbol{p}_2 - \boldsymbol{p}_{13}|} \tag{70}$$

Consider that the vector $\boldsymbol{p}_{123}$ is given by the linear combination of the pure state vectors weighted by the species fractions $x^{(i)}$ ($\sum x^{(i)} = 1$):

$$\boldsymbol{p}_{123} = x^{(1)}\boldsymbol{p}_1 + x^{(2)}\boldsymbol{p}_2 + x^{(3)}\boldsymbol{p}_3 \tag{71}$$

We can rewrite this equation as:

$$\begin{aligned}\boldsymbol{p}_{123} &= \left(x^{(1)}\boldsymbol{p}_1 + x^{(3)}\boldsymbol{p}_3\right) + x^{(2)}\boldsymbol{p}_2 \\ &= \frac{1}{x^{(1)} + x^{(3)}}\left(x^{(1)}\boldsymbol{p}_1 + x^{(3)}\boldsymbol{p}_3\right)\left(x^{(1)} + x^{(3)}\right) + x^{(2)}\boldsymbol{p}_2 \\ &= \boldsymbol{p}_{13}\left(1 - x^{(2)}\right) + x^{(2)}\boldsymbol{p}_2 = \boldsymbol{p}_{13} + x^{(2)}(\boldsymbol{p}_2 - \boldsymbol{p}_{13})\end{aligned} \tag{72}$$

where we have used the following relations:

$$\boldsymbol{p}_{13} = \frac{1}{x^{(1)} + x^{(3)}}\left(x^{(1)}\boldsymbol{p}_1 + x^{(3)}\boldsymbol{p}_3\right) \tag{73}$$

$$\left(x^{(1)} + x^{(3)}\right) = 1 - x^{(2)} \tag{74}$$

We thus obtain the following expression for $\boldsymbol{p}_{123}$:

$$\boldsymbol{p}_{123} = \boldsymbol{p}_{13} + x^{(2)}\boldsymbol{p}_{13\to2} \tag{75}$$

which is equivalent to equation (70) above:

$$\boldsymbol{p}_{123} - \boldsymbol{p}_{13} = \boldsymbol{p}_{13\to123} = x^{(2)}\boldsymbol{p}_{13\to2} \tag{76}$$

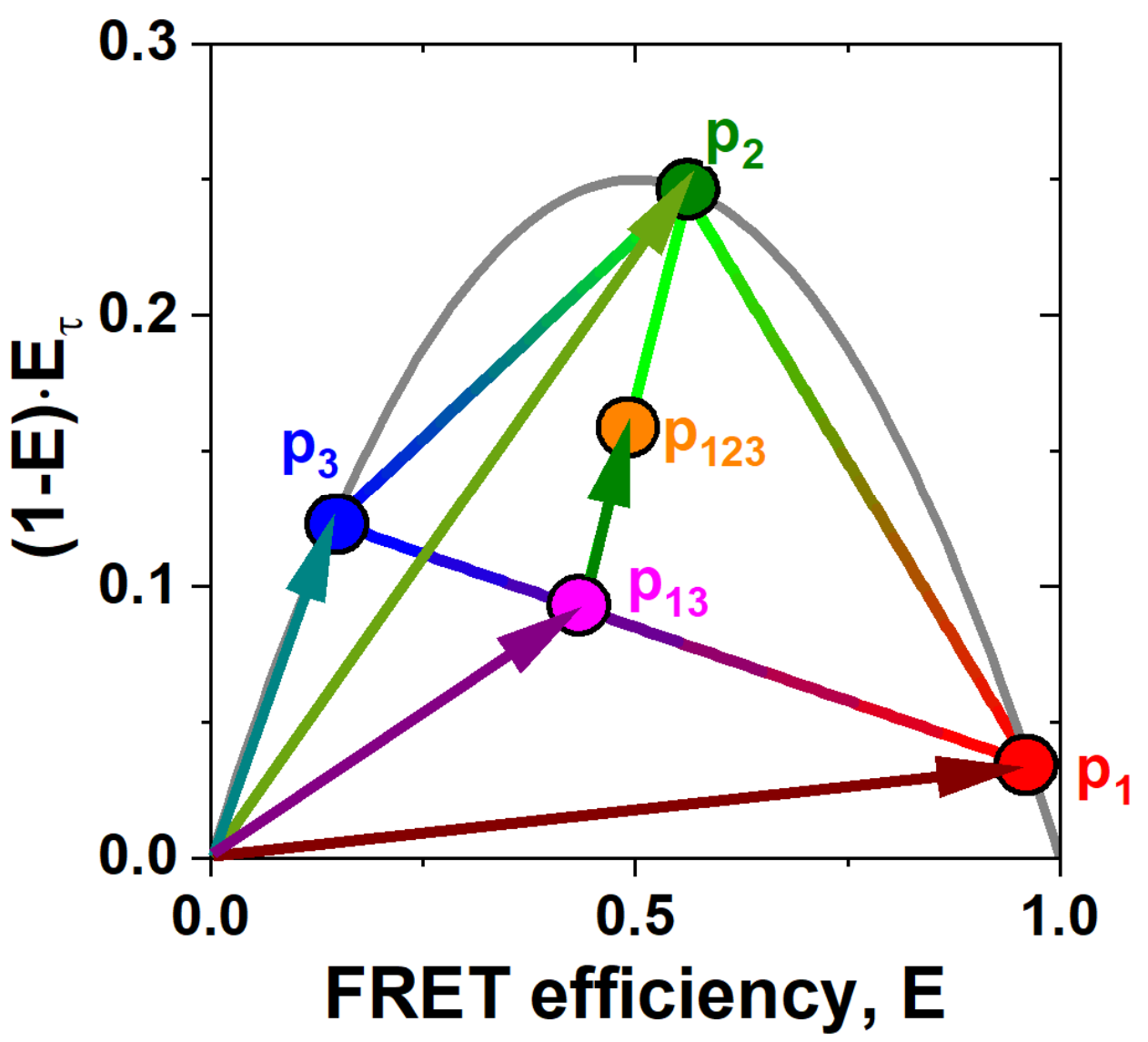

**Figure SN4.1:** Extracting equilibrium fractions for a three-state system. The vector of the mixed population in the moment representation, $p_{123}$, is given by the weighted sum of the individual components $p_1$, $p_2$ and $p_3$.

## Supplementary Note 5: FRET-lines for mixtures of photophysical states

In this section, we derive expressions for FRET-lines for distributions of the photophysical properties of the donor and acceptor fluorophores. Specifically, we consider the case where such a distribution is caused by quenching, i.e. the presence of non-radiative pathways of de-excitation, leading to different donor and acceptor fluorescence lifetimes.

We use the following notation for the involved states:

| | | | |
|---|---|---|---|
| Donor only states: | $\Lambda_D$; | $p(\Lambda_D)$; | $\Phi_{F,D}(\Lambda_D)$; $\tau_{D(0)}(\Lambda_D)$; |
| Acceptor only states: | $\Lambda_A$; | $p(\Lambda_A)$; | $\Phi_{F,A}(\Lambda_A)$; |
| FRET states: | $\Lambda_{DA}$; | $p(\Lambda_{DA})$; | $k_{RET}(\Lambda_{DA})$ |

where $\Phi_F$ is the fluorescence quantum yield, $\tau_{D(0)}$ the donor fluorescence lifetime in the absence of the acceptor and $k_{RET}$ is the FRET rate constant. The donor only lifetime relates to the quantum yield by:

$$\tau_{D(0)}(\Lambda_D) = \tau_{F,D}\Phi_D(\Lambda_D)$$

where $\tau_{F,D} = 1/k_{F,D}$ is the radiative lifetime of the donor. The integrated signals, i.e. the fluorescence intensities, of the donor, $F_{D|D}^{(DA)}$, and acceptor, $F_{A|D}^{(DA)}$, after donor excitation are then given by:

$$F_{D|D}^{(DA)}(\Lambda_D, \Lambda_{DA}) = k_{F,D}\Phi_D(\Lambda_D)\big(1 - E(\Lambda_D, \Lambda_{DA})\big) = k_{F,D}\, \tau_D(\Lambda_D, \Lambda_{DA});$$

$$F_{A|D}^{(DA)}(\Lambda_A, \Lambda_D, \Lambda_{DA}) = k_{F,D}\Phi_A(\Lambda_A)E(\Lambda_D, \Lambda_{DA}) = k_{F,D}\, \gamma'(\Lambda_A, \Lambda_D)\, \big(\tau_{DO}(\Lambda_D) - \tau_D(\Lambda_D, \Lambda_{DA})\big);$$

where:

$$E(\Lambda_D, \Lambda_{DA}) = 1 - \frac{\tau_{D(A)}(\Lambda_D, \Lambda_{DA})}{\tau_{D(0)}(\Lambda_D)}$$

$$\tau_{D(A)}(\Lambda_D, \Lambda_{DA}) = \frac{1}{k_{RET}(\Lambda_{DA}) + \frac{1}{\tau_{D(0)}(\Lambda_D)}}$$

$$\gamma'(\Lambda_A, \Lambda_D) = \frac{\Phi_{F,A}(\Lambda_A)}{\Phi_{F,D}(\Lambda_D)}$$

We define the proximity ratio $E_{PR}$ analogous to the FRET efficiency in the case of a single photophysical state of the donor and acceptor fluorophores (eq. 11 in the main text). For one set of donor, acceptor and FRET states, $\{\Lambda_A, \Lambda_D, \Lambda_{DA}\}$, the proximity ratio $E_{PR}$ is defined as:

$$E_{PR}(\Lambda_A, \Lambda_{DA}) = 1 - \frac{F_{D|D}^{(DA)}(\Lambda_D, \Lambda_{DA})}{F_{D|D}^{(DA)}(\Lambda_D, \Lambda_{DA}) + F_{A|D}^{(DA)}(\Lambda_A, \Lambda_D, \Lambda_{DA})}$$

$$= 1 - \frac{\tau_{D(A)}(\Lambda_D, \Lambda_{DA})}{\tau_{D(A)}(\Lambda_D, \Lambda_{DA}) + \gamma'(\Lambda_A, \Lambda_D)\, \big(\tau_{D(0)}(\Lambda_D) - \tau_{D(A)}(\Lambda_D, \Lambda_{DA})\big)}$$

For a mixture of different states, the corresponding experimental observable is the proximity ratio $E_{PR}$ from calculated from the expectation values of the signals $F_{DD}^{(DA)}$ and $F_{AD}^{(DA)}$:

$$\boxed{\begin{aligned} &E_{PR} = 1 - \frac{\overline{F_{DD}^{(DA)}}}{\overline{F_{DD}^{(DA)}} + \overline{F_{AD}^{(DA)}}} = 1 - \frac{\overline{\tau_D(\Lambda_D, \Lambda_{DA})}}{\overline{\tau'_{D(0)}(\Lambda_A, \Lambda_D, \Lambda_{DA})}}; \\ &\tau'_{D(0)}(\Lambda_A, \Lambda_D, \Lambda_{DA}) = \tau_{D(A)}(\Lambda_D, \Lambda_{DA}) + \gamma'(\Lambda_A, \Lambda_D)\, \big(\tau_{D(0)}(\Lambda_D) - \tau_{D(A)}(\Lambda_D, \Lambda_{DA})\big), \end{aligned}}$$

where $\tau'_{D(0)}$ is the *effective donor-only lifetime* in the presence of quenching. Note that the quantity $E_{PR}$ is defined as the proximity ratio of the expected signals, which differs from the expected proximity ratio. The expression for the intensity-weighted average donor lifetime is identical compared for the single $\tau_{D(0)}$ case. It is given by the ratio of the second and first moments of the lifetime distribution, which now depends both on the donor and FRET states:

$$\boxed{\langle\tau_{D(A)}\rangle_F = \frac{\overline{\tau_{D(A)}^2(\Lambda_D,\Lambda_{DA})}}{\overline{\tau_{D(A)}(\Lambda_D,\Lambda_{DA})}}}$$

***Single γ correctable case***

As an example, we consider the case 1) that the donor, acceptor and FRET states are independent (homogeneous approximation) and 2) that there is only one donor state with $\Lambda_D = \Lambda_D^{(1)}$. In this case, the joint probability of donor, acceptor and FRET states, $p(\Lambda_A,\Lambda_D,\Lambda_{DA})$, is given by:

$$p(\Lambda_A,\Lambda_D,\Lambda_{DA}) = p(\Lambda_A)\delta\left(\Lambda_D - \Lambda_D^{(1)}\right)p(\Lambda_{DA})$$

The expectation of $F_{AD}^{(DA)}$ then takes the form:

$$\overline{F_{AD}^{(DA)}} = k_{F,D}\,\overline{\gamma'}\left(\tau_{D(0)} - \overline{\tau_{D(A)}}\right)$$

where:

$$\tau_{D(0)} = \tau_{D(0)}\left(\Lambda_D^{(1)}\right)$$

$$\overline{\gamma'} = \frac{1}{\Phi_{F,D}\left(\Lambda_D^{(1)}\right)}\int \Phi_{F,A}(\Lambda_A)p(\Lambda_A)d\Lambda_A$$

In this case, we can correct the measured proximity ratio by scaling the measured $F_{DD}^{(DA)}$ signal by the factor $\overline{\gamma'}$. This corrected proximity ratio $E^{(corr)}$ is then equal to the species-weighted average FRET efficiency over the distribution of FRET states $p(\Lambda_{DA})$.

$$E^{(corr)} = 1 - \frac{\overline{\gamma'}\overline{F_{DD}^{(DA)}}}{\overline{\gamma'}\overline{F_{DD}^{(DA)}} + \overline{F_{AD}^{(DA)}}} = 1 - \frac{\overline{\tau_{D(A)}}}{\tau_{D(0)}} = \bar{E} = E$$

The FRET-lines considered in the previous sections correspond to this homogeneous, single donor state approximation.

***Linearization of binary dynamical FRET-lines in the presence of quenching***

In this section, we provide a proof for the linearity of binary dynamic FRET lines in the moment representation in the presence of a distribution of donor states. Analogous to the expression for the moment difference Γ obtained in equation 44 in the main text, we define the variable $\Gamma^*$:

$$\Gamma\left(E^*,\langle\tau_{D(A)}\rangle_F,;\tau_S\right) = (1-E^*)\left(1 - \frac{\langle\tau_{D(A)}\rangle_F}{\tau_S}\right)$$

where $E^*$ may be either a FRET efficiency ($E$) or a proximity ratio ($E_{PR}$) and $\tau_S$ is an arbitrary normalization constant. Then, the binary dynamic lines in both the $(E,\Gamma^*)$ and $(E_{PR},\Gamma^*)$ representations are straight lines, as we will show below. Importantly, in the case of a distribution photophysical states of the dyes, the position of the observed population on the dynamic line does not correspond to the equilibrium fractions of the states anymore. However, the true equilibrium fraction $(x)$ is related to the observed apparent fraction $(\xi)$, as estimated from the position of the population of the dynamic FRET line, by the expression:

$$x(\xi) = \left(1 + \left(\frac{1}{\xi} - 1\right)\frac{\tau'^{(1)}_{D(0)}}{\tau'^{(2)}_{D(0)}}\right)^{-1},$$

where $\tau'^{(1)}_{D(0)}, \tau'^{(2)}_{D(0)}$ are mean values of the effective donor-only lifetime in states 1 and 2.

*Linearization of ratios of linear functions*

Let the functions $a(x), b(x), c(x)$ be linear functions of a mixing parameter (fraction) $x$. Then, they can be written in the form:

$$a(x) = x\,a_1 + (1-x)a_2 = x(a_1 - a_2) + a_2$$

Let the functions $g(x), h(x)$ be ratios of the functions $a(x), b(x), c(x)$ defined as:

$$g(x) = \frac{a(x)}{b(x)}$$

$$h(x) = \frac{c(x)}{a(x)}$$

Obviously, the functions $g(x), h(x)$ are not linear functions of the parameter $x$, and the same applies to the function $h(g)$.

Let's introduce the function $f(x)$ in the form:

$$f(x) = \big(1 - h(x)\big)g(x) = \frac{a(x) - c(x)}{b(x)}$$

Then, the function $f(g)$ is linear.

*Proof:*

Let's compute explicitly the function $f(g)$ by excluding the parameter $x$ from the parametric couple $\{g(x), f(x)\}$.

$$g(x) = \frac{x(a_1 - a_2) + a_2}{x(b_1 - b_2) + b_2} \quad \Rightarrow \quad x(g) = \frac{a_2 - g\, b_2}{(b_1 - b_2) - (a_1 - a_2)}$$

The substitution of $x(g)$ into $f(x)$:

$$f(x) = \frac{x(a_1 - c_1 - a_2 + c_2) + a_2 - c_2}{x(b_1 - b_2) + b_2} \;,$$

gives the function $f(g)$ in the form:

$$f(g) = f\big(x(g)\big) = f_0 + k\, g \;,$$

where

$$k \;=\; \frac{\dfrac{a_1 - c_1}{b_1} - \dfrac{a_2 - c_2}{b_2}}{\dfrac{a_1}{b_1} - \dfrac{a_2}{b_2}} \;=\frac{f_1 - f_2}{g_1 - g_2};$$

$$f_0 = \frac{\left(1 - \frac{c_1}{a_1}\right) - \left(1 - \frac{c_2}{a_2}\right)}{\dfrac{b_1}{a_1} - \dfrac{b_2}{a_2}} = \frac{\dfrac{f_1}{g_1} - \dfrac{f_2}{g_2}}{\dfrac{1}{g_1} - \dfrac{1}{g_2}},$$

which is the equation of the line going through points $\{g_1, f_1\}, \{g_2, f_2\}$.

Note that both functions in the "linearized" pair $g(x), f(x)$ have the same denominator $b(x)$. We can write:

$$g(x) = \frac{x\, a_1 + (1 - x)a_2}{b(x)} = \frac{x\, b_1 g_1 + (1 - x)b_2 g_2}{b(x)} = \underbrace{\frac{x\, b_1}{b(x)}}_{\xi(x)} g_1 + \underbrace{\left(1 - \frac{x\, b_1}{b(x)}\right)}_{1-\xi(x)} g_2$$

Thus, if $x$ is the parameter mixing values of functions $a(x), b(x), c(x)$ in states 1 and 2 , then the parameter mixing corresponding values of ratio functions $g(x), f(x)$ is:

$$\xi(x) = \frac{1}{1 + \left(\frac{1}{x} - 1\right)\frac{b_2}{b_1}}$$

This relation can be inverted:

$$x(\xi) = \frac{1}{1 + \left(\frac{1}{\xi} - 1\right)\frac{b_1}{b_2}}$$

*Application to dynamic FRET-lines*

To prove linearity of binary dynamic FRET-lines in the moment representation, we first observe that we can normalize all lifetimes in the parametric equation for the FRET-lines by any convenient constant $\tau_S$:

$$E^* = 1 - \frac{\overline{\left(\frac{\tau_{D(A)}}{\tau_S}\right)}}{\overline{\left(\frac{\tau'_{D(0)}}{\tau_S}\right)}} = 1 - \frac{\overline{\hat{\tau}_{D(A)}}}{\overline{\hat{\tau}'_{D(0)}}}$$

$$\frac{\langle \tau_{D(A)} \rangle_F}{\tau_S} = \frac{\overline{\left(\frac{\tau_{D(A)}}{\tau_S}\right)^2}}{\overline{\left(\frac{\tau_{D(A)}}{\tau_S}\right)}} = \frac{\overline{\hat{\tau}^2_{D(A)}}}{\overline{\hat{\tau}_{D(A)}}}$$

Then we can take:

$$\begin{aligned}
a(x) &\rightarrow \overline{\hat{\tau}_{D(A)}}(x); \\
b(x) &\rightarrow \overline{\hat{\tau}'_{D(0)}}(x); \\
c(x) &\rightarrow \overline{\hat{\tau}^2_{D(A)}}(x); \\
g(x) &\rightarrow 1 - E^*(x); \\
h(x) &\rightarrow \frac{\langle \tau_{D(A)} \rangle_F(x)}{\tau_S}; \\
f(x) &\rightarrow \Gamma^*(x) = \left(1 - \frac{\langle \tau_{D(A)} \rangle_F(x)}{\tau_S}\right)(1 - E^*(x)),
\end{aligned}$$

and use all results from previous section.

## Supplementary Note 6:Worm-like chain model

Semi-flexible macromolecules are generically well described by the worm-like chain model. However, no closed-form analytical solution for the radial distribution function $q(r)$ of the worm-like chain model is available. Here, we used an approximation as presented in reference [5]:

$$q(r) \propto \left(\frac{1 - c\,r^2}{1 - r^2}\right)^{\frac{5}{2}} \exp\left(\frac{\sum_{i=-1}^{0}\sum_{j=1}^{3} c_{ij}\kappa^i r^{2j}}{1 - r^2}\right) \exp\left(-\frac{d\,\kappa\,a\,b\,(1+b)\,r^2}{1 - b^2 r^2}\right) I_0\left(-\frac{d\,\kappa\,a\,(1+b)\,r}{1 - b^2 r^2}\right) \qquad (77)$$

with

$$a \cong 14.054$$
$$b \cong 0.473$$
$$c \cong 1 - (1 + (0.38\,\kappa^{-0.95})^{-5})^{-1/5}$$
$$d \cong \begin{cases} 1 & , \kappa < 1/8 \\ 1 - \dfrac{1}{\dfrac{0.177}{\kappa - 0.111} + 6.40(k - 0.111)^{0.783}} & , \kappa \geq 1/8 \end{cases}$$
$$\begin{pmatrix} c_{-1,1} & c_{-1,2} & c_{-1,2} \\ c_{0,1} & c_{0,2} & c_{0,3} \end{pmatrix} = \begin{pmatrix} -\frac{3}{4} & \frac{23}{64} & -\frac{7}{64} \\ -\frac{1}{2} & \frac{17}{16} & -\frac{9}{16} \end{pmatrix}$$
$I_0(x)$ – Modified Bessel function of order 0

Here, $a$, $b$, $c$ and $d$ are numerically determined constants, $\kappa = l_p/L$ is the chain stiffness, $l_p$ is the persistence length and $L$ is the chain length. The radial function is defined in the range from 0 to $L$. The distance distribution function $p(R_{DA})$ is obtained from the radial distribution function $q(r)$ above by replacing $r \rightarrow R_{DA}$ and multiplication by $R_{DA}^2$:

$$p(R_{DA}) \propto R_{DA}^2\, q(R_{DA}) \qquad (78)$$

Using this distribution, dynamic FRET lines can be generated for the cases when macromolecule is in either folded or unfolded state.